\documentclass[bibyear]{aa}  

\usepackage{graphicx}
\usepackage{natbib}
\usepackage{amsmath}
\usepackage{nccmath}

\usepackage[flushleft]{threeparttable}

\usepackage[colorlinks, citecolor=blue]{hyperref}
\usepackage{xcolor}
\usepackage{etoolbox}
\setcitestyle{authoryear,open={},close={}}

\hypersetup{
    colorlinks,
    linkcolor={blue!80!black},
    citecolor={blue!80!black},
    urlcolor={blue!80!black}
}

\usepackage{txfonts}
\begin{document}

   \title{Internal water storage capacity of terrestrial planets and the effect of hydration on the M-R relation}

   \author{O. Shah\inst{1,2}
          \and
          Y. Alibert\inst{1}
          \and
          R. Helled\inst{2}
          \and
          K. Mezger\inst{1,3}
          }
 
   \institute{Center for Space and Habitability, Gesellschaftsstrasse 6, Universität Bern, 3012 Bern, Switzerland
         \and
            Institute for Computational Science \& Center for Theoretical Astrophysics and Cosmology, Universität Zürich, Winterthurerstrasse 190, 8057 Zürich, Switzerland
        \and
        Institut für Geologie , Universität Bern, Baltzerstrasse 3, 3012 Bern, Switzerland
             }

   \date{Received July 2020; Accepted December 2020}

 
  \abstract
   {The discovery of low density exoplanets in the super-Earth mass regime suggests that ocean planets could be abundant in the galaxy. Understanding the chemical interactions between water and Mg-silicates or iron is essential for constraining the interiors of water-rich planets. Hydration effects have, however, been mostly neglected by the astrophysics community so far. As such effects are unlikely to have major impacts on theoretical mass-radius relations, this is justified as long as the measurement uncertainties are large. However, upcoming missions, such as the PLATO mission (scheduled launch 2026), are envisaged to reach a precision of up to $\approx 3 \%$ and $\approx 10 \%$ for radii and masses, respectively. As a result, we may soon enter an area in exoplanetary research where various physical and chemical effects such as hydration can no longer be ignored.} 
   {Our goal is to construct interior models for planets that include reliable prescriptions for hydration of the cores and mantles. These models can be used to refine previous results for which hydration has been neglected and to guide future characterization of observed exoplanets.}
   {We have developed numerical tools to solve for the structure of multi-layered planets with variable boundary conditions and compositions. Here we consider three types of planets: dry interiors, hydrated interiors, and dry interiors plus surface ocean, where the ocean mass fraction corresponds to the mass fraction of the $\rm H_2 O$ equivalent in the hydrated case.}
   {We find H and OH storage capacities in the hydrated planets equivalent to $0-6 \rm \ wt\% \ \rm H_{2}O$ corresponding to up to $\approx 800 \rm \  km$ deep ocean layers. In the mass range $0.1 \leq M/M_\oplus \leq 3$, the effect of hydration on the total radius is found to be $\leq 2.5\%$, whereas the effect of separation into an isolated surface ocean is $\leq 5 \ \%$. Furthermore, we find that our results are very sensitive to the bulk composition.}
   {}

   \keywords{planets and satellites: composition -- planet and satellites: oceans -- planets and satellites: interiors -- planets and satellites: terrestrial planets
               }

   \maketitle
%

\section{Introduction}\label{sec:intro}

Water is abundant in the Universe; it is thought to have played an essential role in the onset of biochemistry on Earth and is indispensable for the sustainability of life as we know it (\cite{Kotwicki1991, Kasting1993, Garrido2001, Mottl2007, Jones2010, Guedel2014}). These notions have led to a widely used definition of the classical habitable zone around a star. This zone is defined as the orbital region within which liquid water could exist on a planet's surface in terms of the stellar flux it receives and assuming simple greenhouse effects mediated mainly by $\rm CO_2$ and $\rm H_2 O$ and maintained by the carbon cycle (\cite{Huang1959, Rasool1970, Kasting1993, Kaltenegger2011, Ramirez2018}). This crude definition, although very useful as a first approach, cannot capture the full complexity of possible habitats as many other effects may be equally important for habitability.  
These include more diverse atmospheric compositions and chemical interactions, complex geochemical recycle processes, protection against harmful high energy particles from the host star via for example magnetic fields, and influences of the galactic neighbourhood (\cite{Ramirez2018, Pierrehumbert2011, Gonzalez2001, Gonzalez2005, Morales2011, Gallet2017, Javaux2010}). 
Nevertheless, the presence of liquid water on a planet's surface is still commonly used as a proxy to constrain the orbital region around a star where life could exist. Hence, water is of the utmost interest for both astrophysics and astrobiology and has attracted increasing attention from the planetary science community in recent decades (for example \cite{Kuchner2003, Leger2004, Seager2007, Sotin2007, Noack2016, Alibert2014, Alibert2016, Fu2010, Kitzmann2015, Noack2017, Levi2017}). 
It has been hypothesized that planets that formed beyond the ice line in the protoplanetary disk could migrate inwards into the habitable zone, where the accreted water ices would melt and form gigantic liquid surface oceans (\cite{Kuchner2003, Leger2004}). Indeed, in recent decades a large number of planets in the mass-range $M \lesssim 10 M_{\oplus}$ have been detected with mean densities that are too low to correspond to a purely rocky composition (\cite{Rogers2015, Lozovsky2018}). These objects are not massive enough to have accreted substantial H-He envelopes in their past (\cite{Selsis2007, Sotin2007, Seager2007, Leger2004}). This may suggest that their low densities are the results of extensive amounts of water on their surfaces. 
Such oceans could be maintained over timescales of several gigayears, long enough to allow for the emergence of life (\cite{Kuchner2003, Noack2016}). However, to sustain a biosphere, stable long-term climate conditions are required. For instance, it has been proposed by \cite{Laskar1993} that the stabilization of the Earth's obliquity is crucial to avoid dramatic changes in climate that could have fatal consequences for life. These results were challenged later on by \cite{Lissauer2011}, who found that the presence of a large moon has only a minor influence on the evolution of the obliquity for timescales of hundreds of millions of years. Furthermore, \cite{Armstrong2014} found large variations of obliquity to even extend the outer edge of the habitable zone. Consequently, on Earth, long term climate stability is probably more strongly supported by a negative feedback mediated by the carbon cycle between the atmosphere and the mantle (\cite{Walker1981, Post1990, Kasting1993}). The presence of such a geochemical cycle has often been assumed to be a requirement for the long term sustainability of life on other worlds as well.
However, the interior dynamics of ocean planets could be quite different from those of water depleted planets like the Earth. It is still debated whether geochemical cycles, such as the carbon cycle, could be significant on water worlds. The mantles of these planets are expected to be isolated from the gas envelopes by liquid water layers and further shielded by high pressure ice layers at the bottom of the ocean. As a result, the presence of massive surface oceans could inhibit chemical transport mechanisms between the atmosphere and the interior (\cite{Alibert2014, Kitzmann2015, Noack2016, Abbot2012, Wordsworth2013, Nakayama2019}).
\\

Apart from its astrobiological significance, water is also an important ingredient in many solar system bodies including the ice giants Uranus and Neptune and the icy satellites of Jupiter and Saturn. The exact role of water in internal processes and its distribution in the interiors of these objects are subjects of active research efforts and there are many remaining open questions. The interior structures of Uranus and Neptune, for instance, are still unknown (see review by \cite{Helled2020}). A number of theoretical studies show that various structure models with different assumed compositions can match the gravitational moments measured by the Voyager 2 flybys (for example \cite{Helled2020} and references therein). These studies further suggest that it is possible that the 'ice giants' contain substantially lower ice to silicate ratios than is commonly assumed, and that their interiors could be rock dominated (\cite{Helled2011}). 
Uranus could even have solar ice to rock ratios (\cite{Nettelmann2016}). It is debated whether the rocks and ices are separated into differentiated layers or form a mixed interior with gradual composition gradients (\cite{Helled2011, Nettelmann2013}). In the mixed case chemical interactions between the Mg-silicates in the mantle or the iron in the core and the water ices could be important factors in further constraining the composition and thermal profiles of these planets. However, such hydration effects in the interiors are generally not included in such studies. A detailed investigation of these aspects could significantly improve our understanding of the interiors of water rich solar system planets and would allow for a more reliable exoplanet characterization in the future.
\\

In the context of exoplanets, interior characterization is aided by rather limited data. The planetary radii and masses along with some stellar properties are the primary parameters that are remotely accessible for planets outside the solar system. From these measurements it is possible to gain insight into their compositions and hence formation and evolution histories. Early efforts to link observed masses and radii with theoretical models for the interior compositions have been presented for example by \cite{Leger2004}, \cite{Seager2007} and \cite{Sotin2007} over a decade ago. It is also worth noting the pioneering work by \cite{Zapolsky1969} long before the discovery of the first exoplanet. 
These authors have developed simple one-dimensional structure models to compute the total radii of planets as a function of mass, composition, and surface conditions. It has become common practice over the past years to use such models to interpret measured planetary masses and radii and predict possible interior structures and compositions. 
While such early approaches are quite powerful for inferring the general landscape of planetary properties, these models are highly degenerate. That is, many different internal compositions and profiles could match an observed $M$-$R$ pair. This degeneracy has been noticed early on (for example \cite{Adams2007}) and many researchers have devoted their work to minimize the possible parameter space for planetary interior properties for a given set of observed quantities (for example \cite{Rogers2010, Dorn2015, Dorn2017, Grasset2009, Lozovsky2018}). 
In these studies terrestrial planets are assumed to be fully differentiated into a number of compositionally distinct layers with water being modelled as an isolated surface layer.  However, numerous studies over the past decades have shown that water can be incorporated in many minerals that are likely to be major constituents in the Earth's mantle. 
The upper mantle mineral $\rm (Mg,Fe)_2 Si O_4$
(Olivine polymorphs)  has been found to incorporate up to several wt$\%$ of water (see references in Tables \ref{tab:alpha}-\ref{tab:gamma}) in the correspondingly relevant temperature and pressure regime. 
At pressures $\approx$ 25-30 GPa, Ringwoodite (Rw), a high pressure polymorph of Olivine, dissociates into $\rm (Mg, Fe) O$ (Magnesiowüstite) and $ \rm (Mg, Fe) Si O_3$ (Perovskite) (\cite{Wang1997}). MgO can react with water to form $\rm Mg(OH)_2$ (Brucite) containing as much as  $\approx$ 31 wt $\% \ \rm H_2O$ at moderate temperatures and pressures of $\lesssim$ 1500 K and $\lesssim$ 35 GPa (\cite{Hermann2016}). \cite{Frost1998} has determined the stability of the dense hydrous Mg-silicate Phase D and found it to be stable up to 50 GPa. 
This phase can incorporate up to 18 $\rm wt\%$ of water. \cite{Nishi2014} have shown that Phase D transforms to an assemblage with another hydrous Mg-silicate, Phase H, at pressures above $\approx 48 \ \rm GPa$. They concluded that this phase could transport water in the form of hydrates into the lowermost mantle. Furthermore, Perovskite, which is thought to play a minor role for water storage in the lower mantle of Earth (\cite{Inoue2010, Casanova2005}), transforms into post-Perovskite (pPv) between $\approx$ 100-130 GPa and remains stable up to $\rm \approx 0.8-0.9 \  TPa$ (\cite{Umemoto2011}), approximately the relevant pressure regime in the interiors of Uranus and Neptune. 
Although the hydration behaviour of pPv remains unknown, density-functional theory (DFT) simulations suggest that stable configurations containing at least 2-3 $\rm wt \%$ water exist up to pressures of 150 GPa and possibly higher (\cite{Townsend2015}). Finally, Silica ($\rm SiO_2$), which is stable over the entire pressure range relevant for the interiors of small to giant planets (\cite{Umemoto2011}), has been reported to exhibit significant saturation contents of up to $\rm \approx 3.2 \ wt \%$ below 10 GPa and even up to $\approx \rm 8.4 \ wt \%$ for up to $\approx 100 \ \rm GPa$ (\cite{Nisr2020} and references therein). In addition to the hydration of Mg-Silicates in the mantle, hydrogen solubility in the iron core might be equally important for constraining the total internal storage capacity of $\rm H_2 O $ equivalent in terrestrial planets. Indeed, it has been found that the solubility of hydrogen into iron is strongly enhanced at high pressures, relevant for the interiors of these objects (\cite{Fukai1983, Sugimoto1992}).
Although hydration effects under high pressure conditions have been actively studied by many researchers over the past decades, the bulk water content of the mantle remains one of the most poorly constrained compositional parameters of the Earth (\cite{Townsend2015, Cowan2014}).
\\

The effects of hydration on the equations of state of the core and mantle materials could affect the thermal profiles and total radii of the planets. 
More importantly, the capacity of storing large amounts of H and OH in planetary interiors could lead to the distribution of the total amount of water between an internal reservoir and a surface ocean. For an assumed total mass fraction of $\rm H_2 O$ equivalent this would mean that the surface ocean, and hence the pressure at the mantle-ocean interface, could be considerably smaller than one would expect on a fully differentiated planet. Indeed, \cite{Cowan2014} have derived steady-state solutions to the water partitioning between hydrosphere and mantle on terrestrial planets and find, based on simplified assumptions for the mineral hydration, that super-Earths tend to store large amounts of water in their interiors. 
This can have relevant implications for the results of previous studies. 
For example, \cite{Sotin2007} have presented a model that takes the atomic ratios of Mg, Si and Fe in the interior as input parameters to generate mass-radius relations for given surface conditions. \cite{Grasset2009} have used this model to quantify the effect of a surface water ocean on the total radius of a planet with given bulk composition. They inferred that bulk composition plays only a secondary role for the total radius in comparison to the effect of the ocean mass fraction. The authors estimated that if the planetary mass and radius are precisely known, it is possible to determine the total amount of water with a standard deviation of 4.5 $\%$. \cite{Alibert2014} used a very similar model to compute the maximum radius of habitable planets assuming that habitability is hindered on water worlds by the formation of high pressure ice layers. It was concluded that for planets with 1-12 $M_{\oplus}$ this maximum radius is limited to 1.7-2.2 $R_{\oplus}$ and that larger planets are likely to be not habitable. 
These studies assume dry mantles and isolated surface oceans. However, in the light of the aforementioned hydration behaviour of Mg-silicates and iron, it is possible that these results are affected to a non-negligible extent when hydration is included. Understanding and quantifying these effects is essential given the increasing precision in mass and radius measurements for exoplanets. 
Although the mass and radius will never be precisely known, the uncertainties have decreased drastically over the past decades and will decrease further. To date, one of the most precise measurements of the radius of an exoplanet has been achieved for Kepler-93b (\cite{Ballard2014}). Its radius has been determined from transit observations to with an uncertainty of $\approx \pm 1.3 \ \%$. The PLATO mission, scheduled for launch in 2026, is expected to determine   the radii and masses of terrestrial planets with precisions up to $\approx 3 \ \%$ and $\approx 10 \ \%$, respectively (\cite{Rauer2016}). Hence, second-order effects that were justified to be excluded in the past are expected to become necessary ingredients for future planetary characterization.
\\

In this study we build upon the models presented by \cite{Sotin2007}, \cite{Grasset2009} and \cite{Alibert2014} and include the hydration of the mantle and the core as two separate reservoirs for H and OH and assess upper bounds for the corresponding effects on theoretical $M$-$R$ relations. Furthermore, we apply our model to planets up to $3 \ M_{\oplus}$ with simplified bulk compositions and show how the effects of hydration on $M$-$R$ relations can be estimated. The mass limit of $3 \ M_\oplus$ is dictated by the pressure range over which the core hydration model is valid. The generalization to more complex bulk compositions is briefly discussed.
\\
\begin{figure*}[t]
\includegraphics[width=\textwidth]{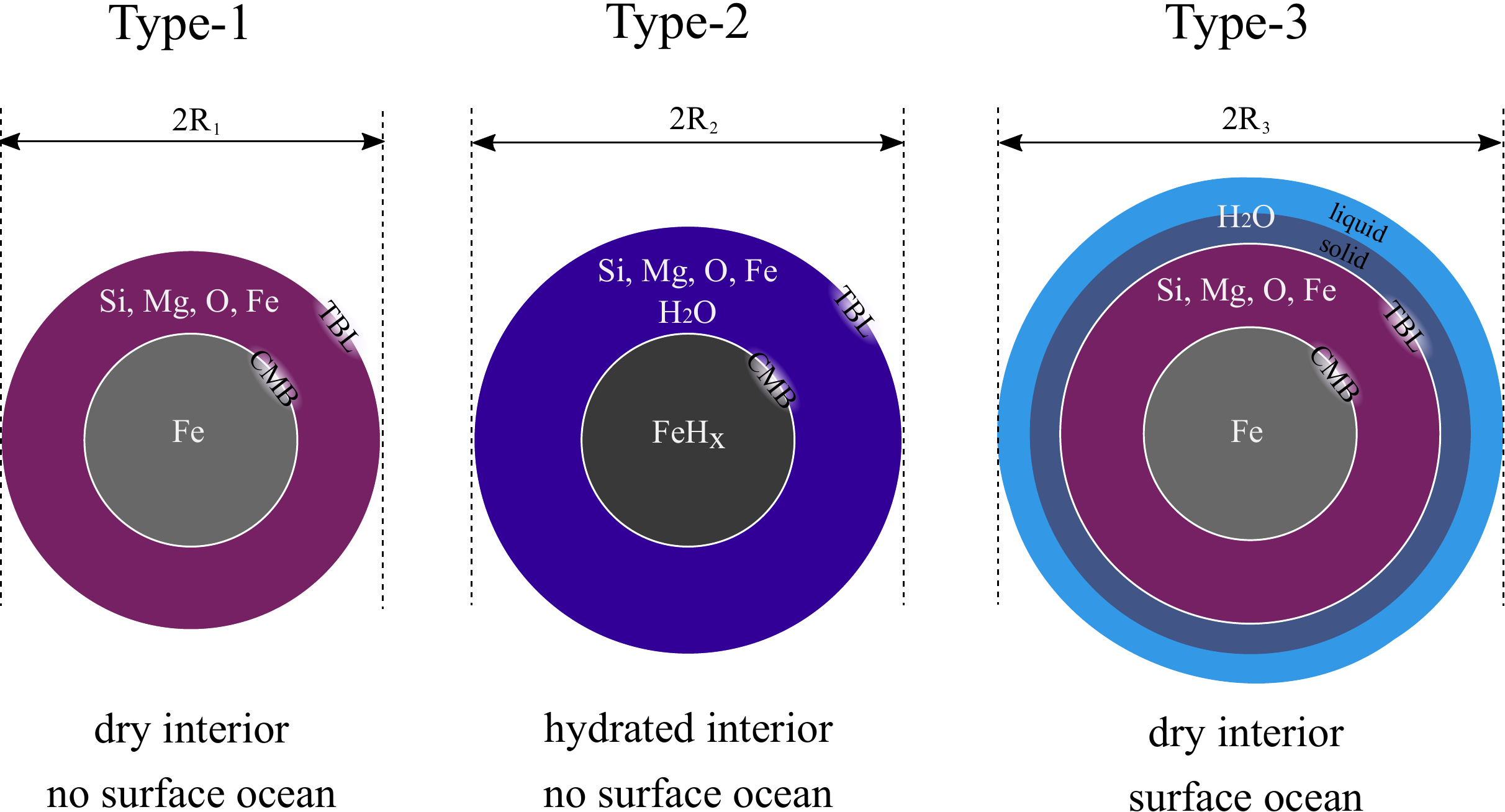}
\caption{Schematic overview of the different planet types modelled in this study. All objects consist of a pure iron or iron hydride core and a dry or hydrated Mg-silicate mantle. We have estimated the effect of hydration by comparing the dry planets (Type-1) with their hydrated counterparts (Type-2) at fixed composition and boundary conditions. We have then extracted the water content in the Type-2 planets and moved it into an isolated surface layer (Type-3) to assess the effect of ocean separation on the total radius at fixed composition and boundary conditions. The oceans of the Type-3 planets can consist of an upper liquid part and a lower solid part if the pressure at the bottom of the oceans is high enough to allow for the formation of high pressure ices.}
\label{fig:compare_counterparts}
\centering
\end{figure*}
\\

This paper is structured as follows: In Section~\ref{sec:planetmodel} we describe how the internal profiles of the planets are constructed for variable boundary conditions and compositions. In Section~\ref{sec:saturationmodel} we explain how the H and OH content in the Mg-silicates and the core is calculated as a function of pressure and temperature. The resulting estimates on the effect of hydration and differentiation on total planetary radii for different bulk compositions as a function of total planetary mass are discussed in Section~\ref{sec:results}. Caveats arising from some simplifying assumptions about the composition and hydration in our model are outlined in Section~\ref{sec:caveats} and a summary of the work is provided in Section~\ref{sec:conclusions}. 
\\
\section{Planetary model}\label{sec:planetmodel}

\subsection{The structure equations}
The internal temperature, pressure, and density profiles need to be constructed to compute theoretical mass-radius relations. To this end the well known one-dimensional structure equations for spherical objects in hydrostatic equilibrium are integrated:
\begin{ceqn}
\begin{align}\label{eq:dPdr}
    \frac{\rm d\mathit{P(r)}}{\rm d\mathit{r}} = - \frac{G m(r) \rho(r)}{r^2}
\end{align}
\begin{align}\label{eq:dmdr}
    \frac{\rm d \mathit{m(r)}}{\rm d\mathit{r}} = 4 \pi \rho(r) r^2
\end{align}

\begin{align}\label{eq:dTdr}
    \frac{\rm d\mathit{T(r)}}{\rm d \mathit{r}} = - \frac{G m(r) \rho(r)}{r^2}\left(\frac{\rm d \mathit{T}}{\rm d\mathit{P}}\right)_{ad}
\end{align}
\begin{equation}\label{eq:rho}
    \rho (r) = \rho(T(r), P(r)).
\end{equation}
\end{ceqn}
Here, $P(r), T(r)$, $\rho(r)$, and $m(r)$ are the pressure, temperature, density, and enclosed mass at radial distance $r$, respectively, $G$ is the gravitational constant and $(dT/dP)_{ad}$ is the adiabatic gradient. To compute the density as a function of pressure and temperature adequate equations of state need to be employed (see Appendix \ref{sec:theos}). The hydration of the mantle and the core, hereafter referred to in terms of mass fraction $X_{\rm H_2O}$ or the molar hydrogen content $\xi_{\rm H}$ in the Mg-silicates and the iron is incorporated into the EoS (see Section \ref{sec:saturationmodel}).
\\

The system of differential equations given by eq. \eqref{eq:dPdr} - \eqref{eq:dTdr} can readily be solved using a standard 4th order Runge-Kutta scheme with adaptive integration step size $\rm d\mathit{r}$:
\begin{ceqn}
\begin{equation}\label{eq:dr}
    \rm d \mathit{r \equiv \epsilon_r min\left( \frac{P(r)}{\rm{d}\mathit{P(r)}/\rm d\mathit{r}}, \frac{T(r)}{\rm d\mathit{T(r)}/\rm d\mathit{r}}, \frac{m(r)}{\rm d\mathit {m(r)}/\rm d \mathit{r}}\right)},
\end{equation}
\end{ceqn}
where $\epsilon_r$ is the refinement parameter. For an optimal trade off between stability and performance we recommend to use $0.1 \leq \epsilon_r \leq 0.5$. All results presented here have been obtained with $\epsilon_r = 0.25$. The gradients have to be updated in each integration step to define the subsequent integration step size. With this procedure the planets are typically divided into a total of $\sim 100-1000$ individual shells. For simplicity, we set the water content $X_{\rm H_2O}(P, T)$ constant in each integration step. This induces a small error as $P$ and $T$ are of course different at the bottom of a shell and the top of the same shell. However, this effect can be controlled by the refinement parameter and can be neglected for our choice of 
$\epsilon_r$.
\subsection{Thermal model}\label{sec:thermal_model}
The adiabatic gradient in eq.~\eqref{eq:dTdr} is related to the material properties according to:
\begin{ceqn}
\begin{equation}\label{eq:adgrad}
    \left(\frac{\rm d \mathit{T}}{\rm d\mathit{P}}\right)_{ad} = \frac{\gamma (T,P) T}{{K}_S(T, P)},
\end{equation}
where $\gamma(T,P)$ and $K_S(T,P)$ are the Gr{\"u}neisen parameter and the adiabatic bulk modulus, respectively. Eq.~\eqref{eq:adgrad} is commonly written in terms of the logarithmic temperature derivative $\nabla_{ad}$ as:
\begin{equation}\label{eq:dTdP}
    \left(\frac{\rm d \mathit{T}}{\rm d\mathit{P}}\right)_{ad} = \frac{T}{P}\nabla_{ad},
\end{equation}
with
\begin{equation}\label{eq:dTdP_log}
    \nabla_{ad} \equiv \frac{\rm{d} \mathit{log(T)}}{\rm{d} \mathit{log(P)}}.
\end{equation}
\end{ceqn}
The adiabatic bulk modulus can be related to the isothermal bulk modulus $K_T$ via the Gr{\"u}neisen parameter $\gamma$ and the thermal expansion coefficient $\alpha_{th}$ using:

\begin{ceqn}
\begin{equation}\label{eq:KSfromEOS}
K_S(P,T) = \left[1+\gamma(T,P) \alpha_{th}(T, P) T\right] K_T(T, P).
\end{equation}
$K_T$ and $\alpha_{th}$ can be computed from the EoS (see Appendix \ref{sec:theos}) directly:
\begin{equation}\label{eq:KTfromEOS}
K_T(P,T) \equiv \rho(P, T) \left(\frac{\partial\rho(P, T)}{\partial P}\right)^{-1} 
\end{equation}
\begin{equation}\label{eq:alphafromEOS}
\alpha_{th}(P,T) \equiv -\frac{1}{\rho(P, T)}\frac{\partial \rho(P, T)}{\partial T}.
\end{equation}
The $P$ and $T$ dependance of the Gr{\"u}neisen parameter can be expressed in terms of the density:

\begin{equation}\label{eq:gamma}
    \gamma (T,P) = \gamma_0 \left(\frac{\rho (T, P)}{\rho_0}\right)^{-q}.
\end{equation}
\end{ceqn}
In eq.~\eqref{eq:adgrad} - \eqref{eq:gamma}, $T$ and $P$ are functions of the radial distance $T(r)$ and $P(r)$, which we have omitted to write explicitly for clarity. 
Eq.~\eqref{eq:adgrad} is only used for the iron in the core and the Mg-silicates in the mantle as for the pure water the adiabatic gradient $(\rm d \mathit T / \rm d \mathit P)_{ad}$ can self-consistently be extracted directly from the EoS used in this study (see Appendix \ref{sec:theos} for details). The corresponding values for the Gr{\"u}neisen parameter for the different layers are summarized in Table \ref{tab:layer_props}.
\\

In the case of mixtures of $N$ materials denoted by the index $i$ the density is obtained using linear mixing:
\begin{ceqn}
\begin{equation}\label{eq:mixlaw}
    {\rho}(T,P) = \left(\sum_{i=1}^N \frac{X_i(T, P)}{\rho_i(T,P)}\right)^{-1},
\end{equation}
\end{ceqn}
where $X_i$ is the weight fraction of material $i$. The weight fractions are normalized to one, such that $\sum_{i} X_{i} = 1$. The isothermal bulk modulus and thermal expansion are then given by:
\begin{ceqn}
\begin{equation}\label{KT}
    {K}_{T}(T, P) = - \sum_{i=1}^N \frac{X_i}{\rho_i} \cdot \left(\sum_{i=1}^{N} \frac{1}{\rho_i} \frac{\partial X_i}{\partial P}-\frac{X_i}{\rho_i^2}\frac{\partial \rho_i}{\partial P}\right)^{-1}
\end{equation}
\begin{equation}\label{alphaT}
    {\alpha}_{th}(T,P) = \left( \sum_{i=1}^N \frac{X_i}{\rho_i} \right)^{-3} \cdot \left(\sum_{i=1}^{N} \frac{1}{\rho_i} \frac{\partial X_i}{\partial T}-\frac{X_i}{\rho_i^2}\frac{\partial \rho_i}{\partial T}\right)^{-1},
\end{equation}
\end{ceqn}
where we have omitted to write the $T-P$ dependence of $X_i$ for clarity. Here we do not account for composition gradients and hence the derivatives $\partial X_i/\partial P$ and $\partial X_i / \partial T$ vanish. 
\subsection{Composition}\label{sec:comp}
The model planets are composed of the elements Fe, Si, Mg, O and H. Minor elements such as C, Ni, Ca, S or Al are excluded since they would significantly increase the complexity of the composition beyond the scope of this study (see also \cite{Sotin2007}). 
We point out that the presence of these elements could affect the hydration behaviour of the Mg-silicates and therefore change their water content (see also Section~\ref{sec:caveats}). Based on the assumed elemental composition the modelled objects are divided into two to four layers: a pure iron core, an upper and lower Mg-silicate mantle with variable Fe content and possibly a surface water ocean. The transition from the upper to the lower mantle is defined via the dissociation of Ringwoodite (Rw) into Magnesiowüstite (Mw) and Perovskite (Pv) occurring at $\approx 25-30 \rm \  GPa$ (\cite{Wang1997}):
\begin{ceqn}
\begin{equation}\label{eq:OlivineDissoc}
    \rm (Mg_{1-\xi_{Fe}}, Fe_{\xi_{Fe}})_2 Si O_4 \rightarrow{} (Mg_{1-\xi_{Fe}}, Fe_{\xi_{Fe}})SiO_3 + (Mg_{1-\xi_{Fe}}, Fe_{\xi_{Fe}})O. 
\end{equation}
\end{ceqn}
If the pressure at the bottom of the mantle does not exceed this transition pressure, only an upper mantle is present. We do not account for a possible differentiation of the core into a liquid and solid part and neglect the possible presence of lighter elements other than hydrogen in the core for simplicity.
\\

For the purpose of illustrating how our model can be applied to estimate the water contents in planetary interiors, we imposed three simplifying assumptions: (1) The mantles of the modelled objects consist of pure $\rm(Mg_{(1-\xi_{Fe})},Fe_{\xi_{Fe}})_2SiO_4$ or  $\rm (Mg_{1-\xi_{Fe}}, Fe_{\xi_{Fe}})SiO_3 + (Mg_{1-\xi_{Fe}}, Fe_{\xi_{Fe}})O$ in the upper or lower mantle, respectively. The choice of these minerals was motivated by the fact that they are major constituents in the Earth's mantle (\cite{Valdez2013, Wu2012, Jacobsen2006, Sinmyo2014, Li2014}). Furthermore, Olivine (and its high pressure polymorphs) can incorporate considerable amounts of water and MgO can react with $\rm H_2O$ to form $\rm Mg(OH)_2$ (see references in Table \ref{tab:alpha}-\ref{tab:gamma} and also Section~\ref{sec:saturationmodel}). (2) The minerals in the interiors of the hydrated planets are fully saturated. (3) Since higher Fe content in Olivine polymorphs generally leads to an increase in the water saturation level, we chose $\xi_{\rm Fe}=0.25$ as it roughly marks the upper limit for which our hydration model is valid (see Section \ref{sec:saturationmodel}). We point out that, for the simplified bulk composition, the density in the lower mantle can be lower than in the upper mantle in the fully water saturated case if only little Fe is present. This is due to the high water storage capacity of $\approx \rm 31 \ wt \%$ in Brucite ($\rm Mg(OH)_2$ (Br)). This further motivates the choice of high Fe contents in the mantle. For our choice of $\xi_{\rm Fe}$ this spurious behaviour could be avoided for all modelled cases and would naturally vanish in the Fe-free case if more Si-rich phases were present in the lower mantle.
\\

In order to estimate the effects of hydration and ocean separation on $M$-$R$ relations we consider here three types of planets: Dry planets (Type-1), hydrated planets (Type-2) and ocean planets (Type-3). The surface oceans are assumed to consist of pure water. An overview of the different types is provided in Fig.~\ref{fig:compare_counterparts}. The Type-1 planets have fully OH and H depleted interiors and no surface oceans. For the Type-2 planets the same boundary conditions, that is bulk composition and surface conditions, as for the Type-1 planets have been employed, but the mantles and cores are hydrated. A comparison between Type-1 and Type-2 planets allows for the assessment of the effect of hydration on total planetary radii. To estimate the effect of ocean differentiation we computed the amount of $\rm H_2 O$ equivalent in the Type-2 planets and added it as an isolated surface ocean on top of a dry mantle for the Type-3 planets. Since we are interested in estimating maximum effects in this study, intermediate cases where the water is partially distributed between an internal reservoir and a surface ocean, were not considered.

\subsection{Boundary conditions}\label{sec:boundary_cond}

\begin{figure*}[t]
\includegraphics[width=\textwidth]{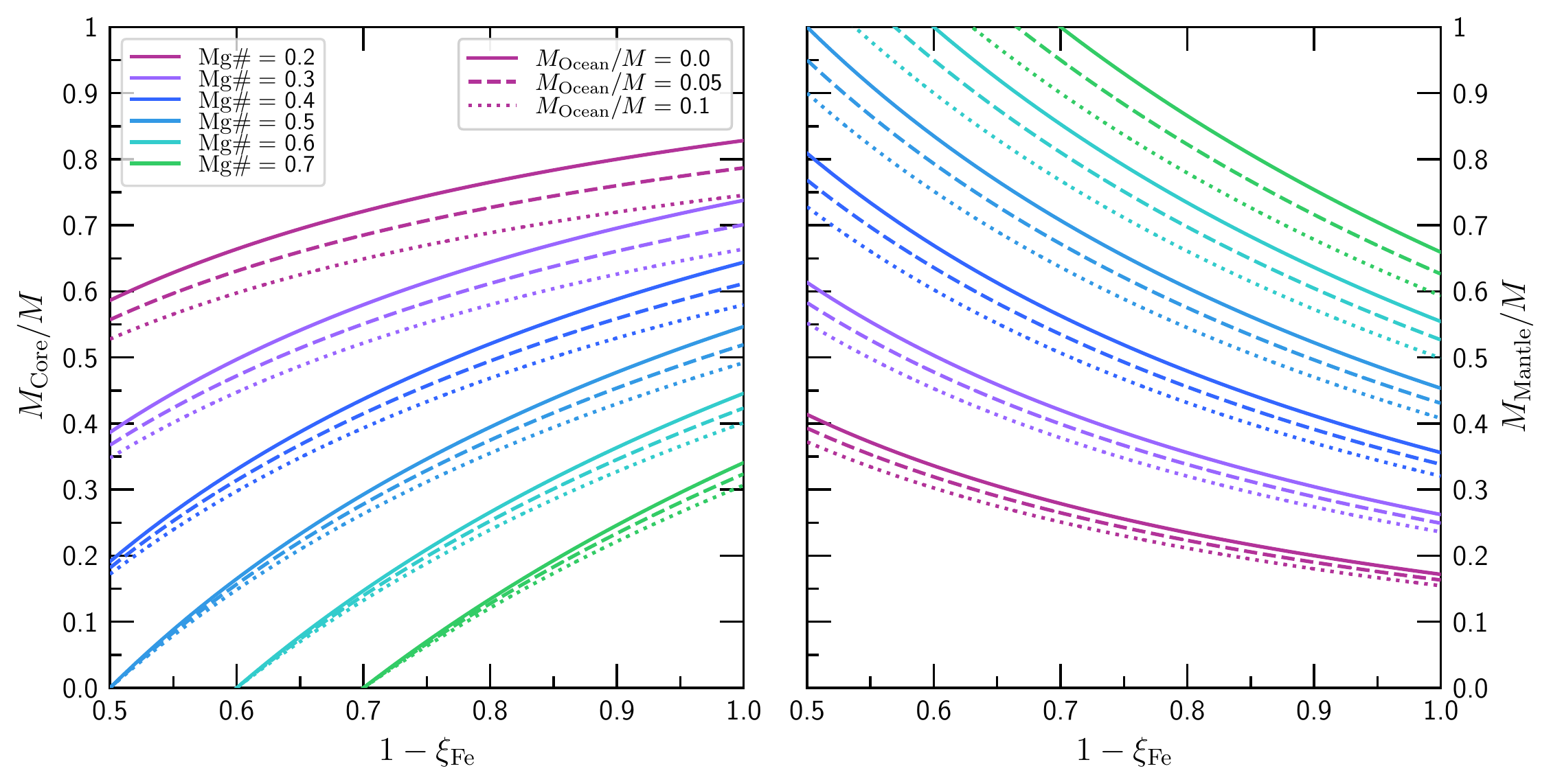}
\caption{Core mass fractions (left) and mantle mass fractions (right) as a function of the mantle composition $\xi_{\rm Fe}$ according to eq. \ref{eq:core_mass_frac} for different bulk compositions $\rm Mg \#$ (colours) for $M_{\rm Ocean}/M = 0$ (solid), 0.05 (dashed), and 0.1 (dotted).}
\label{fig:core_masses}
\centering
\end{figure*}
After defining the compositions of the layers the interior structure was determined by integrating from the centre to the outermost layer using initial guesses for the central pressure $P_{\rm C}$ and temperature $T_{\rm C}$. For differentiated planets, the layer transitions must be assessed numerically using one of the integrated variables $Q$ as a constraint, where $Q$ is either the pressure, temperature, enclosed mass or radial extent of a layer or the total magnesium number of the planet. We used the core mass $M_{\rm Core}$ to define the core mantle boundary (CMB) and the surface temperature $T_{\rm S}$ to constrain the central temperature $T_{\rm C}$. The boundary conditions used in this study are: The total magnesium number $\rm Mg \# \equiv [Mg]/([Mg] + [Fe])$ including core and mantle, and the surface conditions $P_{\rm S}$, $T_{\rm S}$ and $M$. The surface pressure is automatically matched as it defines the surface of the planet.  $T_{\rm S}$, $M$ and $\rm Mg \#$ were iteratively probed by adjusting $T_{\rm C}$, $P_{\rm C}$ and $M_{\rm Core}$ until the desired precision for the boundary conditions was achieved. Since these quantities are not independent, we employed the following prescription for predicting a new set of initial values $\vec{S}$ in each iteration:
\begin{ceqn}
\begin{equation}\label{eq:iteration_scheme_M_Core}
 S_j(Q_i) = a_{ij} Q_i
\end{equation}
\end{ceqn}
with
\begin{ceqn}
\begin{equation}\label{eq:iteration_scheme_P_C}
   \vec Q \equiv (1, {\rm Mg} \#, {\rm log}(P_{\rm S}), T_{\rm S}),
\end{equation}
\end{ceqn}
\begin{ceqn}
\begin{equation}\label{eq:iteration_scheme_T_C}
    \vec{S} \equiv ({\rm log}(M_{\rm Core}/M), {\rm log}(P_{\rm C}), T_{\rm C}).
\end{equation}
\end{ceqn}
We found eq.~\ref{eq:iteration_scheme_M_Core} to be sufficient in most cases to obtain the desired precision for all parameters and have hence omitted mixed terms in the $Q_i$. In each iteration the coefficients $a_{ij}$ are obtained by solving the following matrix equation for $k,i = 0,1,2,3$ and $j = 0,1,2$:
\begin{ceqn}
\begin{equation}\label{eq:matrix_eq}
    M_{ki} \cdot a_{ij} = x_{kj}
\end{equation}
\end{ceqn}
\begin{ceqn}
\begin{equation}\label{eq:matrix_elements}
    M_{ki} \equiv (Q_i)_k
\end{equation}
\end{ceqn}
\begin{ceqn}
\begin{equation}\label{eq:x_vector}
     x_{kj} = (S_j)_k. 
\end{equation}
\end{ceqn}
The subscript $k$ refers to the $k$th data point that is needed to constrain all $a_{ij}$. In our case, four independent data points are required before the iterative prediction for the subsequent $x_{kj}$ can be carried out. These data points were obtained by roughly estimating the values $S_j$ for a given set of boundary conditions $\vec{Q}$ from a pre-constructed grid and then adopting slight variations to the $S_j$. This procedure results in a very robust and efficient evaluation of the values $S_j$ with very high accuracy for a given $\vec{Q}$. In the case of hydrated planets, the hydrogen content in the core $\xi_{\rm H}$ is a priori unknown since it depends on the water content in the mantle and the temperature at the core mantle boundary (see Section \ref{sec:core_hyd} for details). However, as $\xi_{\rm H}$ is a required input for the $\rm FeH_x$ EoS in the core, it must be estimated prior to the integration. 
Here, we treated $\xi_{\rm H}$ as a passive parameter in addition to the actively probed parameters $S_{j}$. In each iteration the two previous data points were used to predict the new value for $\xi_{\rm H}$ as a function of the new value for $T_{\rm C}$ solely. With the procedure described above, $\xi_{\rm H}$ could be matched with high accuracy if $T_{S}$ was probed with sufficient accuracy. We note that in general a less sophisticated iteration scheme, where the values $S_j$ are probed alternately via simple bisection, can be sufficient to achieve the same accuracy for the dry planets and the ocean planets. However, it failed for a considerable fraction of models for the hydrated planets. Furthermore, the procedure outlined above was found to be robust and more efficient than bisection.
To determine when the quantity $Q_i$ has reached the desired precision we evoked the following criterion:
\begin{ceqn}
\begin{equation}\label{eq:criterion}
    \frac{|Q_{i,is}-Q_{i,should}|}{Q_{i,should}} \leq \epsilon_{Q_i}, 
\end{equation}
\end{ceqn}
where $\epsilon_{Q_i}$ is the desired precision for the parameter $Q_i$. The same condition is used to probe the layer transitions using either the enclosed mass, the transition pressure or transition temperature as constraints. As the layer transitions in all our models depend on only one parameter, we probed the transition using simple bisection during the structure integration. The core-mantle transition has been probed using a fixed value for $M_{\rm Core}$, the upper-lower mantle transition via a fixed value for the dissociation pressure of Olivine, $P_{\rm UM \rightarrow LM}$ and the mantle-ocean transition using the enclosed mass $M-M_{\rm Ocean}$ at the bottom of the ocean. The corresponding values for the different parameters that have been used to define the boundary conditions and layer transitions are summarized in Table \ref{tab:NumPrec}. We set higher accuracy for $M$ of the Type-3 planets and $T_{\rm S}$ of the Type-2 planets. This was necessary to achieve the desired precision for $M_{\rm Ocean}/M$ and $\xi_{\rm H}$, respectively. The transition criteria for the different layers are listed in Table \ref{tab:layer_props}. The specified accuracy was reached in about $80-90 \ \%$ of the modelled cases. For the rest of the planets the achieved accuracy for $M$,  $\rm Mg \#$, $\xi_{\rm H}$, and $M_{\rm Ocean}/M$ were at least $10^{-3}$, $5 \cdot 10^{-3}$, $3 \cdot 10^{-2}$, and $2 \cdot 10^{-2}$, respectively. These numerical uncertainties have, however, minor effect on the main results presented in this study. To obtain the mass-radius relations it is sufficient to vary the central pressure $P_{\rm C}$ to cover a desired mass range. However, since we aim to compare different mass-radius curves directly we explicitly fixed $M$ via $P_{\rm C}$ for each of the modelled planets individually. This allows a direct comparison of the planetary properties for a given mass, composition, and surface conditions. In the case of dry mantles, where the composition in all layers is constant with depth, the core mass fraction can be computed analytically as a function of the total bulk and mantle compositions $\rm Mg \#$ and $\xi_{\rm Fe}$, the total mass $M$ and the ocean mass $M_{\rm Ocean}$:
\begin{ceqn}
\begin{equation}\label{eq:core_mass_frac}
\begin{split}
    \frac{M_{\rm Core}}{M} = & \left( 1- \frac{M_{\rm Ocean}}{M} \right)\left(\frac{\rm 1-\xi_{\rm Fe}}{\rm Mg \#}-1 \right) \\
    & \times \left[ \xi_{\rm Fe} \frac{ \tilde{m}}{m_{\rm Fe}}+\frac{1-\xi_{\rm Fe}}{\rm Mg \#}-1\right]^{-1},
\end{split}
\end{equation}
\end{ceqn}
where $m_{\rm Fe}$ the molar mass of iron. $\tilde{m}$ is the mass of a portion of the material in the dry mantle containing one mole of Fe. It is defined as: 
\begin{ceqn}
\begin{equation}\label{eq:mean_molecular_weight_mantle}
\begin{split}
\tilde{m} \equiv  & \frac{\rm1-\xi_{\rm Fe}}{\xi_{\rm Fe}} m_{\rm Mg}+\frac{\rm Si \#}{1-\rm Si \#} m_{\rm Si} \\
& \frac{\rm O \#}{1-\rm O \#} m_{\rm O} + m_{\rm Fe}. 
\end{split}
\end{equation}
\end{ceqn}
The oxygen number $\rm O \#$ is given by:
\begin{ceqn}
\begin{equation}\label{eq:O_number}
\rm O \# = \frac{1 + 2 \small{\frac{\rm Si}{\rm Mg}} (1-\xi_{\rm Fe})}{2 +2 \small{\frac{\rm Si}{\rm Mg}}(1-\xi_{\rm Fe})-(1-\xi_{\rm Fe})}.
\end{equation}
\end{ceqn}
The mantle mass fraction can then simply be computed for a given total mass and ocean mass fraction as $M_{\rm Mantle} \mathit = M - M_{\rm Core} \mathit - M_{\rm Ocean}$. In these cases the iteration for the core mass to match $\rm Mg \#$ can be omitted. The core mass fractions and corresponding mantle mass fractions as a function of the mantle composition are plotted in Fig.~\ref{fig:core_masses} for different bulk compositions and ocean mass fractions.
\\

For differentiated planets each individual layer can have a different convection behaviour based on the composition and the $P-T$-profiles. Hence, the heat transport via convection from the interior can be partially interrupted at a layer transition. This leads to a temperature drop upon transiting from a convective layer into a less convective one. \cite{Stixrude2014} have presented models for the temperature profiles of super-Earths considering thermal regulation controlled by silicate melting. It was found that the temperature drop at the CMB  denoted as ${\rm \Delta} T_{\rm CMB}$, can be represented by a simple scaling law as a function of the total planetary mass $M$:
 \begin{ceqn}
\begin{equation}\label{eq:delta_T_CMB}
\Delta T_{\rm CMB} = 1400 \ {\rm K} \cdot \left(\frac{M}{M_{\oplus}} \right)^{3/4}.
\end{equation}
\end{ceqn}

 Here, we use eq. \ref{eq:delta_T_CMB} to compute $\Delta T_{\rm CMB}$ over the modelled mass range and for all bulk compositions. We did not impose additional temperature jumps in the mantle transition zone (MTZ).  We treat the temperature drop $\Delta T_{\rm TBL}$ in the thermal boundary layer (TBL) at the top of the Mg-silicate mantle (mantle-surface or mantle-ocean interface) as a free parameter and have considered values of 200 K, 700 K, 1200 K and 1700 K. The value $\Delta T_{\rm TBL} = 1200 \ \rm K$ roughly corresponds to modern day Earth with $T_{\rm S} \approx 300 \ \rm K$ (for example \cite{Sotin2007}). Since the water saturation in the mantle Mg-silicates drops with increasing temperature (see Section \ref{sec:saturationmodel}), the maximum effect at given composition and surface conditions is reached when the temperature discontinuity in the MTZ vanishes. We note, however, that a non-vanishing temperature jump in the MTZ could in principle increase the hydration of the iron core (see Section \ref{sec:core_hyd}). Furthermore, we only consider one value $T_{\rm S} = 300 \ \rm K$ because different values for $T_{\rm S}$ could, in principle, be interpreted as different values for $\Delta T_{\rm TBL}$. For instance, $T_{\rm S} = 300 \ \rm K$ and $\Delta T_{\rm TBL} = 1700 \ \rm K$ could also be interpreted as $T_{\rm S} = 800$ and $\Delta T_{\rm TBL} = 1200 \ \rm K$. Furthermore, at elevated surface temperatures the definition of the Type-3 analogues would be less straightforward as part of the ocean mass fraction would enter the vapor phase. Since we do not account for atmospheres in this study, we explicitly fix $T_{\rm S} = 300 \ \rm K$ for the Type-3 planets.
 \\

In this study, we computed the mass-radius curves for fixed surface pressures and temperatures of $P_{\rm S}$ = 1 bar and $T_{\rm S} = 300 \ \rm K$. The $\rm Mg\#$ has been varied between $0.2-0.7$. This is roughly the range covered by estimates for elemental ratios of planet hosting stars presented by \cite{Grasset2009} (see their Fig. 2). We point out that for their presented value range of Mg/Si and Fe/Si, the magnesium number could be as high as $\approx 0.8$. However, in this study $\rm Mg \#$ must be strictly smaller than $1-\xi_{\rm Fe} = 0.75$ for the assumed Fe content in the mantles of $\xi_{\rm Fe} = 0.25$ if a core is present.
 
\begin{table}
  \begin{center}
    \caption[]{\label{tab:layer_props} Layer properties.}
    \begin{tabular}{lcc} 
      \hline
       layer & $\gamma$ & transition criterion \\
      \hline
         core & 2.43 & $M_{\rm Core}$\\
         lower mantle & 1.96 & $P_{\rm UM-LM}$\\
         upper mantle & 1.26 &  $M-M_{\rm Ocean}$ or $P_{\rm S}$\\
         ocean & EoS & $P_{\rm S}$\\
    \hline
    \end{tabular}
    
  \end{center}
\end{table}

\begin{table}
  \begin{center}
    \caption[]{\label{tab:NumPrec} Numerical precision for planetary parameters.}
    \begin{tabular}{lccc} 
    \hline
      $Q$  & & $\epsilon_Q$ & \\
      \hline
       & Type-1 & Type-2 & Type-3 \\
      \hline
         $M$ & $10^{-4}$ & $10^{-4}$ & $10^{-5}$ \\
         $M_{\rm Core}$ & $10^{-6}$ & $10^{-6}$ & $10^{-6}$ \\
         $P_{\rm S}$ & $10^{-3}$ &  $10^{-3}$ & $10^{-3}$ \\
         $T_{\rm S}$ & $10^{-2}$ & $10^{-4}$ & $10^{-2}$\\
         $\rm Mg \#$ & $10^{-3}$ & $10^{-3}$ & $10^{-3}$\\
         $P_{\rm UM-LM}$ & $10^{-6}$ & $10^{-6}$ &  $10^{-6}$\\
         $M-M_{\rm Ocean}$ & ... & ... & $ 10^{-6}$\\
         $M_{\rm Ocean}/M$ & ... & ... & $5\cdot 10^{-3}$ \\
         $\xi_{\rm H}$ & ... & $5 \cdot 10^{-3}$ & ... \\
    \hline
    \end{tabular}
    
  \end{center}
\end{table}
\subsection{Updating the shell contents}\label{sec:shell_cont}
We assume the hydrogen content in the core to be homogeneous. For a given composition the water content in the mantle, however, is a function of temperature and pressure (and hence of radius). The water content was updated in each shell for the hydrous phases before integration according to the water storage capacity of the Mg-silicates (see Section \ref{sec:saturationmodel}). A change in the molar water content, $\xi_{{\rm H_2O}}$, leads to a change of the ratios Si/Mg. Therefore, in order to keep Si/Mg at a fixed value in each shell, the molar fractions of the different minerals need to be updated according to the water content. Here, this is irrelevant for the upper mantle since it consists of only one stoichiometry, the Olivine polymorphs, and hence its molar abundance, denoted by $\xi_{\rm Ol}$, is always equal to unity. In the lower mantle two minerals are present: Pv or pPv and Mw with molar Fe content $\rm \xi_{Fe}$. If water is added, the mole fractions of Pv or pPv and Mw $\xi_{\rm Pv}$ and $\xi_{\rm Mw}$ must be updated. 
The molar abundance of Mw is given by: 
\begin{ceqn}
\begin{equation}\label{eq:xi_Mw}
\begin{split}
\xi_{\rm Mw} = \frac{(1-\small{\frac{\rm Si}{\rm Mg}})(1-{\xi}_{\rm  H_2O, Pv})
}{\small{\frac{\rm Si}{\rm Mg}} (1-{\xi}_{\rm H_2O, Mw}) + (\small{1-\frac{\rm Si}{\rm Mg}}) (1-{\xi}_{\rm H_2O,  Pv})}.
\end{split}
\end{equation}
\end{ceqn}
The molar abundance of Pv/pPv is then simply given by:
\begin{ceqn}
\begin{equation}\label{eq:xi_Pv}
\begin{split}
\xi_{\rm Pv} = 1- \xi_{\rm Mw}.
\end{split}
\end{equation}
\end{ceqn}
Here $\xi_{\rm H_2 O, Mw}$ is either 0 or $(1-\xi_{\rm Fe})/2$, depending on the location in the phase diagram of $\rm Mg(OH)_2$ (see Section~\ref{sec:saturationmodel}). For Pv/pPv we assume constant water content of $0.1 \ \rm wt \%$ (Pv) and $3.0 \ \rm wt \%$ (pPv) (see Section \ref{sec:saturationmodel}). We have assumed the hydrogen content in the core to be homogeneous with depth and hence $\xi_{\rm H}$ remains constant throughout the integration of the core.
\subsection{Extracting the deep water reservoirs}\label{sec:extractwater}

\subsubsection{In the core}\label{sec:core_reservoir}

The total amount of $\rm H_2O$ equivalent in the core is given by the hydrogen content in $\rm FeH_x$, which is assumed to be homogeneous in the entire core. We discuss how the parameter x can be computed in Section~\ref{sec:core_hyd}. Once the stoichiometric hydrogen content $\rm x$ in the core is known, it can easily be converted to a molar abundance $\xi_{\rm H}$ as:

\begin{ceqn}
\begin{align}\label{eq:xi_H}
\xi_{\rm H} = \frac{x}{1+x}.
\end{align}
\end{ceqn}

The total atomic hydrogen content in the core is then given by:

\begin{ceqn}
\begin{align}\label{eq:M_H_Core}
N_{\rm H, Core} =  \frac{M_{\rm Core}\xi_{\rm H}}{(1-\xi_{\rm H})m_{\rm Fe} + \xi_{\rm H} m_{\rm H}}.
\end{align}
\end{ceqn}
The $\rm H_2 O$ equivalent mass in the core can then easily be obtained:

\begin{ceqn}
\begin{align}\label{eq:M_H2O_core}
M_{\rm H_2O, Core} = \frac{N_{\rm H, Core}}{2} m_{\rm H_2 O}.
\end{align}
\end{ceqn}
\subsubsection{In the mantle}\label{sec:mantle_reservoir}
The local water weight fraction $X_{\rm H_2O}(P, T)$ in the mantle depends on the radial distance via $P(r)$ and $T(r)$. As outlined in Section \ref{sec:shell_cont}, we assumed $X_{\rm H_2O}$ to be constant in each shell. The total mantle reservoir in a planet is then simply given by summation over all water containing shells. For a given shell, the water content $M_{\rm H_2O,Shell}$ was extracted from the shell properties using the following statement:
\begin{ceqn}
\begin{equation}\label{eq:MH2O}
    M_{\rm H_2O,Shell} = N_{\rm H_2O, Shell} m_{\rm H_2O},
\end{equation}
\end{ceqn}
where $N_{\rm H_2O, Shell}$ and $m_{\rm H_2O}$ are the molar amount of water in the shell and the molar mass of water, respectively. $N_{\rm H_2O, Shell}$ is related to the shell mass $M_{\rm Shell}$ and the molar water contents $\xi_{\rm H_2O,i}$ in the Mg-silicates (hereafter denoted as Mg-Si):
\begin{ceqn}
\begin{equation}\label{eq:NH2O}
N_{\rm H_2O,Shell}=\frac{M_{\rm Shell}}{\tilde{m}_{\rm Mg-Si}+\tilde{m}_{\rm H_2O}}\sum_{i=1}^{n_{\rm Mg-Si}} \xi_{i}\xi_{\rm H_2O, \mathit i}, 
\end{equation}
\end{ceqn}
with the relative molar masses given by:
\begin{ceqn}
\begin{align}\label{eq:tildemasses}
\tilde{m}_{\rm Mg-Si} & = \sum_{i=1}^{n_{\rm Mg-Si}}m_{i}\xi_{i}(1-\xi_{\rm H_2O,\mathit i})
\\
\tilde{m}_{\rm H_2O} & =\sum_{i=1}^{n_{\rm Mg-Si}}m_{\rm H_2O}\xi_{i}\xi_{\rm H_2O,{\mathit i}}.
\end{align}
\end{ceqn}
Here, $n_{\rm Mg-Si}$ denotes the number of coexisting Mg-Si phases in the shell. $\xi_{\rm H_2O ,{\mathit i}}$ can directly be computed from $X_{\rm H_2O,{\mathit i}}$ by invoking mass balance:
\begin{ceqn}
\begin{equation}\label{eq:massbalance}
    \xi_{\rm H_2O, \mathit i}m_{\rm H_2O} = X_{\rm H_2O, \mathit i} \left[ (1-\xi_{\rm H_2O, \mathit i}) m_{i} + \xi_{\rm H_2O,\mathit i} m_{\rm H_2O} \right]
\end{equation}
\end{ceqn}
and hence:
\begin{ceqn}
\begin{equation}\label{eq:x_to_xi}
    \xi_{\rm H_2O,\mathit i}(X_{\rm H_2O, \mathit i}) = \frac{X_{\rm H_2O,\mathit i} m_{i}}{m_{\rm H_2O}(1-X_{\rm H_2O,\mathit i}) + m_{i} X_{\rm H_2O,\mathit i}}.
\end{equation}
\end{ceqn}
\begin{figure*}
    \centering
    \includegraphics[width=\textwidth]{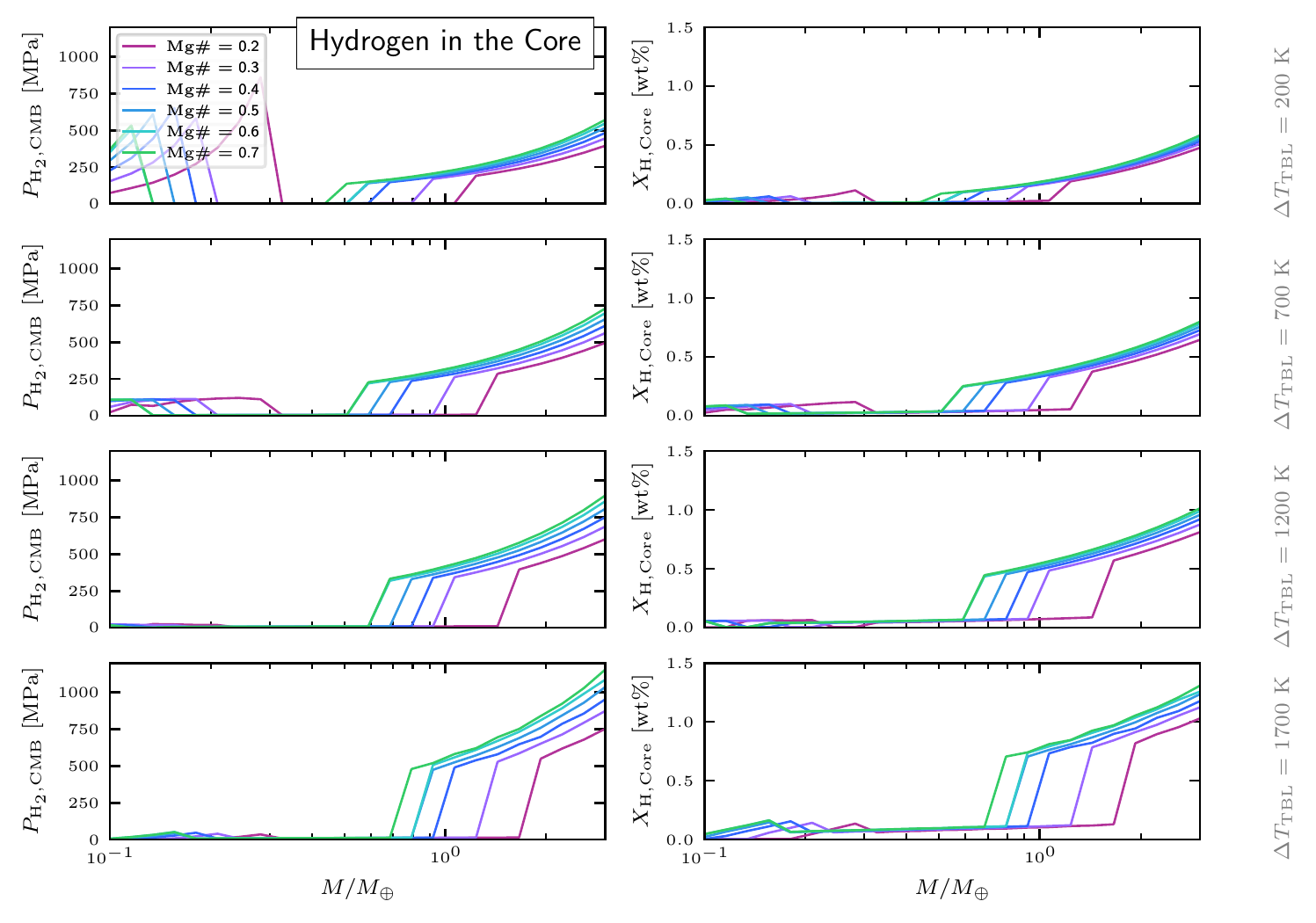}
    \caption{Partial hydrogen pressure $P_{\rm H_2}$ (left column) and resulting hydrogen content (right column) in the cores of the modelled Type-2 planets. We have limited our models to $M \leq 3 \ \rm M_\oplus$ which keeps $P_{\rm H_2}$ below $1.2 \ \rm GPa$ as the hydrogen solubility from eq. \ref{eq:Sievert} has been found to be considerably underestimated at higher pressures. The hydrogen content is computed from chemical equilibrium between the core and the mantle. It depends on the temperature at the CMB and the partial hydrogen pressure in the lowermost part of the mantle (see text for details).}
    \label{fig:core_content}
\end{figure*}
\begin{figure*}
    \centering
    \includegraphics[width=\textwidth]{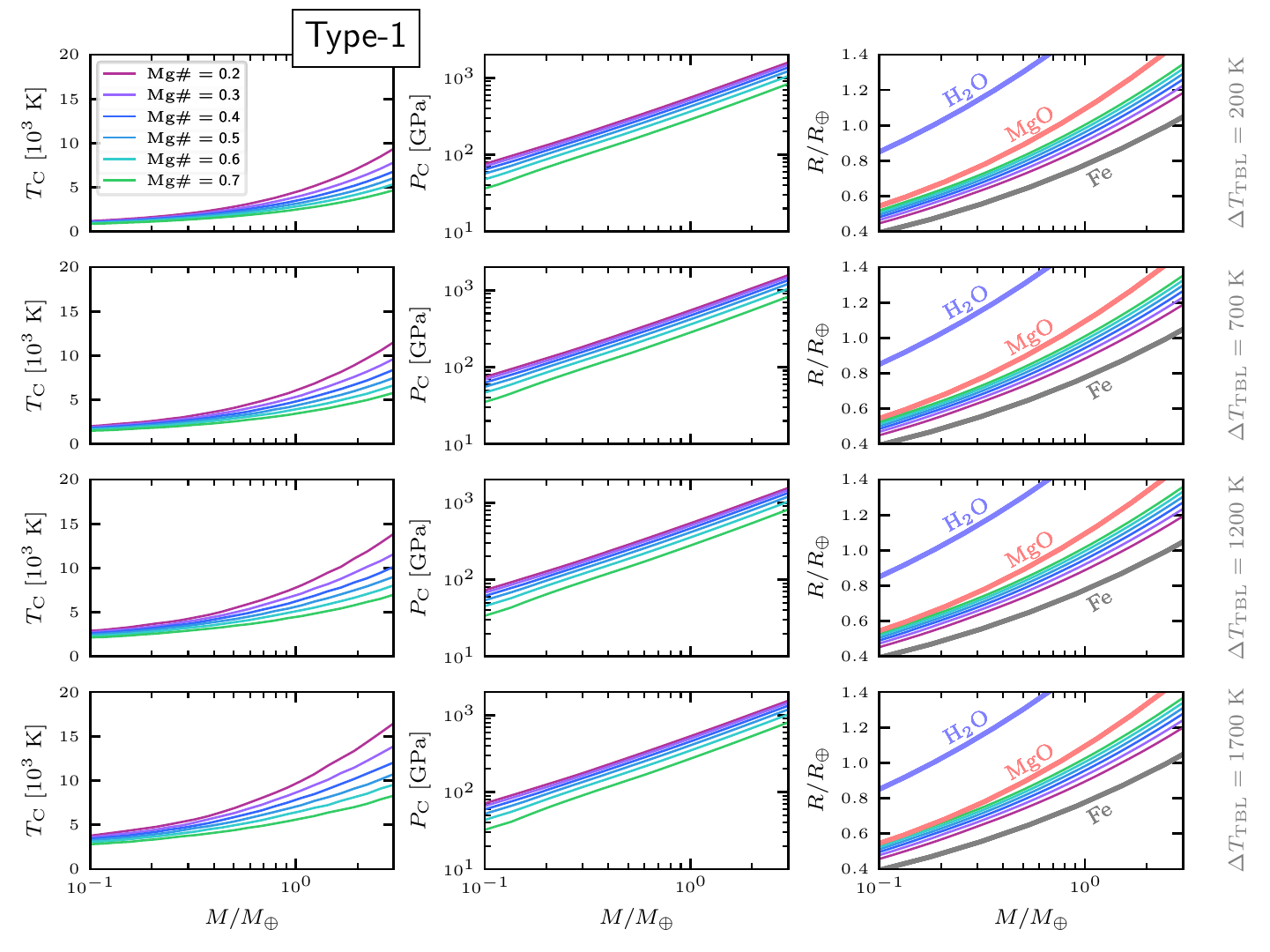}
    \caption{Central temperature (left column), central pressure (middle column), and M-R relation (right column) for the Type-1 planets. The different rows correspond to different assumed values for $\Delta T_{\rm TBL}$ (see annotations at the right). While the core temperature increases significantly for higher $\Delta T_{\rm TBL}$ the central pressure and radius remain vastly unaffected. The mass-radius relations for pure $\rm H_2 O$, MgO and Fe are shown for reference.}
    \label{fig:internals_type_1}
\end{figure*}
\begin{figure*}
    \centering
    \includegraphics[width=\textwidth]{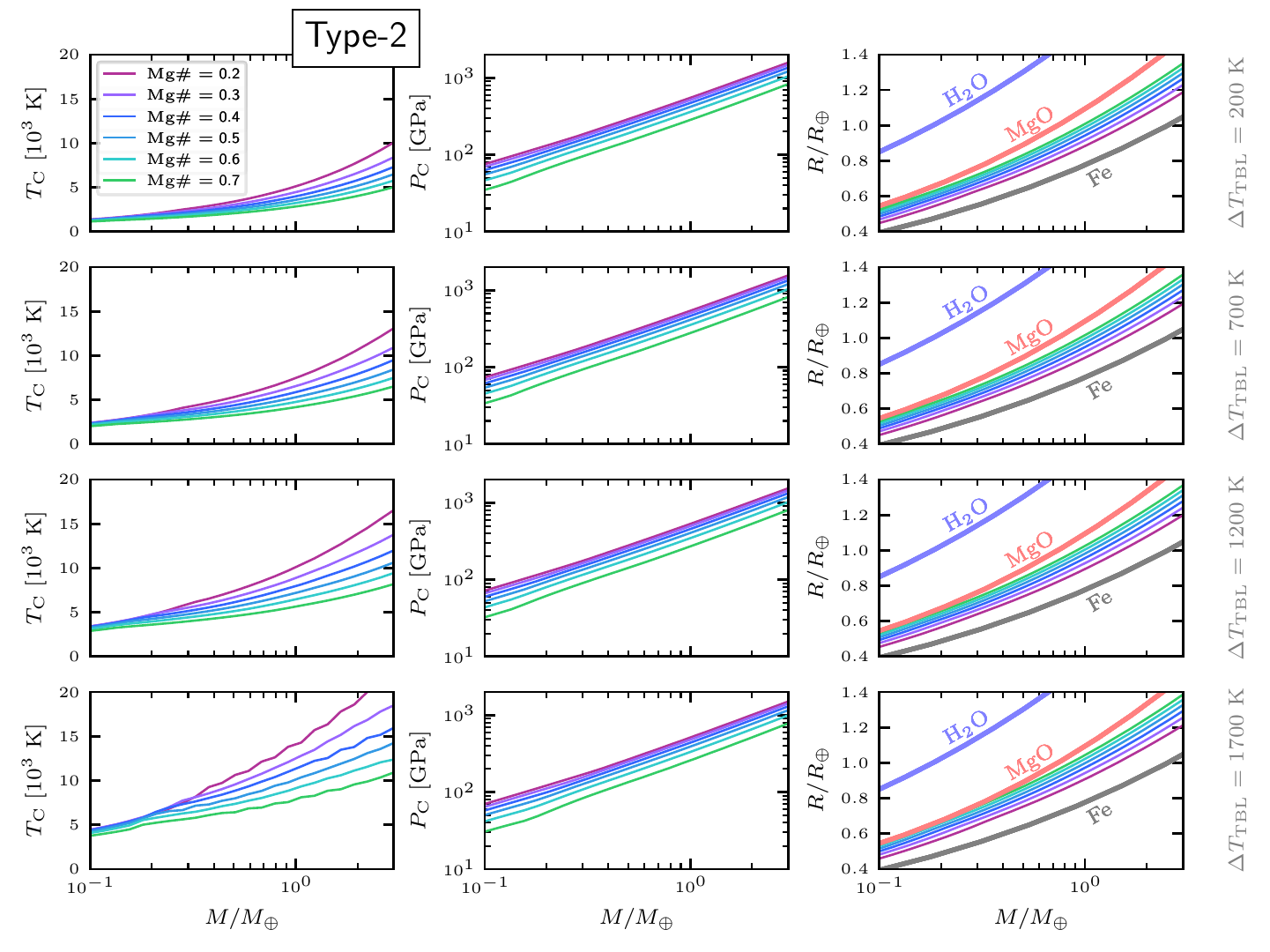}
    \caption{Central temperature (left column), central pressure (middle column), and M-R relation (right column) for the Type-2 planets. The different rows correspond to different assumed values for $\Delta T_{\rm TBL}$ (see annotations at the right). While the core temperature increases significantly for higher $\Delta T_{\rm TBL}$ the central pressure and radius remain vastly unaffected. The central temperatures are strongly enhanced with respect to the Type-1 planets as the adiabatic gradient tends to get steepened by hydration. The mass-radius relations for pure $\rm H_2 O$, MgO and Fe are shown for reference.}
    \label{fig:internals_type_2}
\end{figure*}
\begin{figure*}[t]
\includegraphics[width=\textwidth]{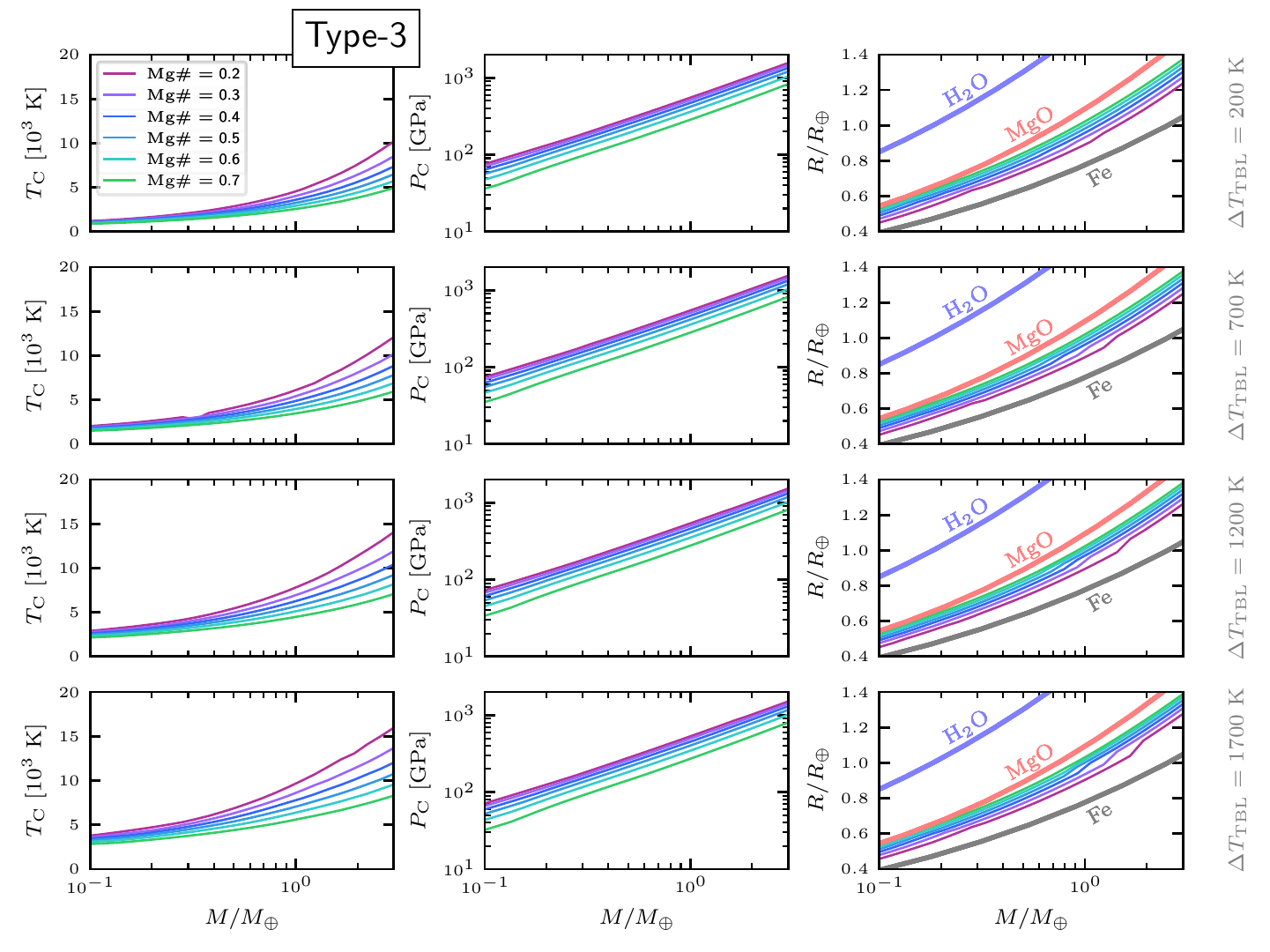}
\caption{Central temperature (left column), central pressure (middle column), and M-R relation (right column) for the Type-3 planets. The different rows correspond to different assumed values for $\Delta T_{\rm TBL}$ (see annotations at the right). While the core temperature increases significantly for higher $\Delta T_{\rm TBL}$ the central pressure and radius remain vastly unaffected. The kinks in the mass-radius curves correspond to the Pv-pPv transition, which leads to a drastic increase in the core reservoirs in the Type-2 planets. Accordingly, the ocean mass fractions in the Type-3 planets increase in the same mass range, which decreases the mean density of the palents. The mass-radius relations for pure $\rm H_2 O$, MgO and Fe are shown for reference.}
\label{fig:internals_type_3}
\centering
\end{figure*}
\begin{figure*}[t]
\includegraphics[width=\textwidth]{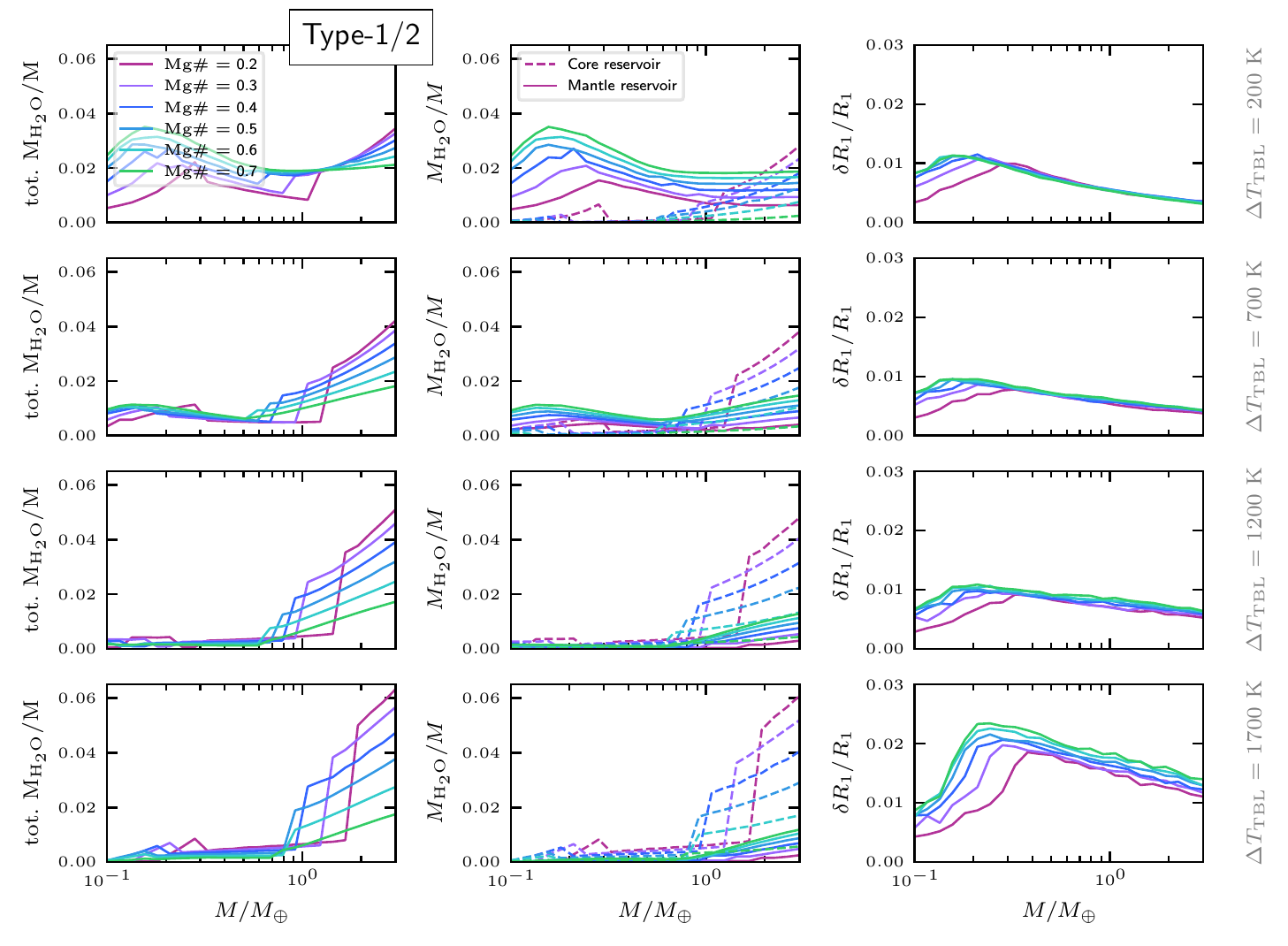}
\caption{Comparison between the Type-1 and Type-2 planets. Left column: Total $\rm H_2 O$ equivalent in the interior assuming saturated Mg-silicates in the mantle and hydrogen content in the core computed from Sieverts law. Middle column: Partitioning into separate contributions from the mantle- and core reservoirs. Right column: Relative effect $\delta R_1/R_1 \equiv (R_2 - R_1)/R_1$ of hydration on total planetary radius at given mass, composition and surface temperature (see also Fig. \ref{fig:compare_counterparts}).}
\label{fig:Delta_R1}
\centering
\end{figure*}
\begin{figure*}[t]
\includegraphics[width=\textwidth]{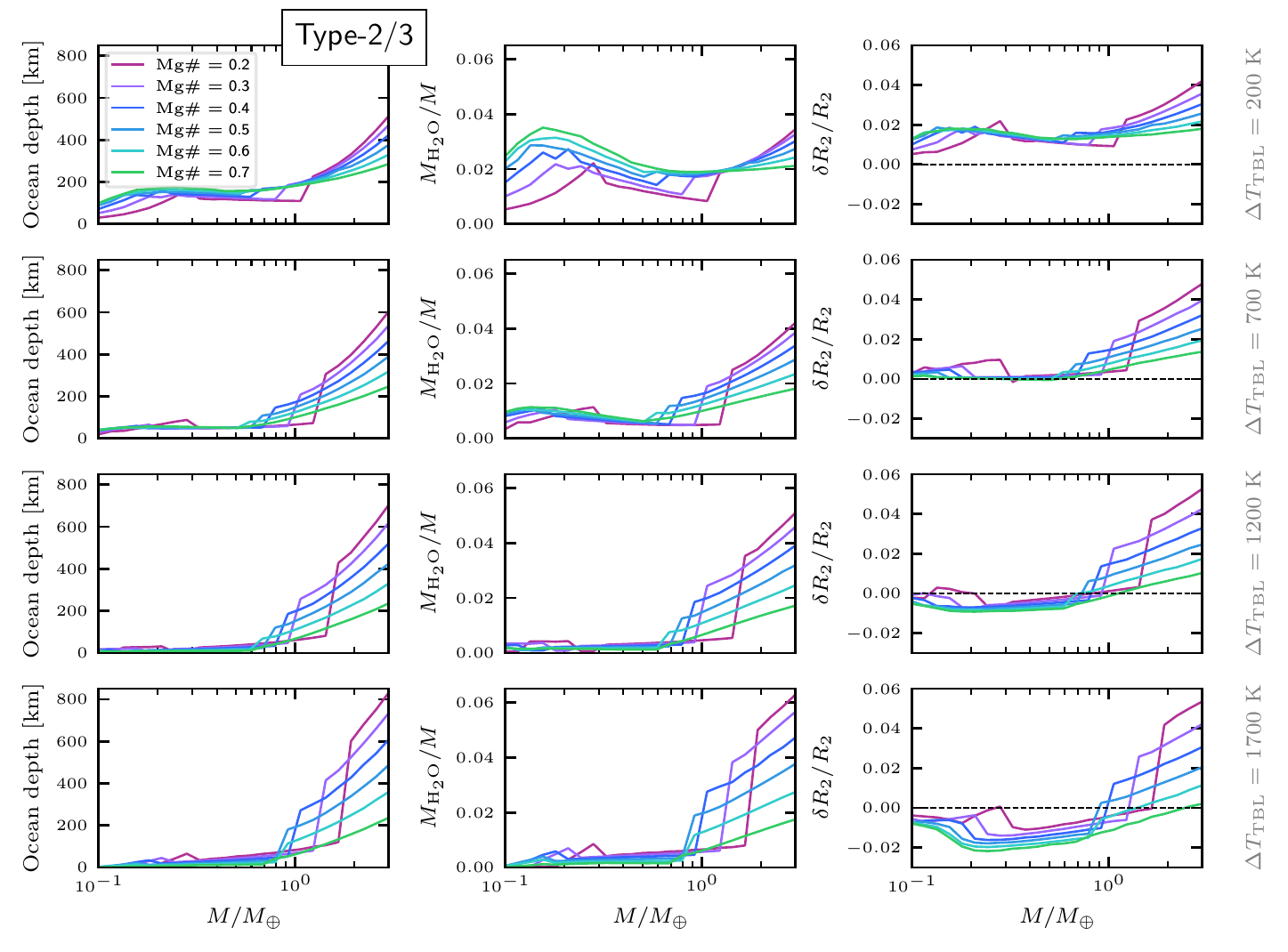}
\caption{Comparison between the Type-2 and Type-3 planets. Left column: Depth of the surface oceans. Middle column: Ocean mass fraction in the Type-3 planets. Of course, $M_{\rm H_2 O}/M$ in the Type-3 planets corresponds to the values of the Type-2 planets. Right column: Relative effect $\delta R_2 \equiv (R_3 - R_2)/R_2$ of ocean differentiation on total planetary radius at given mass, composition and surface temperature (see also Fig.~ \ref{fig:compare_counterparts}). For high values of $\Delta T_{\rm TBL}$ and low masses, separation into an isolated surface ocean can lead to a net increase in the planetary density in comparison to the Type-2 planets. This is a result from the combined effects of ocean differentiation on the EoS parameters.}
\label{fig:Delta_R2}
\centering
\end{figure*}
\begin{figure*}[t]
\includegraphics[width=\textwidth]{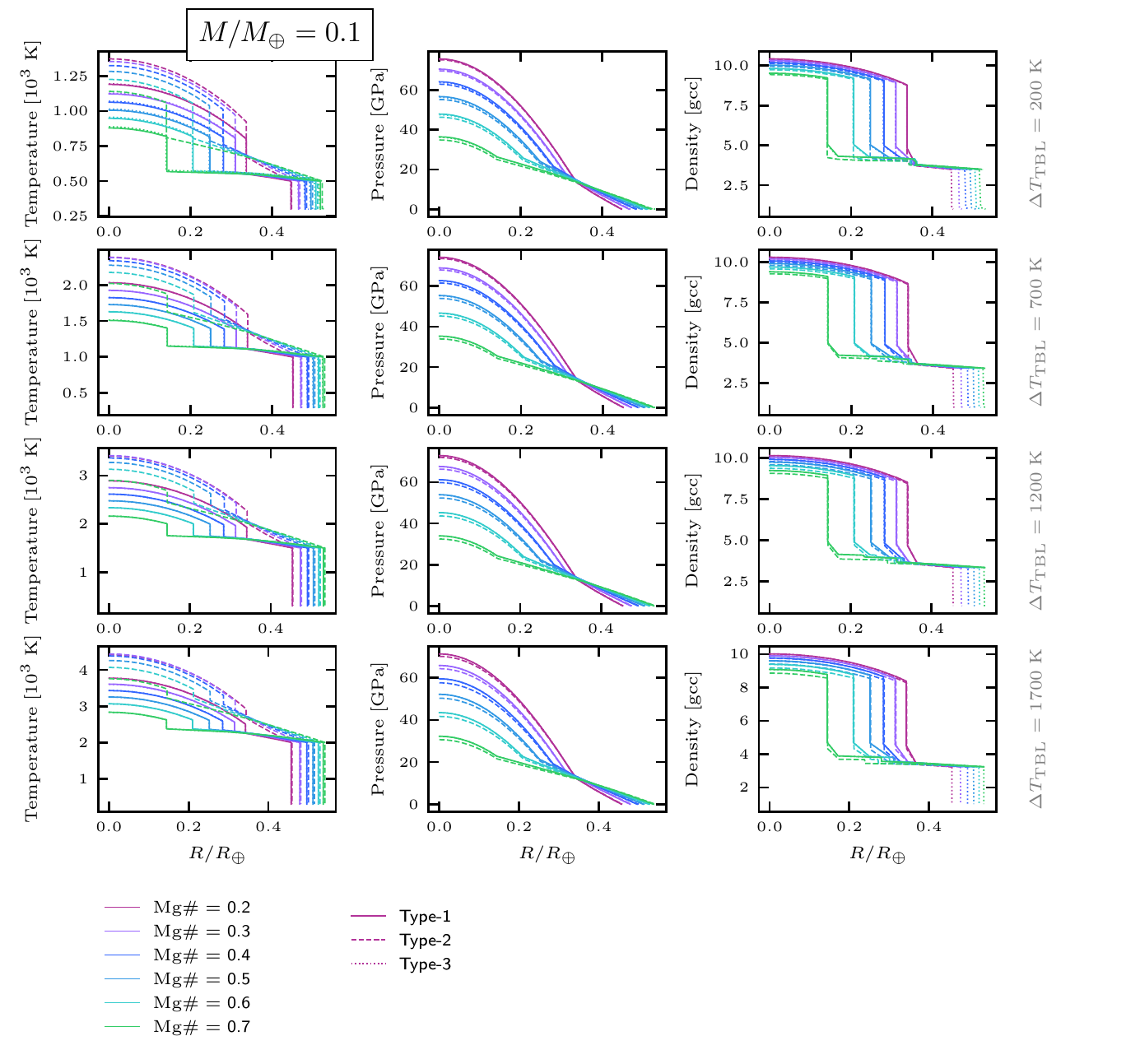}
\caption{Internal profiles as functions of radial distance from the centre for $M/M_{\oplus} = 0.1$ for all modelled $\Delta T_{\rm TBL}$ and compositions for the temperature (left column), pressure (middle column), and density (right). The solid curves correspond to Type-1 planets, the dashed curves to Type-2 planets and the dotted curves to Type-3 planets (see text for details). Since hydration tends to steepen the temperature gradient in the mantle, the Type-2 planets exhibit generally higher temperatures in the interiors than their Type-1 and Type-3 analogues.}
\label{fig:profiles1}
\centering
\end{figure*}
\begin{figure*}[t]
\includegraphics[width=\textwidth]{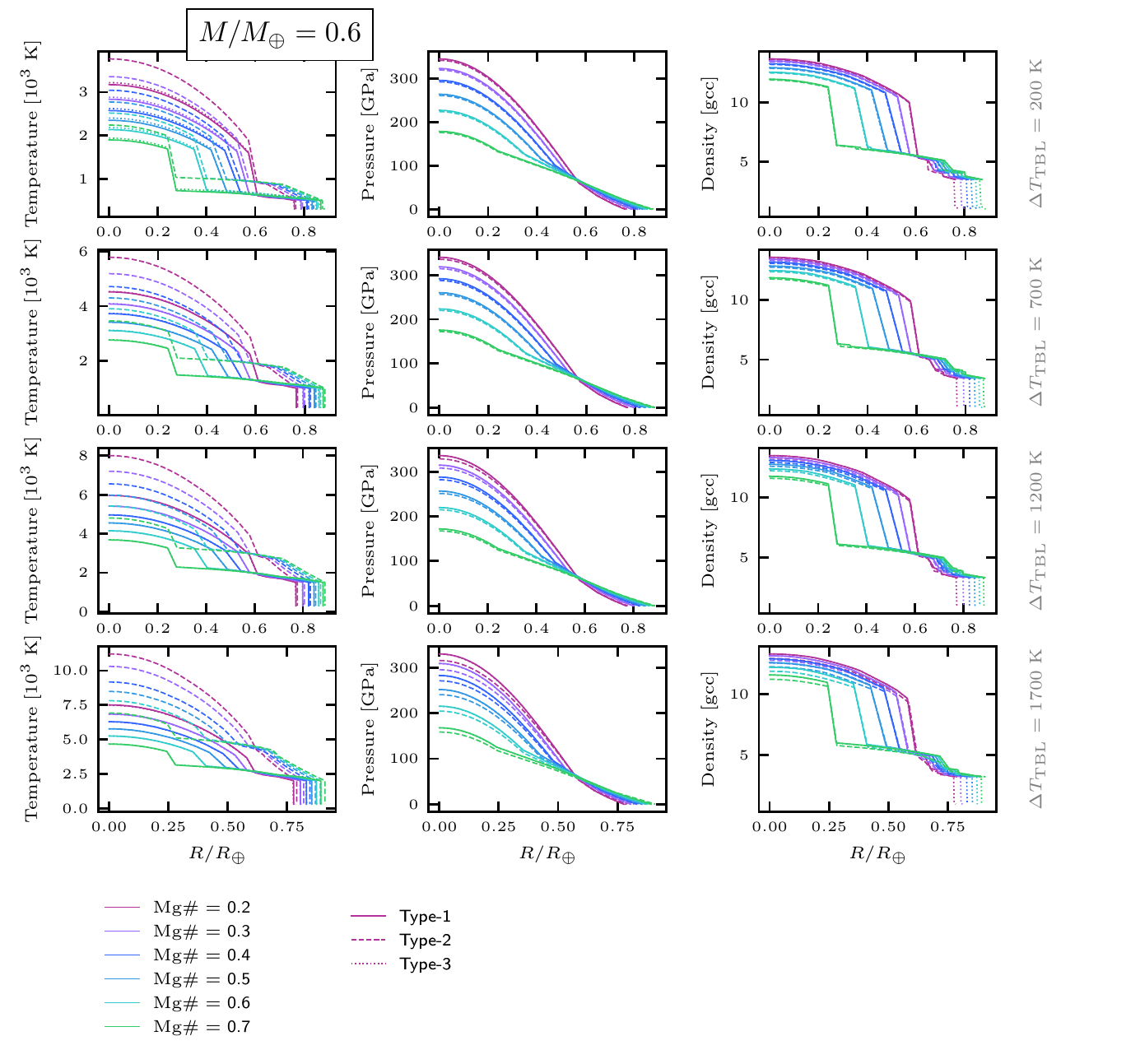}
\caption{Internal profiles as functions of radial distance from the centre for $M/M_{\oplus} = 0.6$ for all modelled $\Delta T_{\rm TBL}$ and compositions for the temperature (left column), pressure (middle column), and density (right column). The solid curves correspond to Type-1 planets, the dashed curves to Type-2 planets and the dotted curves to Type-3 planets (see text for details). Since hydration tends to steepen the temperature gradient in the mantle, the Type-2 planets exhibit generally higher temperatures in the interiors than their Type-1 and Type-3 analogues.}
\label{fig:profiles2}
\centering
\end{figure*}
\begin{figure*}[t]
\includegraphics[width=\textwidth]{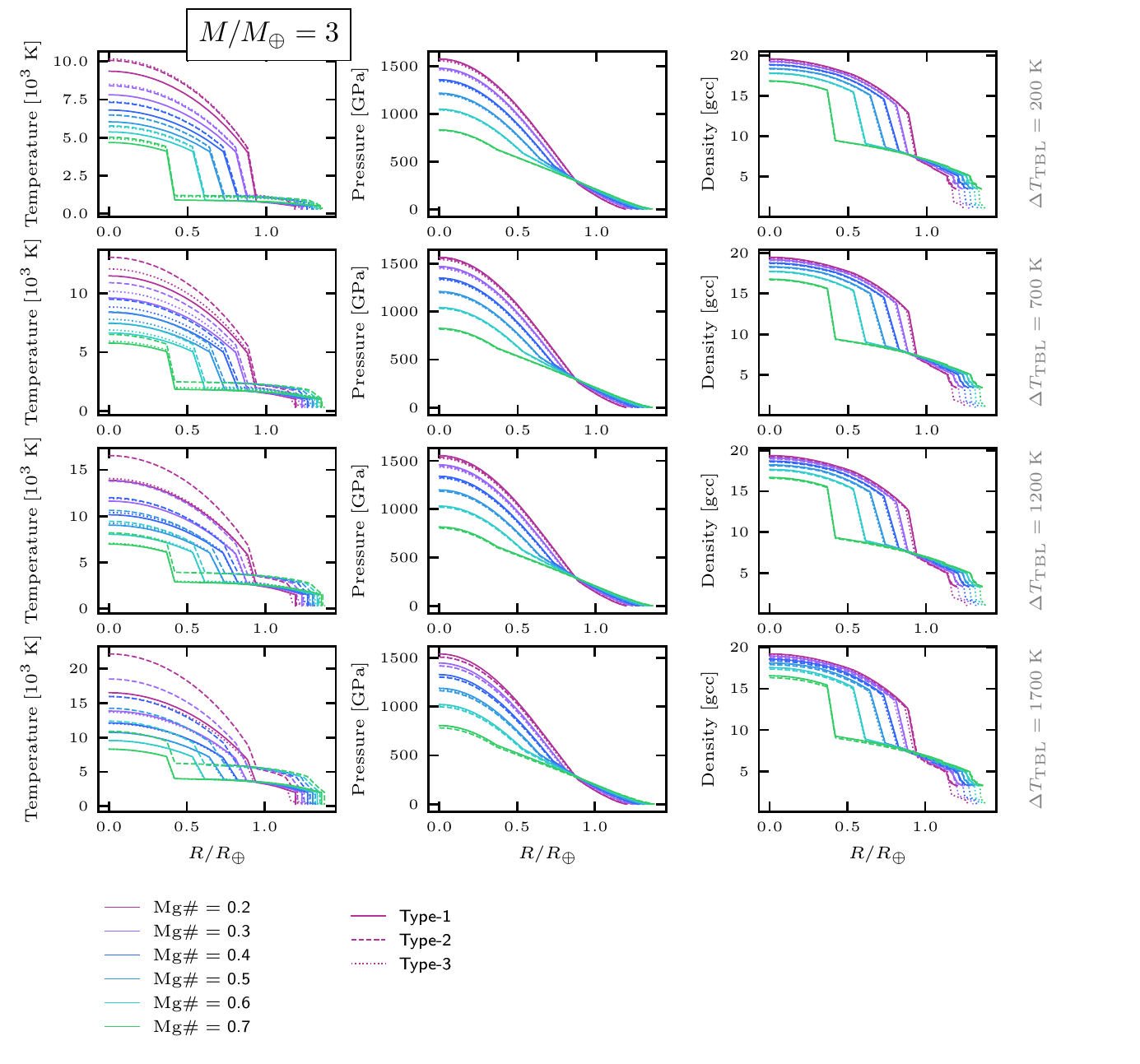}
\caption{Internal profiles as functions of radial distance from the centre for $M/M_{\oplus} = 3$ for all modelled $\Delta T_{\rm TBL}$ and compositions for the temperature (left column), pressure (middle column), and density (right column). The solid curves correspond to Type-1 planets, the dashed curves to Type-2 planets and the dotted curves to Type-3 planets (see text for details). Since hydration tends to steepen the temperature gradient in the mantle, the Type-2 planets exhibit generally higher temperatures in the interiors than their Type-1 and Type-3 analogues. For larger masses, the temperature profiles of the Type-3 planets can, however, exceed the temperature of the Type-2 planets for low values of $\Delta T_{\rm TBL}$ (see top row).}
\label{fig:profiles3}
\centering
\end{figure*}
\begin{figure*}[t]
\includegraphics[width=\textwidth]{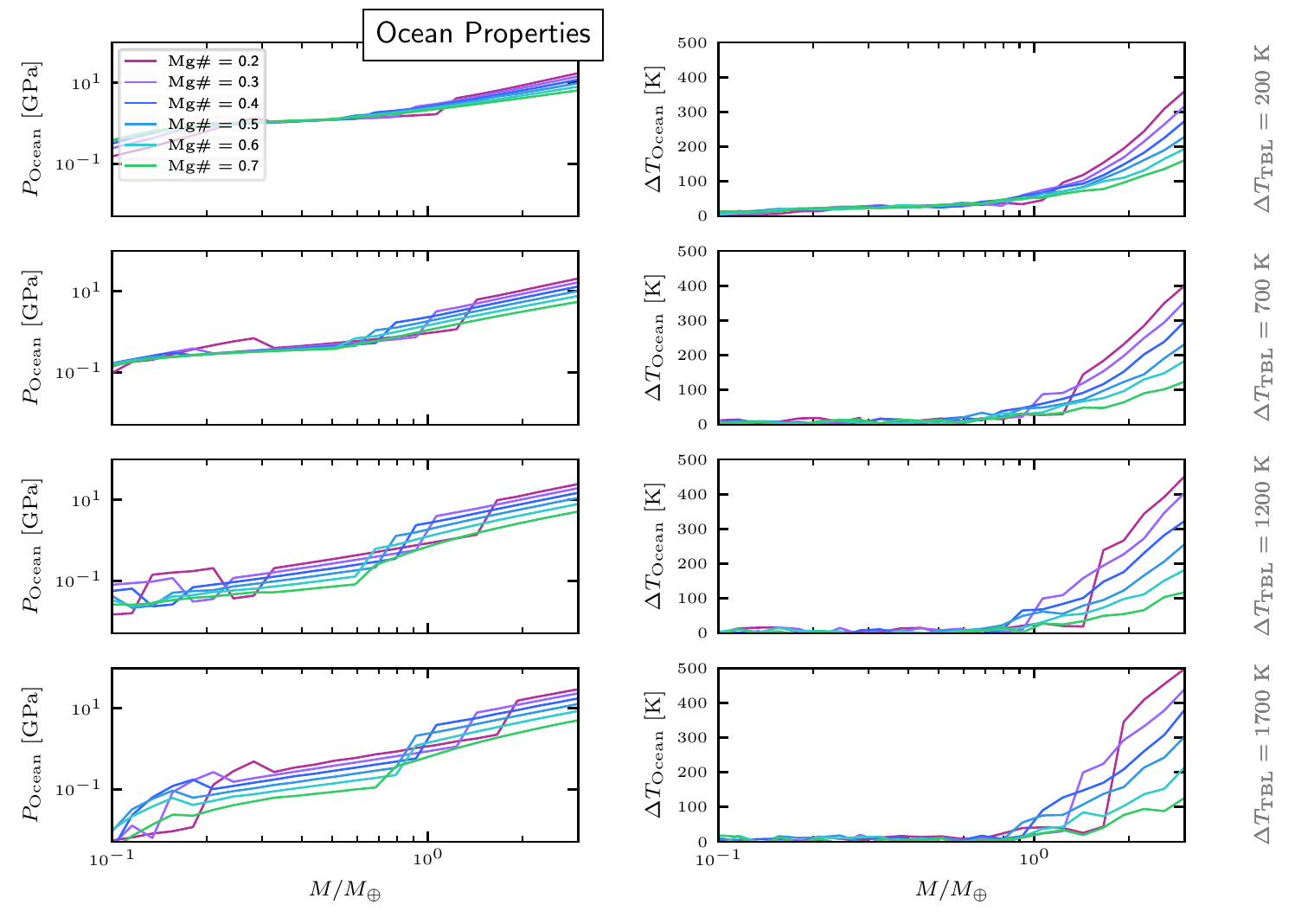}
\caption{Pressure at the bottom of the ocean (left column) and the temperature difference between the surface and the bottom of the ocean (right column). For large masses, the ocean mass is sufficient to reach pressures at the bottom of the oceans that exceed the transition pressure from liquid water to high pressure ices. This can also be seen in the density profiles in Fig. \ref{fig:profiles3}.}
\label{fig:ocean_props}
\centering
\end{figure*}
\section{The hydration model}\label{sec:saturationmodel}
\subsection{The mantle}\label{sec:mantle_hyd}
Here we present a simple model for water saturation in Fe bearing Olivine polymorphs (upper mantle) and Brucite and Perovskite or post-Perovskite (lower mantle). We choose these minerals because they are known to be major constituents of the Earth's mantle, where they are thought to play significant roles as hydrated Mg-silicates (see Section \ref{sec:comp}). The upper limit of hydration in the Mg-silicates is in general given by the saturation water content, which must be either measured in experiments or computed using molecular dynamics simulations. 
In an astrophysical framework, however, the extent of internal hydration of a planet cannot be assessed directly. That is, we cannot constrain it from stellar composition and total mass and radius alone. Thus, the water content in the mantle is written as $X_{\rm H_2O}(P,T) = \epsilon_{\rm H_2O}(P,T) X_{\rm H_2O, sat}(P,T)$, where $\epsilon_{\rm H_2O} \in [0, 1]$ is a free model parameter. It can be constrained from geophysical studies of terrestrial planets. Indeed, numerous authors have investigated the influence of water on the interior evolution of small objects and super-Earths and how the internal water content in mantle Mg-silicates evolves under the influence of a variety of geochemical and physical processes. For Mars, for instance, it has been shown that water loss from the interior over its evolution has been minor and that most of the initial water content in the mantle should still be present today (\cite{Hauck2002, Ruedas2013}). 
\cite{Nakagawa2015} modelled water circulation mechanisms for the Earth and found that the water content in an initially anhydrous mantle increases over time as a result of hydration of subducting slabs. They found that a balance between regassing and degassing at shallow depths is reached, which dictates the total water content in the mantle. \cite{Nakagawa2017} showed that in the case of strongly water-dependent viscosity water transport into the lower mantle is more effective, which can lead to higher internal water contents. \cite{Tikoo2017} modelled magma ocean solidification with variable initial water contents for the young Earth and showed that the water can be retained at saturation levels in the mantle upon cooling. Furthermore, \cite{Cowan2014} investigated water cycling between mantle and ocean and found that super-Earths tend to absorb more water in their mantles as the seafloor pressure is proportional to the surface gravity. 
These studies suggest that whenever a sufficient amount of water is present during planetary evolution, it is likely that the Mg-silicate mantle of the planet either remains or becomes strongly hydrated, which means $\epsilon_{\rm H_2O} \approx 1$. 
We are mainly interested in ocean planets that contain between a few and tens of wt$\%$ of water for which the condition of 'sufficient' amount of water is likely met. For the cases of lower water contents we still adopted $\epsilon_{\rm H_2O}=1$ to estimate the maximum possible effects of hydration. Therefore, whenever we refer to hydrated planets we shall assume fully water saturated Mg-silicates in the mantle, that is $X_{\rm H_2O}(P,T) = X_{\rm H_2O, sat}(P,T)$, unless otherwise stated.
\\

Our prescription for water saturation in Olivine polymorphs in the upper mantle is based on a collection of experimental data that has been acquired by various groups over the last decades. We construct a simple tri-linear model from a least squares fit to compute the saturation water content $X_{\rm H_2O, sat}(T,P, \xi_{\rm Fe})$ as a function of temperature $T$, pressure $P$ and iron content $\xi_{\rm Fe}$:
\begin{ceqn}
\begin{equation}\label{eq:bilinearmodel}
X_{\rm H_2O, sat}(T,P, \rm \xi_{Fe}) =  max \left( \sum_{\mathit{i}=0}^1 \sum_{\mathit{j}=0}^1\sum_{\mathit{k}=0}^1 \mathit {c_{ijk} T^i P^j} \rm \xi_{Fe}^{\mathit{k}}, 0 \right).
\end{equation}
\end{ceqn}
A full list of the collected data that has been used for the fit, along with the corresponding references, is provided in Table \ref{tab:alpha}-\ref{tab:gamma} for the three different phase regions of $\rm (Mg_{(1-\xi_{Fe})}, Fe_{\xi_{Fe}})_2 Si O_4$. The parameter range spanned by the data is $T \approx 1200 - 2400$ K, $P \approx 0 - 30$ GPa and $\rm{\xi_{Fe}}\approx 0 - 0.2$. The fit has been performed such that the input parameters for eq. \eqref{eq:bilinearmodel} are $T \ [\rm{K}]$, $P \ [\rm{GPa}]$, $\rm {\xi_{Fe}}$ [mol fraction] and the output is $X_{\rm H_2O, sat}$ [wt fraction]  (we note that these units are different from the original data set given in Table \ref{tab:alpha}-\ref{tab:gamma}). The resulting set of fitting parameters for eq.~\eqref{eq:bilinearmodel} are listed in Table \ref{tab:coeffTable1}. We find that, to avoid overfitting, the coefficients $c_{111}$ for all phases and $c_{011}$ and $c_{101}$ for the $\beta$- and $\gamma$-phase (for which the data sets are considerably smaller) have to be omitted. 
The same fit has been performed by omitting $c_{011}$ and $c_{101}$ also for the $\alpha$-phase. However, since the water content in the $\alpha$-phase spans a large range depending primarily on pressure and iron content, the effect of $\xi_{\rm Fe}$ on $X_{\rm H_2O, sat}$ is misrepresented at elevated pressures if $c_{011}$ and $c_{101}$ are excluded from the fit. Furthermore, at very high temperatures or very low pressures our fit can yield negative values for $X_{\rm H_2O, sat}$. We therefore invoke an artificial lower threshold value of $X_{\rm H_2O, sat} \geq 0$ to avoid a non-physical behaviour.
\\

In the lower mantle we chose Brucite ($\rm Mg(OH)_2$) and Perovskite ($\rm MgSiO_3$) as the water bearing phases. We fix the water content in hydrous Perovskite at 0.1 $\rm wt \%$ and neglect effects on the EoS for such low water contents (\cite{Inoue2010, Casanova2005}). For post-Perovskite it has been found that configurations containing more than 2 $\rm wt \%$ of water are stable (\cite{Townsend2015}). However, the pressure and temperature dependence of the water content has not been constrained. For this reason we adopt a constant value of 3 $\rm wt \%$ (see also Appendix \ref{sec:pv_per_wüs}). The stability field of Brucite is limited to rather low temperatures and pressures. It dissociates into MgO + $\rm H_2O$ above $\rm \approx 30-35 \ GPa$ (\cite{Hermann2016}). We extracted the stability field of $\rm Mg(OH)_2$ in the $P-T$ plane from Fig.~2 in \cite{Hermann2016}. As a result, the lower mantle can itself be divided into two parts: an upper part where $\rm Mg(OH)_2$ is present and a lower part where only MgO is present.

\subsection{The core}\label{sec:core_hyd}
We assume that the main source of hydrogen that is dissolved in iron and eventually deposited into the core comes from the Mg-silicates and enters the iron via the reaction (\cite{Wu_2018}):
\begin{ceqn}
\begin{align}\label{eq:hydrolysis_reaction}
\rm \left( \frac{x}{2} + 1 \right) Fe + \frac{x}{2} H_2 O \rightarrow \frac{x}{2} FeO + FeH_x. 
\end{align}
\end{ceqn}
In an early magma ocean hydrogen dissolves in liquid iron droplets and is incorporated into the core during the segregation process. If we assume that iron hydride, $\rm FeH_{\rm x}$, droplets are in chemical equilibrium with the Mg-silicates at all times, then the hydrogen content in the iron is dictated by the conditions at the CMB. This notion allows us to compute the hydrogen content in the core by adopting chemical equilibrium between the core and the lowermost mantle. If the water content in the mantle changes with time, either through external sources of water or the thermal evolution of the mantle, the hydrogen content in the core would adapt in order to maintain chemical equilibrium. Here we employ a simple model for the H-solubility as a function of the temperature at the CMB and the partial pressure of $\rm H_2$ in the lowermost part of the silicate mantle following \cite{Wu_2018}. These authors have used Sieverts' law for gas solubility in metals to estimate the hydrogen concentration $\xi_{H}$ in the iron cores of planetary embryos. We refer the interested reader to their Section 2.3 for further details. For our purpose it is sufficient to state the prescription for the hydrogen solubility in pure iron, which is given by:

\begin{ceqn}
\begin{align}\label{eq:Sievert}
\xi_{\rm H} = \left( \frac{P_{\rm H_2}}{P_0}\right)^{1/2} \cdot {\rm exp} [(\Delta H^0 - T_{\rm CMB} \Delta S^0)/RT_{\rm CMB}],
\end{align}
\end{ceqn}
where $P_0 = 1 \  \rm bar$ is the reference pressure, $P_{H_2}$ is the partial pressure of $\rm H_2$ in the silicates at the CMB, $R$ is the universal gas constant and $\Delta H^0 = +31.8 \ \rm kJ \ mol^{-1}$ and $\Delta S^0 = -38.1 \  \rm kJ \ mol^{-1} \ K^{-1}$ are the enthalpy and entropy for dissolution of hydrogen into iron, respectively. $P_{\rm H_2}$ is given by the amount of hydrogen present in the silicate mantle at the CMB. Here we used the van-der-Waals (VDW) EoS to relate the density of $\rm H_2$ to the partial pressure at given temperature. The Sieverts law is well suited to compute gas solubility in metals over an extended pressure range. However, \cite{Sugimoto1992} suggested that the Sieverts law underestimates the hydrogen solubility in iron at partial pressure $P_{\rm H_2} > 1 \ \rm GPa$. Therefore, we restrict our study to planets with $M \leq 3 \ M_\oplus$, which keeps $P_{\rm H_2}$ strictly below $\approx$ 1.2 GPa for all modelled cases (see. Fig. \ref{fig:core_content}).

\section{Results and discussion}\label{sec:results}
In the following we present the results for the total storage capacities of $\rm H_2 O$ equivalent (including core- and mantle reservoirs) in the mass range $0.1 \leq M/M_\oplus \leq 3$ for different bulk compositions, 1 bar surface pressure, 300 K surface temperature and $\Delta T_{\rm TBL}$ between 200 K and 1700 K. We also compare the different types of planets as described in Fig. \ref{fig:compare_counterparts} and  estimate the net effect on the total radii as well as the internal density, pressure and temperature profiles. Finally, we discuss possible implications of our results to exoplanet characterization.
\subsection{Water storage capacity and mass-radius relations}\label{sec:waterstoragecapacity}
Here we address three main questions to illustrate the application of our model:
\begin{enumerate}
	\item What is the maximum amount of $\rm H_2O$ equivalent that a terrestrial object of a given size, composition, and surface conditions can store in its interior?
	\\
	\item How different is the radius of such a hydrated object in comparison to its dry counterpart?
	\\
	\item How much does the radius change if the internal reservoir of $\rm H_2O$ equivalent is moved into an isolated surface ocean?
	
\end{enumerate}
To answer these questions we modelled three types of planets in the mass range $0.1 \leq M/M_{\oplus} \leq 3$ by solving the structure equations for adequate boundary conditions (see Section \ref{sec:planetmodel} for details). An overview of the different types of planets is given in Fig.~\ref{fig:compare_counterparts}. We modelled a total of 1728 planets for four different values of $\Delta T_{\rm TBL}$, six different $\rm Mg \#$, and twenty four values for $M$ uniformly distributed in logarithmic space in the range $0.1 \leq M/M_\oplus \leq 3$. Fig.~\ref{fig:internals_type_1}-\ref{fig:internals_type_3} show the central temperatures $T_{\rm C}$ (first column), central pressures $P_{\rm C}$ (second column) and the mass-radius relations (third column) obtained over the entire range of compositions and masses. The mass-radius relations for pure $\rm H_2 O$, MgO and Fe at $P_{\rm S}$ = 1 bar, $T_{\rm S}$ = 300 K are also shown for reference. We find that while the central temperature changes significantly between the different types, the central pressure is relatively  unaffected. This is because the adiabatic temperature gradient is steeper in hydrated Mg-silicates than in the anhydrous case, which leads to a net increase in the internal temperature profile for the Type-2 planets with respect to the Type-1 and Type-3 planets. Furthermore, Fig.~\ref{fig:internals_type_2} shows a slight change in the slope of the central temperature between $\approx 0.1-0.3 \ M_\oplus$ corresponding to the positions of local maximum in the internal water content (see also discussion below and Fig. \ref{fig:Delta_R1}). Larger core mass fractions generally lead to  larger central temperatures because the temperature gradient is steeper in the core than in the Mg-silicates. A similar behaviour is obtained for the central pressure. The central pressure decreases with $\rm Mg \#$ for all planets because the pressure gradient is steeper in the core than in the Mg-silicate mantle (see also Section \ref{sec:internalprofiles}). That is, for a given surface pressure and total mass, increasing the core mass fraction (or equivalently decreasing $\rm Mg \#$) must be compensated by an increase in the central pressure. The local maximum of the total water content are also reflected in the curves for the central pressure by slight changes of the slope between $\approx 0.1-0.3 \ M_\oplus$. The mass-radius relations are very similar for all planet types and show consistently larger radii for larger $\rm Mg \#$ due to  the lower density of the Mg-silicates with respect to the iron in the core. Furthermore, in Fig. \ref{fig:internals_type_3} small kinks in the radius between $\approx 0.8-2 \ M_\oplus$ are visible, corresponding to the Pv-pPv transition where the core reservoir in the Type-2 planets starts to increase drastically (see also next paragraph). This leads to a significant increase in the ocean mass fraction and hence reduction of the mean density in the Type-3 planets in the same mass range.
\\

\subsubsection{Comparison between Type-1 and Type-2 planets}\label{sec:1vs2}
In Fig.~\ref{fig:Delta_R1} we show the total water content (left column), the core- and mantle reservoirs (middle column) of the Type-2 planets and the resulting relative change in radius between Type-1 and Type-2 planets (right column). As expected, the fractional water content in the mantle increases strongly with the total $\rm Mg \#$. This is because larger mantle mass fractions have the effect of increasing the pressure range in the mantle, which leads to higher water saturation contents in Olivine polymorphs (see Fig.~\ref{fig:ParamMaps}). In addition, higher mantle mass fractions increase the amount of Mg-silicates with respect to the total planetary mass. Furthermore, the mantle reservoirs scale with the planetary mass only as long as an upper mantle is present. If the mass is sufficiently large that the transition pressure $\rm Ol \rightarrow Pv + Mw$ is reached, more material from the lower mantle with lower water content is added. This is reflected in the distinct peaks in the total water content and the mantle reservoirs between $0.1 \lesssim M/M_{\oplus} \lesssim 0.3$. The peak is highest for $\rm Mg \# = 0.7$ and $\Delta T_{\rm TBL} = 200 \ \rm K$ at $\approx 3.8 \rm \ wt\%$ and decreases strongly for lower $\rm Mg \#$ or higher $\Delta T_{\rm TBL}$. In contrast, at higher masses the total water content is dominated by the core reservoir, which leads to an opposite  trend. This is because the core reservoir scales with the core mass fraction and hence scales inversely with the $\rm Mg \#$. The core reservoirs themselves exhibit local maximum between $0.1 \lesssim M/M_{\oplus} \lesssim 0.3$ where the transition from the water rich upper mantle to the water poor lower mantle occurs. These maxima coincide with the local maxima in the mantle reservoirs (we note that this is only roughly true in the Fig. \ref{fig:Delta_R1} because of the finite resolution of the curves). At higher masses $0.6 \lesssim M/M_{\oplus} \lesssim 2$ the pressure at the bottom of the lower mantle exceeds the transition pressure between Pv and pPv leading to higher water contents in the Mg-Silicates at the CMB and hence higher equilibrium hydrogen content in the core. This is reflected in an increase in both core and mantle reservoirs for $M \gtrsim 0.6-2 \ M_\oplus$ and results in maximum total $\rm H_2 O$ equivalent reservoirs at $3 \ M_\oplus$ of up to $\approx 3 \  \rm wt \%$ for $\Delta T_{\rm TBL} = 200 \ \rm K$ and $\approx 6 \  \rm wt \%$ for $\Delta T_{\rm TBL} = 1700 \ \rm K$. The relative change in radius upon hydration is shown in the third column of Fig.~\ref{fig:Delta_R1} and exhibits three major trends:
\\

First, independent of temperature, the effect of hydration on the mean density is always strongest roughly in the range $0.1 \lesssim M/M_{\oplus} \lesssim 0.3$ where the mantle reservoirs peak despite the fact that for higher temperatures, the total fractional water content is much larger at higher masses. This means that hydration manifests more strongly for the upper mantle minerals than the iron core.
\\

Second, the heights of the peaks clearly depend on $\Delta T_{\rm TBL}$. Between 200 K and 700 K the maximum $\delta R_1 / R_1$ decreases from $\approx 0.012$ to $\approx 0.01$, remains roughly constant between 700 K and 1200 K and strongly increases between 1200 K and 1700 K up to $\approx 0.025$. To understand this behaviour several effects must be taken into account: (a) The water content in the upper mantle typically decreases with temperature. (b) The hydrogen content in the core increases with the water content in the upper mantle, provided there is no lower mantle. (c) The hydrogen solubility in the core increases with temperature. (d) Hydration tends to steepen the adiabatic temperature gradient leading to generally hotter interiors for the same boundary conditions, which increases the radius. Effects (a) and (b) lead to a decrease in the total fractional water content with temperature, which generally leads to a smaller change in the radius. Effect (c), on the other hand, leads to an increase in the total fractional water content with temperature. Hence, both effects (c) and (d) contribute positively to the relative change in radius while effects (a) and (b) have a negative contribution. The observed trend in the heights of the peaks therefore means that below 700 K effects (a) and (b)  dominate while above 1200 K effects (c) and (d) are more profound. In the intermediate range between $\sim$700-1200 K, the negative and positive contributions balance each other.
\\

Finally, the third trend is that the $\rm Mg \#$ for which $\delta R_1/R_1$ is largest at given mass and temperature corresponds to the value of $\rm Mg \#$ for which the total fractional water content is largest.  
Exceptions of this trend can be seen for $\Delta T_{\rm TBL} = 200 \ \rm K$ below $\approx 0.8 \ M_\oplus$. However, these small fluctuations
are likely to have a numerical rather than a physical origin.
\\

\subsubsection{Comparison between Type-2 and Type-3 planets}\label{sec:2vs3}
The comparison of Type-2 and Type-3 planets is summarized in Fig.~\ref{fig:Delta_R2}. The ocean mass fractions (middle column) for the Type-3 planets are calculated from the total water content in the Type-2 planets (compare first column in Fig.~\ref{fig:Delta_R1}). The corresponding ocean depths are shown in the left column of Fig.~\ref{fig:Delta_R2}. As expected, the ocean depths exhibit overall similar behaviour as the ocean mass fractions. But for lower masses a fixed ocean mass fraction results in generally lower ocean depths in comparison to the same ocean mass fraction at higher total mass. This is because $\left< \rho \right>R^3 \propto M$ for spherical objects (see Appendix \ref{sec:estimate_ocean_depth} for further discussion). We find that the estimated maximum $\rm H_2 O$ equivalent mass for all modelled planets corresponds to ocean layers with depths up to $\approx 800 \rm \ km$ for $M/M_{\oplus} = 3$ and $\rm Mg\# = 0.2$. The effect of $\Delta T_{\rm TBL}$ on the ocean depth is twofold: higher temperatures reduce the water content in the mantle but can increase the core reservoir of the Type-2 planets (see Fig.~\ref{fig:ParamMaps} and \ref{fig:core_content}). Because the water content in the low mass planets is dominated by the upper mantles, this results in a downwards trend of the ocean depth with increasing $\Delta T_{\rm TBL}$. Since $X_{\rm H_2 O}$ in pPv is constant, at higher masses only the positive effect of $\Delta T_{\rm TBL}$ on the core reservoir manifests. This explains why the ocean depths scale upwards monotonously with temperature for $M/M_\oplus \gtrsim 0.8$ but not for lower masses. However, we remind the reader that the water content in pPv is in reality likely to change with $P$ and $T$. A more realistic model for the hydration of pPv could therefore change the aforementioned behaviour.
\\

The relative difference in radius $\delta R_2 /R_2 \equiv (R_3 - R_2)/R_2$ between the hydrated Type-2 planets and the fully differentiated Type-3 planets is in general larger than $\delta R_1/R_1$ for a given mass. The maximum $\delta R_2/R_2$ can reach up to $\approx 5 \ \%$ for $M/M_\oplus = 3$ and $\rm Mg \# = 0.2$. For temperatures $\Delta T_{\rm TBL} > 700 \ \rm K$ the $\delta R_2 / R_2$ can in our model framework even become negative. This means, that differentiation into a surface ocean results in a net decrease in the planetary radius. Although counterintuitive at first glance, this feature does in fact arise intrinsically from the properties of the EoS used in this study. The temperature gradient in pure water tends to be shallower than in the hydrated silicates. As a result there is a considerable reduction of the interior temperature profiles between the Type-2 and Type-3 planets of up to a few thousand Kelvin (compare Figs.~\ref{fig:internals_type_2} and \ref{fig:internals_type_3} or see Section~\ref{sec:internalprofiles}). The densities of the cores and mantles of the Type-3 planets are hence enhanced due to the much lower temperatures and the absence of H or OH, respectively. For the 200 K and 700 K planets, this effect is, however, very small, explaining why in this case all Type-3 planets are indeed inflated with respect to their Type-2 counterparts due to the presence of the low density surface ocean. At higher values of $\Delta T_{\rm TBL}$ the effect becomes large enough to counteract the decreasing effect on the mean density of the surface ocean. This can lead to a decrease in the total radius unless enough water is present. For large ocean mass fractions the decreasing contribution of the low density ocean dominates and the Type-3 planets are  always larger than the Type-2 planets in the super-Earth regime. It is interesting to note that the change of sign of $\delta R_2/R_2$ roughly lies at $1 \ M_\oplus$ for solar composition ($\rm Mg \# \gtrsim 0.5$). This is because the Pv-pPv transition occurs at roughly the temperature and pressure conditions near the bottom of the Earth's lower mantle.
\\

Our findings indicate that the change in the total radius due to hydration is very likely below the detection limit of current or near future instrumentation for characterizing exoplanets. Furthermore, the maximum effect of hydration on the radius occurs in a mass range that is well below $1 \ M_\oplus$. The partitioning of water between internal and surface reservoirs, on the other hand, can indeed affect a planet's radius within plausible limits of observations in the foreseeable future. As this effect is enhanced for larger planetary masses, it is of particular interest for future characterization of planets in the super-Earth regime. However, here we used a relatively simple model for the hydration of the cores and mantles and have assumed a rather simple mantle composition. We discuss the limitations of our model in Section \ref{sec:caveats}.
%
\subsection{Internal profiles}\label{sec:internalprofiles}

In Figs.~\ref{fig:profiles1}-\ref{fig:profiles3} we show the internal temperature (left columns), pressure (middle columns), and density (right columns) profiles as functions of radial distance from the centre for three selected planetary masses. The solid, dashed, and dotted lines correspond to the Type-1, Type-2, and Type-3 analogues, respectively. The pressure profiles are very similar for all planet types. They are shifted to slightly lower pressures between Type-1 and Type-2 planets but are essentially indistinguishable by eye between Type-1 and Type-3 planets. The reason is the steeper pressure gradient in the core. As the mantle is hydrated, the ratio between total mantle mass (Mg-silicates plus water) and the core mass becomes larger. Therefore, for a fixed total mass and surface pressure, the central pressure needs to be reduced. Since the pressure difference in the surface oceans is very small in comparison to the central pressures no similar effect occurs between the Type-1 and Type-3 planets.
The temperature profiles are much more sensitive to the model type and this sensitivity increases with increasing mass. The profiles shift to higher temperatures between Type-1/Type-2 planets. This is because hydration tends to steepen the temperature gradient in the mantle. Furthermore, the temperature profiles of the Type-1 and Type-3 planets are indistinguishable for $0.1 \ M_\oplus$ but not for $ 0.6 \ M_\oplus$ and $3 \ M_\oplus$. The reason for this is that the oceans are very shallow at $0.1 \ M_\oplus$ and become deeper for higher masses. As deeper oceans generally lead to larger temperature differences between the surface and the oceans, the internal temperature profiles at given surface temperature are more affected for deeper oceans (compare also Fig.~\ref{fig:ocean_props}). An interesting feature in Fig. \ref{fig:profiles3} is that the temperature profiles in the Type-3 planets are higher for $\Delta T_{\rm TBL} = 200 \ \rm K$ than even in the Type-2 planets. For higher values of $\Delta T_{\rm TBL}$, however, the difference between the temperature profiles of the Type-3 planets and the Type-2 or Type-1 planets decreases. At very high temperatures the profiles of the Type-3 planets eventually fall even below the Type-1 planets. Considering the effect of the surface ocean on the temperature profiles alone cannot explain this behaviour. It would be intuitive to assume that larger oceans (larger temperature differences between surface and mantle) would monotonically increase the temperature in the mantle and the core. However, large ocean mass fractions also increase the pressure at the ocean-mantle interface. While the temperature gradient in the mantle becomes generally steeper for higher temperatures, it becomes shallower for higher pressures. For lower values for $\Delta T_{\rm TBL}$ the temperature effect dominates, rendering the temperature gradient steeper in the interior and hence resulting in somewhat higher temperature. For larger values of $\Delta T_{\rm TBL}$, however, the pressure effect takes over, leading to a net decrease in the temperature gradient. The density profiles are very similar for all planet types. The density of the mantles and cores of the Type-2 planets are reduced compared to the Type-1 planets because of the hydration effects on the EoS. The Type-1 and Type-3 planets have, apart from the surface oceans, nearly  identical density profiles.

Fig.~\ref{fig:ocean_props} shows the pressure at the bottom of the oceans (left column) and the temperature difference between the surface and the bottom of the oceans (right column) for Type-3 planets.  The ocean pressures reach up to $\sim 0.01-10 \ \rm GPa$ where $P_{\rm Ocean} \gtrsim 1 \ \rm GPa$ is only reached for $M \gtrsim 1 \ \rm M_\oplus$. This means that for the majority of the modelled objects the oceans are fully liquid (compare also Fig.~\ref{fig:water_phase_diagram}). For higher masses both the pressure and the temperature at the bottom of the ocean significantly increase. The temperature increase between the surface and the bottom of the ocean reaches up to $\approx 500 \ \rm K$, corresponding to temperatures at the bottom of the ocean of $\approx 800 \ \rm K$. The transition pressures from liquid water to the high pressure ices in the temperature range $\approx 300 - 800 \ \rm K$ roughly lie in the range $\approx 1-10 \ \rm GPa$. This suggests that for larger masses the oceans of the Type-3 planets can posses an upper liquid layer and a lower solid layer (this is also evident in the density profiles shown in Fig.~\ref{fig:profiles3}). These results indicate that the amount of $\rm H_2 O$  that can be stored in the interiors of terrestrial planets could reach up to ocean mass fractions sufficient for the formation of high pressure ice layers below the liquid parts of the oceans. This means, we find that large total water mass fractions of up to a few $\rm wt \ \%$ could in fact easily be partitioned into interior and surface reservoirs so that the remaining surface oceans, that would in isolation be sufficiently large to form high pressure ice layers, would be liquid throughout. For low enough surface temperatures it is in principle also possible to have an ocean that is frozen throughout, regardless of the total ocean mass fraction. This scenario has, however, not been considered in this study.
\\

\subsection{Implications for previous studies}\label{sec:connection_previous_work}
We have found that planets with considerable water content of at least a few $\rm wt \%$ could still have only very small or no surface oceans. This could have important implications in the context of previously published studies. The maximum radii for habitable planets derived by \cite{Alibert2014}, for instance, are likely to be underestimated. That is, the maximum total amount of $\rm H_2 O$ equivalent for which no high pressure ice layer forms at the bottom of the ocean could be considerably larger. This would lead to larger maximum radii for habitable planets.

Furthermore, our results could alter the findings of \cite{Grasset2009}. These authors have estimated that the total ocean mass fraction could be derived from the bulk composition, the mass and the radius of a planet alone with a standard deviation of no more than $4.5 \ \%$ if the mass and radius were to be measured precisely. However, our study shows that the partitioning of the total water content into internal and surface reservoirs can have considerable effects on the radius of a planet. In particular, for super-Earths the partitioning between surface and internal reservoirs would decrease the total radius with respect to a fully differentiated ocean planet. That means that the mean density of such a planet would become more similar to the mean density of a fully water depleted planet and it becomes more difficult to distinguish between these objects. In this light, it has to be assumed that the determination of the water mass fraction present on an observed exoplanet from mass-radius measurements is less precise than suggested by \cite{Grasset2009}.
\section{Caveats}\label{sec:caveats}

First, a number of simplifying assumptions for the bulk composition of the planets have been made (see Section~\ref{sec:comp}). Including a larger variety of mantle minerals could induce a different overall hydration behaviour. For instance, Pyroxene plays a dominant role along with Olivine in the upper mantle of the Earth (\cite{Jacobsen2006}). The coexistence of different minerals can change the water content in the end member phases (\cite{Grant2006}). Furthermore, $\rm SiO_2$ has been found to have large water storage capacity over a very wide pressure range and could be a major agent for water storage under lower mantle conditions. \cite{Nisr2020} have studied hydrous $\rm SiO_2$ samples up to $\approx 100 \ \rm GPa$ and find water contents of $0.4-8.4 \rm \ wt \%$. We have not included the possible presence of $\rm SiO_2$ in our models. In addition to the exclusion of some minerals that are thought to be important constituents in the Earth's mantle, the hydration behaviour of some of the minerals are very weakly constrained. Most importantly, the saturation content in pPv is only poorly studied and to our knowledge, no data on the hydration behaviour as a function of pressure and temperature is currently available. The simplified assumption of constant water content of $3 \ \rm wt \%$ over the entire pressure and temperature range is likely to be unrealistic. The water content in pPv dictates the equilibrium hydrogen content in the core and is therefore a key parameter in our models and presumably one of the major sources of uncertainty at this stage.

These simplifying assumptions on the composition limit the interpretability of our results with respect to specific objects. For example, when assuming roughly solar $\rm Mg \# \approx 0.5$ for the bulk Earth, our results indicate that a maximum of $\approx 1.4 \ \rm wt \%$ of water equivalent could be stored in the Earth's interior (see Fig.~\ref{fig:Delta_R1}). This is much higher than standard estimates for the actual water content. From isotopic mass balance the total amount of water in the Earth's interior at present is found to be $\approx 0.06-0.3 \ \rm wt\%$ (see review by \cite{Marty20120}). Geochemical and geophysical considerations lead to generally lower estimates for the mantle reservoirs of $\approx0.006-0.07 \ \rm wt\%$ (see review by \cite{Ohtani2019}). Furthermore, \cite{Cowan2014} have developed simple two-box models for the water cycling between ocean and mantle and derived the internal water content of terrestrial planets based on the equilibrium between degassing and regassing. They reported a steady state solution for Earth with an internal water content of $\approx 0.08 \ \rm wt \%$.
One reason for the large discrepancy between our model and these results is, of course, the fact that we assumed water saturated Mg-silicates in the mantle. However, the Earth's mantle is likely not fully water saturated (\cite{Ohtani2019}) and hence we do not expect our model to reproduce the actual water content but rather to give an upper limit. Another important factor is the fact that the ratio [Si]/[Mg] takes a fixed value of [Si]/[Mg]=2/3 for our simplified composition. This is much lower than the solar value of $\rm [Si]/[Mg] \approx 0.9$ (\cite{Sotin2007}). Furthermore, it is not clear if the Earth's core contains significant amounts of hydrogen.
Adjusting the composition to roughly solar [Si]/[Mg] requires the inclusion of more Si-enriched phases in the mantle. We performed preliminary tests by adding dry $\rm (Mg,Fe)_2 Si_2 O_6$ to the upper mantle and increase the amount of $\rm (Mg,Fe)SiO_3$ in the lower mantle accordingly in order to increase the ratio [Si]/[Mg] to roughly solar. Furthermore, we have used a lower Fe content in the Mg-silicates of $\xi_{\rm Fe} \approx 0.1$, which is thought to approximately represent the average Fe content in the Earth's mantle (\cite{Sotin2007}). Finally, to mimic the Earth as closely as possible we adopted temperature discontinuities in the mantle transition zone (MTZ) at the top of the upper mantle of $\Delta T_{\rm MTZ} = 300 \ \rm K$ (\cite{Sotin2007}). With this the maximum water content in the mantle for an Earth analogue is $\approx 0.28 \ \rm wt\%$, in good agreement with the value range obtained from \cite{Marty20120}. However, since we ignored the hydration of Pyroxene for this preliminary test, the actual predicted value for the maximum water content in the Earth's mantle from our model would be somewhat higher. Furthermore, we have not included an $\rm SiO_2$ phase in the mantle composition, which could further increase the overall water storage capacity. In addition, the water content should be expected to be rather sensitive to $\Delta T_{\rm MTZ}$ and $\Delta T_{\rm TBL}$ as these parameters strongly affect the thermal profile in the mantle. These aspects indicate that the water content in the planetary interior is extremely sensitive to the bulk composition and the thermal state of the interior. Therefore, our model is suited for estimating the  upper limits for the hydration effects for a given planetary mass and for $\rm Si/Mg = 2/3$. 
Currently our model cannot be used to accurately predict the maximum water reservoirs in the interior of an object with a different composition. To this end, more diverse compositions need to be considered.
\\

Second, in addition to excluding a number of mineral phases in the mantle that could play an import role for water storage, we also have not included possible effects on hydration of other elements s.a. C, Ni, Ca, S or Al. 
Various studies show, however, that water solubility can be rather sensitive to the incorporation of some of these elements. For example, Al has been found to affect the stability field of hydrous phases (\cite{Hermann2016}) and could lead to the stability of hydrous phases up to higher pressures. It furthermore has been found to increase the water solubility in Olivine polymorphs and Pyroxene (\cite{Grant2006, Ferot2012}).

Accounting for more complex compositions as outlined above would also rise the question of how adequate our assumption of uniform composition in the mantle is (see secs.~\ref{sec:comp} and \ref{sec:shell_cont}). 
For example the Al content in Olivine is found to decrease significantly with increasing pressure and decreasing temperature (\cite{Ferot2012}). As the Al content can strongly influence the hydration behaviour, its pressure and temperature dependence might be an important factor. 
Similar issues could arise for the Fe content but has not been further investigated by the others at this point. 
However, we argue that the generalization to more complex compositions could inevitably lead to the necessity of allowing for composition gradients at least in the mantles of the planets. 
This would require a more sophisticated geochemical model for the phase assemblages as functions of pressure and temperature and the corresponding partitioning of water between the coexisting phases. We plan to address some of these topics in a subsequent study.
\\

Third, our model is restricted to $M \leq 3 \ M_\oplus$ due to the simple model for the chemical equilibrium between the core and mantle. As a result, for higher masses the partial hydrogen pressure in the lower parts of the mantle can  exceed 1 GPa where the Sieverts law underestimates the hydrogen solubility in iron (\cite{Sugimoto1992}). In the context of exoplanets, however, the super-Earth regime $1 \lesssim M/M_\oplus \lesssim 10$ is of great  interest. 
Therefore, it is desirable to employ more sophisticated chemical models in the future so the model can be  extended to higher masses.
\\

Fourth, we treated $\Delta T_{\rm TBL}$ and $\xi_{\rm Fe}$ as free parameters. Furthermore, the estimate of $\Delta T_{\rm CMB}$ via eq.~\ref{eq:delta_T_CMB} might not be universally applicable to planets of different sizes, compositions, and evolution histories. This renders these three aforementioned parameter as important sources of uncertainty. In Appendix \ref{sec:sensitivity_study} we present a brief sensitivity study of our model with respect to $\xi_{\rm Fe}$ and $\Delta T_{\rm CMB}$. Both parameters could have significant effects on the H and OH reservoirs and the resulting effects on the mean densities of the planets. In the super-Earth regime, the role of $\xi_{\rm Fe}$ is difficult to estimate at this stage. This is because the effect of the iron content on the hydration behaviour of pPv could not be included in our model due to a lack of available data. Furthermore, the roles of pressure and temperature on the water content in pPv are, to our knowledge, not further constrained neither by experiments nor numerical simulations. However, the exact behaviour of $X_{\rm H_2 O}(P, T, \xi_{\rm Fe})$ in pPv could significantly affect the predicted mantle and core reservoirs of super-Earths. We therefore emphasize the importance of further investigation of these minerals to allow more reliable predictions of internal H and OH reservoirs of terrestrial planets.
\\

\section{Summary and conclusions}\label{sec:conclusions}
We have presented planetary structure models that account for the hydration of the Mg-silicate mantles and the iron cores. This allows for assessing the corresponding effects of water-interior interactions and ocean differentiation on the mass-radius relation and the partitioning of water between surface and internal reservoirs. 

We considered three types of planets with simple bulk compositions (see Fig. \ref{fig:compare_counterparts}): Dry planets (Type-1), hydrous planets (Type-2) and fully differentiated ocean planets (Type-3). For a first set of models we only considered Olivine polymorphs, Perovskite, Magnesiowüstit and Brucite in the mantle with a uniform Fe content of 25 mol$\%$. The core-to-mantle mass ratio was derived from the total magnesium number Mg$\#$, which is a free parameter. We then computed mass-radius curves for different bulk compositions at 300 K surface temperature and 1 bar surface pressure in the mass range $M = 0.1-3 \ M_{\oplus}$ and for different values for the temperature drop in the thermal boundary layer.
Our main conclusions can be summarized as follows:

\begin{enumerate}

    \item The storage capacity of $\rm H_2 O $ equivalent in Type-2 planets is $\approx 0-6 \ \rm wt\%$ for $0.2 \leq \rm Mg\# \leq 0.7$ and $M \leq 3 \ M_{\oplus}$. This corresponds to ocean depths in Type-3 planets of $\approx 0-800 \ \rm km$.
	\\
	\item The effect of hydration on the planetary radius peaks at $\approx 0.1-0.3 \ M_{\oplus}$.
	\\
	\item The radii of super-Earths can be considerably smaller if water is stored in the interior instead of an isolated surface ocean.
	\\
	\item For the effects of hydration and ocean separation on the total radius, we find $\delta R_1/R_1 \leq 2.5 \ \%$ and $\delta R_2/R_2 \leq 5 \ \%$, respectively.
	\\
	\item The temperature drop at the TBL has a great influence on the H and OH storage capacity and the resulting total radii.
	\\
	\item The results are very sensitive to the bulk composition and planet mass.

\end{enumerate}

These results reveal the necessity of accounting for internal hydration for a more reliable characterization of the compositions and structures of exoplanets, in particular in the super-Earth regime. This will become even more essential with the increased precision of the mass-radius measurements that are expected in the near future. Here we have laid the foundations for further guiding the interpretation of observational data by incorporating these aspects into theoretical predictions of mass-radius relations.

\begin{acknowledgements}
This work has been carried out within the framework of the National Centre of Competence in Research PlanetS supported by the Swiss National Science Foundation. The authors acknowledge the financial support of the SNSF. We thank Morris Podolak and the anonymous referee for valuable comments.   
\end{acknowledgements}

%
%

\bibliography{library}

\appendix
\section{The sensitivity of the results to $\Delta T_{\rm CMB}$ and $\xi_{\rm Fe}$}
\label{sec:sensitivity_study}
In this section we investigate the sensitivity of our results to the assumed temperature drop at the CMB ($\Delta T_{\rm CMB}$) and the iron content in the Mg-silicates in the mantle ($\xi_{\rm Fe}$). 
We ran three comparison cases: $\Delta T_{\rm CMB}=0.8\Delta T_{\rm CMB, nom}$, $\Delta T_{\rm CMB}=1.2\Delta T_{\rm CMB, nom}$ and $\xi_{\rm Fe} = 0$, where the subscript 'nom' refers to the nominal case as presented in the main text. For simplicity, we only ran the bracketing cases of $\rm Mg \# = 0.2, 0.7$ and $\Delta T_{\rm TBL} = 200, 1700 \ \rm K$. $\xi_{\rm Fe}$ is kept at the nominal value when $\Delta T_{\rm CMB}$ was changed and vice versa. Furthermore, the convergence criteria for the planets were somewhat relaxed as we only aim to give a rough estimation of the effects of these parameters.

\begin{figure*}[t]
\includegraphics[width=\textwidth]{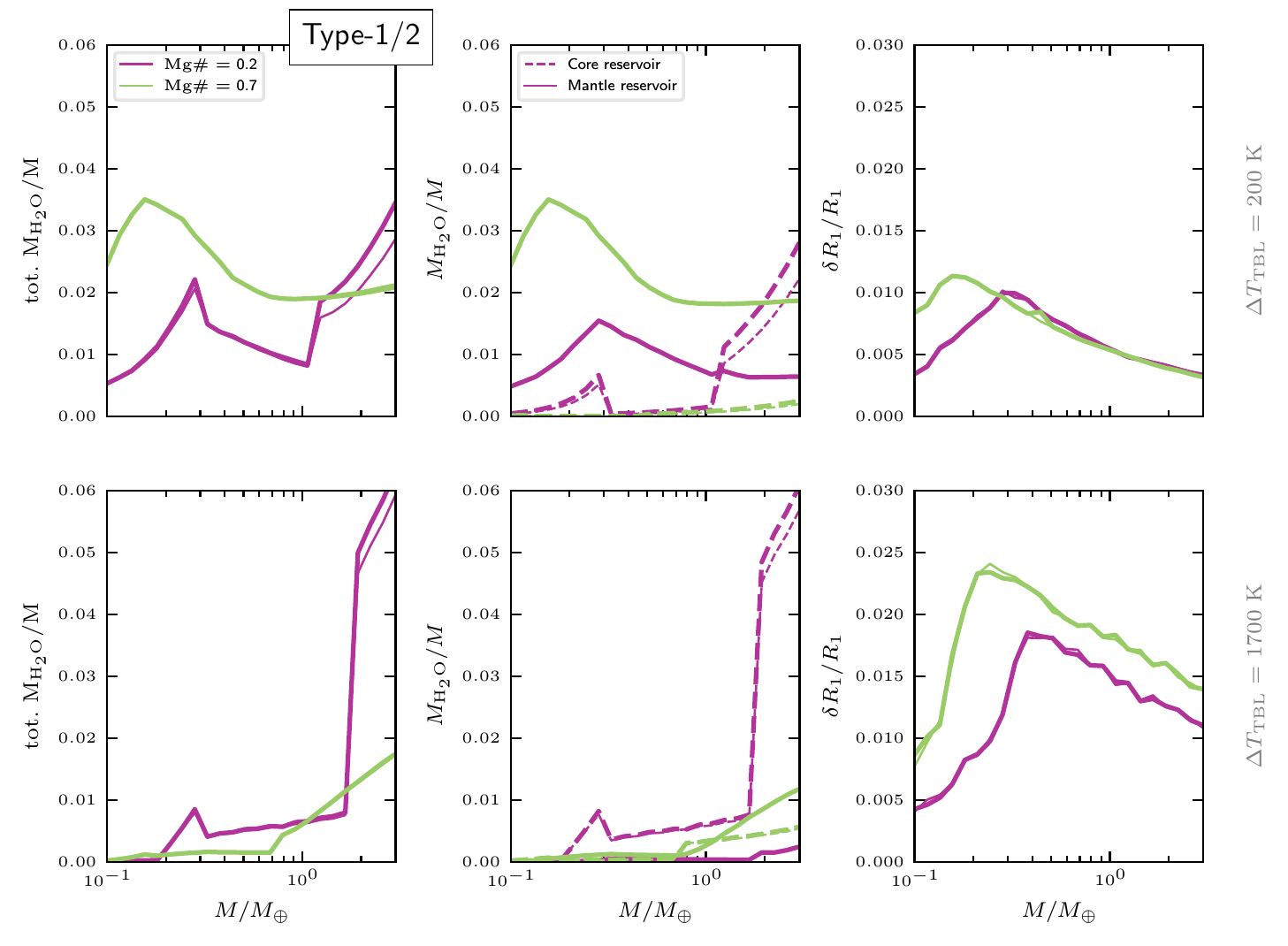}
\caption{Comparison of the nominal case (thick curves) and $\Delta T_{\rm CMB} = 0.8 \cdot \Delta T_{\rm CMB, nom}$ (thin curves) for the bracketing cases $\rm Mg \# = 0.2, 0.7$ and $\Delta T_{\rm TBL} = 200, 1700 \ \rm K$.}
\label{fig:Delta_R1_CMB_0_8}
\centering
\end{figure*}

\begin{figure*}[t]
\includegraphics[width=\textwidth]{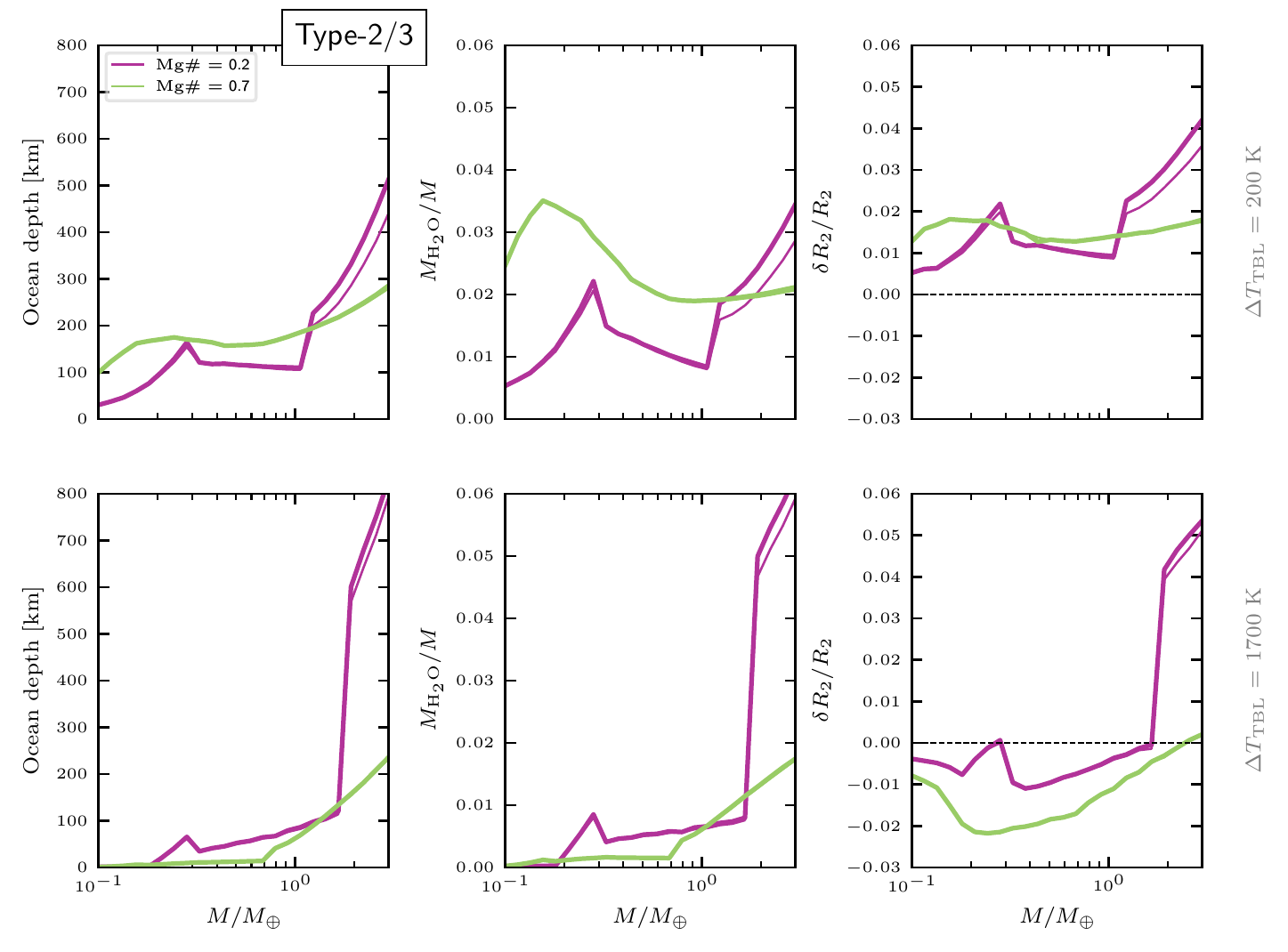}
\caption{Comparison of the nominal case (thick curves) and $\Delta T_{\rm CMB} = 0.8 \cdot \Delta T_{\rm CMB, nom}$ (thin curves) for the bracketing cases $\rm Mg \# = 0.2, 0.7$ and $\Delta T_{\rm TBL} = 200, 1700 \ \rm K$.}
\label{fig:Delta_R2_CMB_0_8}
\centering
\end{figure*}

\begin{figure*}[t]
\includegraphics[width=\textwidth]{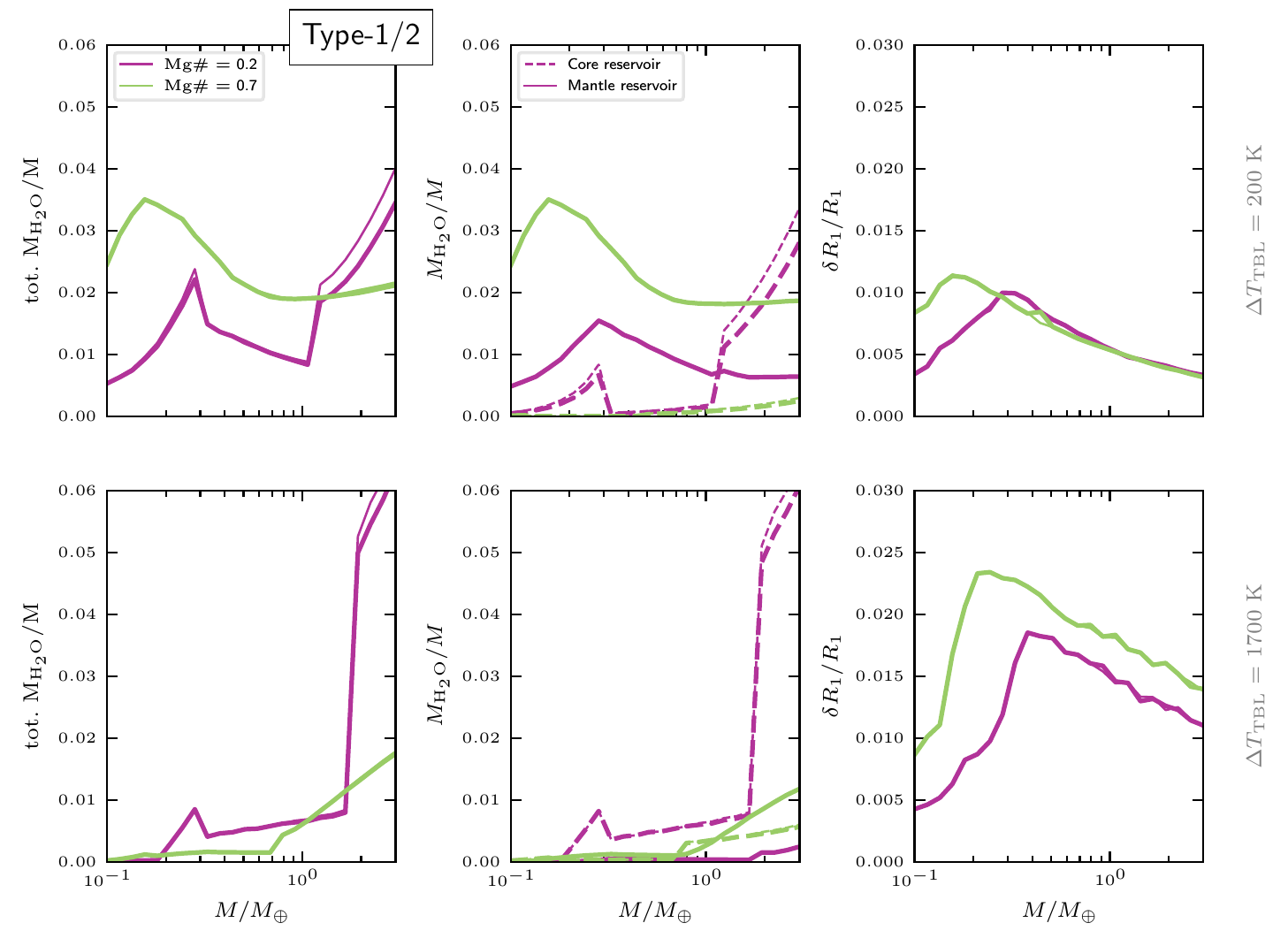}
\caption{Comparison of the nominal case (thick curves) and $\Delta T_{\rm CMB} = 1.2 \cdot \Delta T_{\rm CMB, nom}$ (thin curves) for the bracketing cases $\rm Mg \# = 0.2, 0.7$ and $\Delta T_{\rm TBL} = 200, 1700 \ \rm K$.}
\label{fig:Delta_R1_CMB_1_2}
\centering
\end{figure*}

\begin{figure*}[t]
\includegraphics[width=\textwidth]{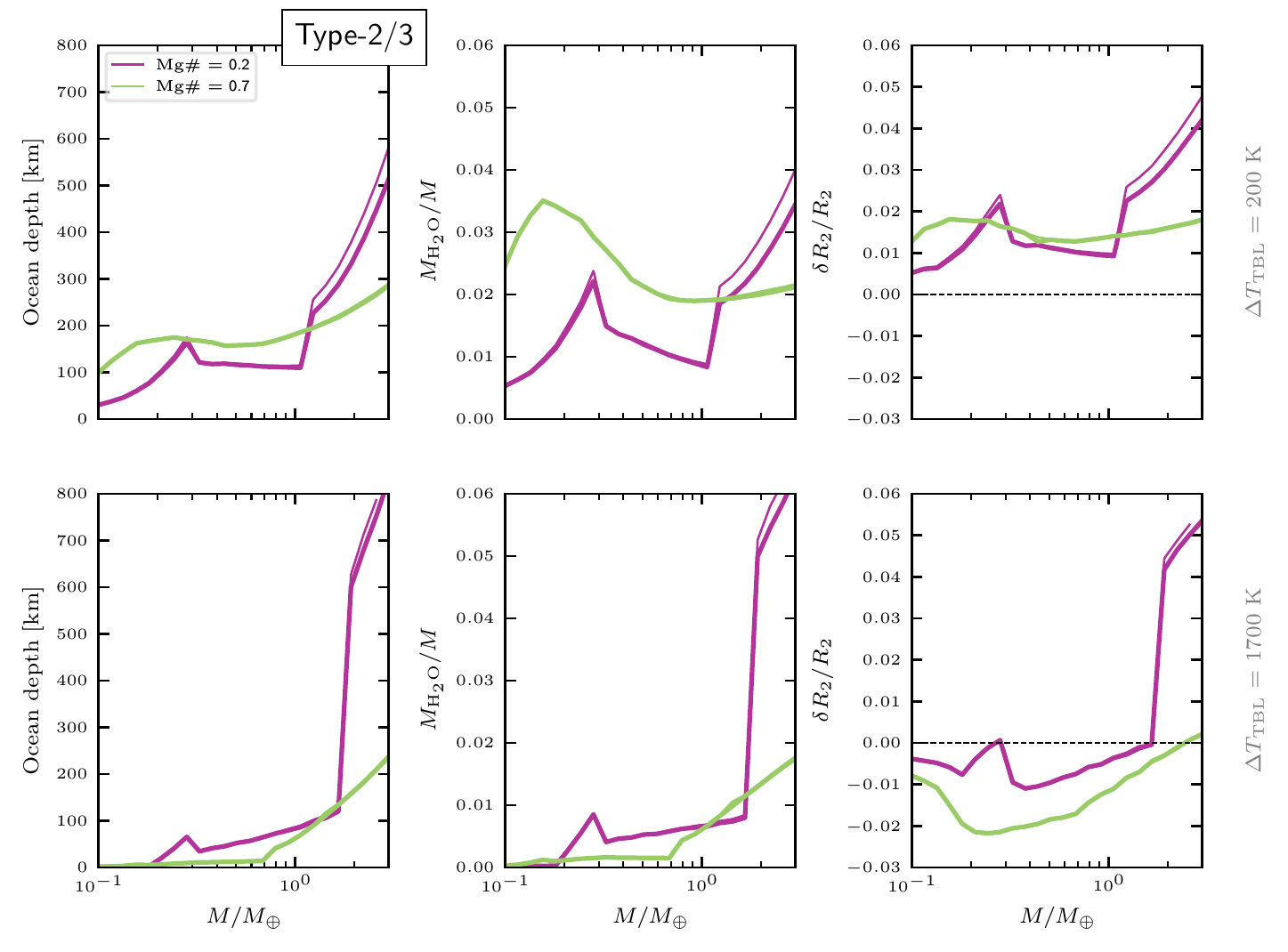}
\caption{Comparison of the nominal case (thick curves) and $\Delta T_{\rm CMB} = 1.2 \cdot \Delta T_{\rm CMB, nom}$ (thin curves) for the bracketing cases $\rm Mg \# = 0.2, 0.7$ and $\Delta T_{\rm TBL} = 200, 1700 \ \rm K$.}
\label{fig:Delta_R2_CMB_1_2}
\centering
\end{figure*}

\begin{figure*}[t]
\includegraphics[width=\textwidth]{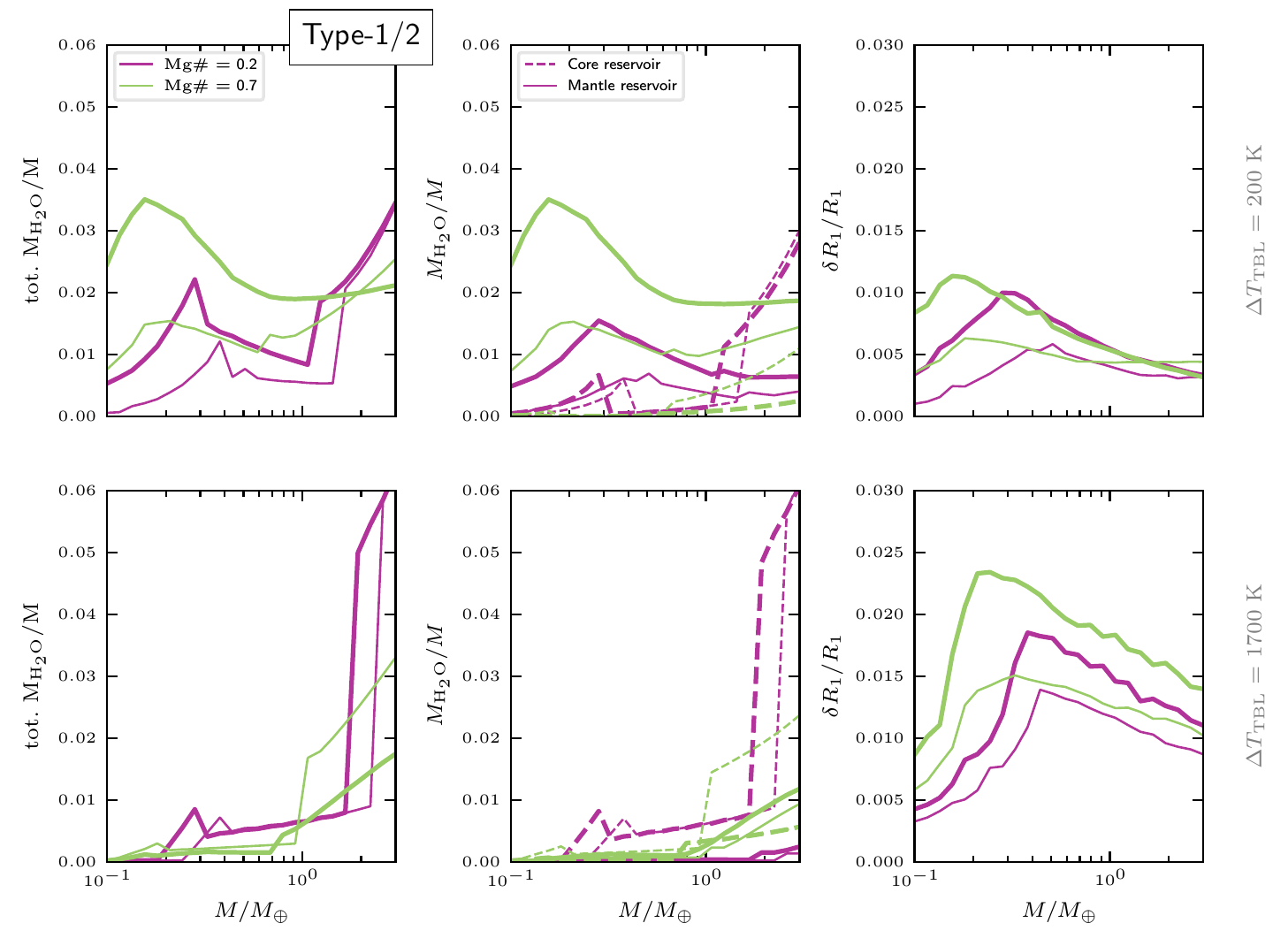}
\caption{Comparison of the nominal case (thick curves) and $\xi_{\rm Fe} = 0.0$ (thin curves) for the bracketing cases $\rm Mg \# = 0.2, 0.7$ and $\Delta T_{\rm TBL} = 200, 1700 \ \rm K$.}
\label{fig:Delta_R1_Fe_0}
\centering
\end{figure*}

\begin{figure*}[t]
\includegraphics[width=\textwidth]{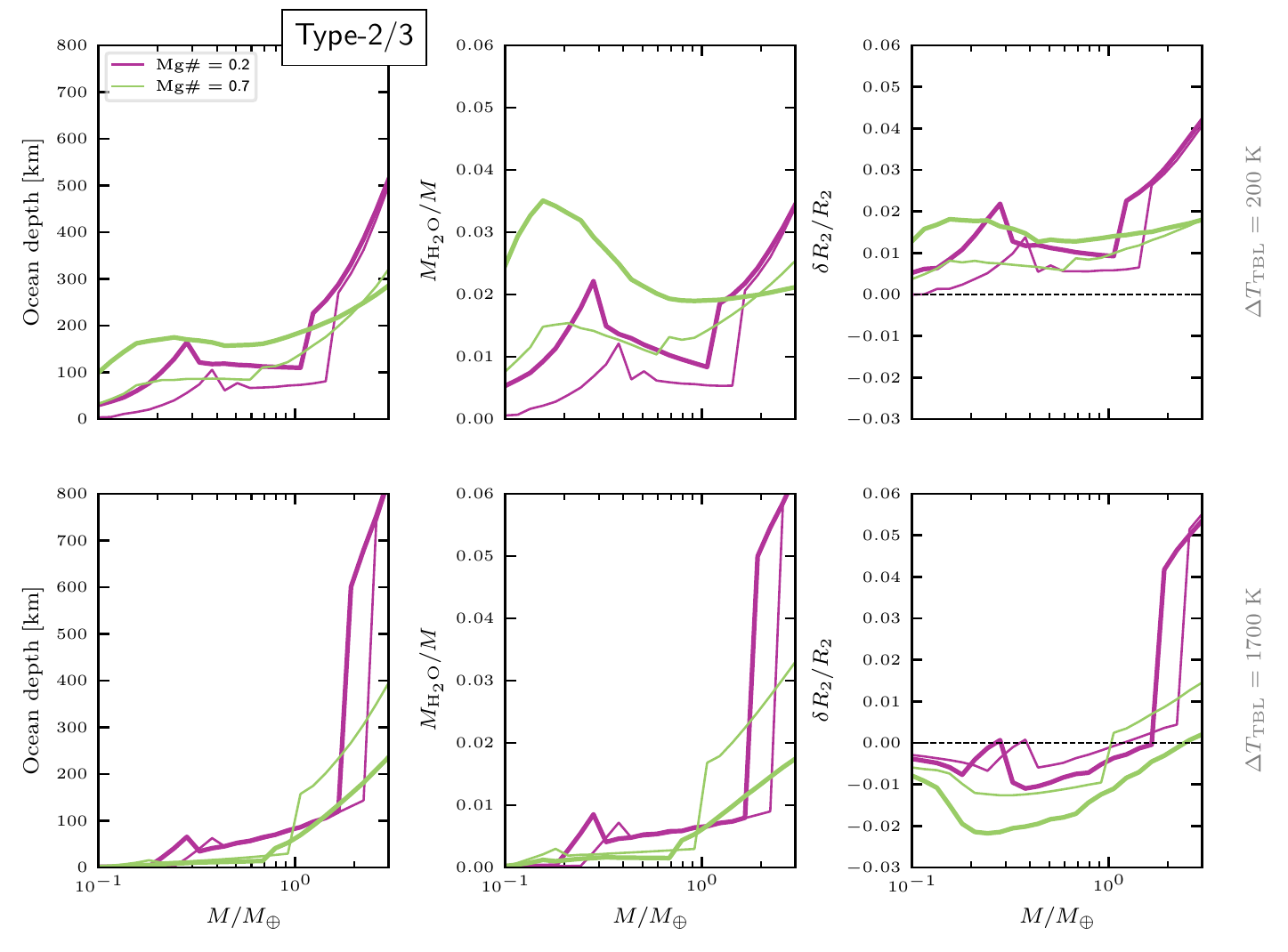}
\caption{Comparison of the nominal case (thick curves) and $\xi_{\rm Fe} = 0.0$ (thin curves) for the bracketing cases $\rm Mg \# = 0.2, 0.7$ and $\Delta T_{\rm TBL} = 200, 1700 \ \rm K$.}
\label{fig:Delta_R2_Fe_0}
\centering
\end{figure*}

Similarly to Figs.~\ref{fig:Delta_R1} and \ref{fig:Delta_R2} of the main text, Figs.~\ref{fig:Delta_R1_CMB_0_8} - \ref{fig:Delta_R2_Fe_0} show the comparisons between the Type-1 and Type-2 planets and the Type-2 and Type-3 planets. The thick and thin curves correspond to the nominal and comparison cases, respectively. We find that changing $\Delta T_{\rm CMB}$ by $\pm 20 \ \%$ has a rather minor effect on the overall results. 

The differences between the comparison and nominal cases are most prominent at higher masses where the core reservoir dominates. This is expected since for fixed surface conditions the temperature drop at the CMB predominantly affects the temperature in the core and not in the mantle. Therefore, the mantle reservoirs are rather  unaffected by $\Delta T_{\rm CMB}$. Furthermore, an increase or decrease in $\Delta T_{\rm CMB}$ leads to an increase or decrease in the hydrogen content in the core, which is consistent with eq. \ref{eq:Sievert}.

The effects of $\Delta T_{\rm CMB}$ on $\delta R_1/R_1$ are very weak while $\delta R_2/R_2$ is shifted to slightly smaller/larger values for smaller/larger values of $\Delta T_{\rm CMB}$ for high masses and low $\rm Mg \#$. This indicates that $\Delta T_{\rm CMB}$ plays an important role for iron-rich super-Earths. The iron content in the mantle $\xi_{\rm Fe}$ has a considerable effect on the main results, especially at lower masses, where the mantle reservoir dominates. This reflects the fact that the iron content in the mantle strongly influences the saturation water content in the upper mantle. If no iron is present in the mantle, the mantle reservoir is significantly decreased.

Furthermore, less iron in the mantle leads to larger cores at fixed bulk composition, which is reflected in slightly larger core reservoirs. In addition, the Pv-pPv transition is shifted to somewhat higher masses as the core to mantle ratio is enhanced for lower iron contents in the mantle. This is reflected in an offset of the core reservoirs between the nominal and the comparison case $\xi_{\rm Fe}=0$. The effects on $\delta R_1 / R_1$ and $\delta R_2/R_2$ are most prominent at lower masses and diminish at higher masses. It is still, however, unclear how reliably this behaviour reflects reality. This is because the water content in the lower mantle is intrinsically independent of $\xi_{\rm Fe}$ in our model because no sophisticated model for the hydration of pPv could be employed. In reality, the water storage capacity in pPv could, however, very well be influenced by the iron content, which would render our results more sensitive to $\xi_{\rm Fe}$ also at higher masses. These comparisons indicate that the iron content in the mantle is a crucial factor for determining internal H and OH reservoirs for low mass planets but  its role is somewhat unclear for super-Earths at this stage.

The opposite is true for $\Delta T_{\rm CMB}$, which mainly affects the core reservoirs of super-Earths. In the context of exoplanets most of the observed terrestrial planets are in the super-Earth regime. For these planets both $\Delta T_{\rm CMB}$ and $\xi_{\rm Fe}$ could play an important role for determining the internal H and OH reservoirs and corresponding effects of hydration and ocean differentiation on mass-radius relations. 

\section{Thermal equation of state}\label{sec:theos}
\begin{table*}
    \caption[]{\label{tab:coeffTable2} EoS parameters for for eq. \eqref{eq:BMEOS} collected from existing literature for $\rm (Mg,Fe)_2 SiO_4$. The coefficients have been scaled such that $X_{\rm H_2O}$ takes values between 0 and 1.}
    \begin{center}
        \begin{tabular}{ cccccccccccc }
         \hline
         phase & $K_0$ & $a_P$ & $a_T$ & $b_T$ & $c_T$ & $\partial a_T/\partial X_{\rm H_2O}$ & $\partial b_T/\partial X_{\rm H_2O}$ & $\partial c_T/\partial X_{\rm H_2O}$ & $\partial a_P / \partial X_{\rm H_2O}$ & $q$ & $\gamma_0$\\
          & $\rm [GPa]$ &  $\rm [GPa/K]$ & $\rm [K^{-1}]$ & $\rm [K^{-2}]$ & $\rm [K]$ & $\rm [K^{-1}]$ & $\rm [K^{-2}]$ & $\rm [K]$  & $\rm [GPa / K]$\\
         \hline
         $\alpha$ & $128.8^a$ & $-0.016^b$ & 2.19e-5$^d$ & 2.4e-8$^d$ & 0.0$^d$ & -4.6e-4$^{d*}$ & 0.9e-6$^{d*}$ & 0.0$^{d}$ & 0.0$^{**}$ & 2.9$^f$ & 1.26$^f$\\
         $\beta$ & $170.0^a$ &  $-0.018^b$ & 1.6e-5$^d$ & 2.2e-8$^d$ & 0.0$^d$ & -6.7e-4$^{d*}$ & 2.5e-6$^{d*}$ & 0.0$^{d}$ & 0.0$^{**}$ & 2.9$^f$ & 1.26$^f$\\
         $\gamma$ & $185.0^a$ & $-0.021^c$ & 2.54e-5$^e$ & 1.22e-8$^{e}$ & 0.0$^d$ & -9.7e-4$^{d*}$ & 2.6e-6$^{d *}$ & 0.0$^{d}$ & 0.0$^{**}$ & 2.9$^f$& 1.26$^f$
          \\ 
         \hline
        \end{tabular}
    \begin{tablenotes}
    \small
    \textit{a} fit parameters of \cite{Mao2016} \\
    \textit{b} extracted from Figure 2 of \cite{Valdez2013} \\
    \textit{c} widest temperature range in Table 3 of \cite{Mayama2005} \\
    \textit{d} Table 2 \& 3 of \cite{Ye2009} \\
    \textit{e} Katsura 2004 (larger $T$-range than \cite{Ye2009})\\
    \textit{f} \cite{Alibert2014} (Table 5, outer mantle)\\
    \textit{*} extra- or interpolated values \\
    \textit{**} No sources found \\
    \end{tablenotes}
    \end{center}
\end{table*}
\begin{table*}
    \caption[]{\label{tab:coeffTable3} Coefficients for density and bulk modulus of $\rm (Mg,Fe)_2 Si O_4$ as a function of the water and Fe content from \cite{Mao2016}. $\rm Fe \#$ and $X_{\rm H_2 O}$ both take values between 0 and 1.}
    \begin{center}
        \begin{tabular}{ ccccccccc }
         \hline
         phase & $d_1$ & $d_2$ & $d_3$ & $d_4$ & $k_1$ & $k_2$ & $k_3$ & $k_4$\\
         & $\rm [kg /m^{3}]$ & $\rm [kg /m^{3}]$ & $\rm [kg /m^{3}]$ & $\rm [kg /m^{3}]$ & $\rm [GPa]$ & $\rm[GPa]$ & $\rm[GPa]$ & $\rm[GPa]$\\
         \hline
         $\alpha$ & 3222 & 1200 & -4900 &  0 & 128.8 & 10 & -380 & -4000 \\
         $\beta$ & 3468 & 1300 & -4700 &  0 & 170 & 0 & -1240 & -5000 \\
         $\gamma$ & 3562 & 1500 & -4800 &  0 & 185 & 36 & -1180 & 0 
          \\ 
         \hline
        \end{tabular}
        
    \end{center}
\end{table*}
\begin{table*}
    \caption[]{\label{tab:coeffs_EoS} EoS parameters for the different materials.}
    \begin{center}
        \begin{tabular}{ ccccccccccccc }
         \hline
         material & abbrev. &  EoS & $\rho_0$ & $K_{T,0}$ & ${K^\prime}_{T,0}$ & $a_P$ & $a_T$ & $b_T$ & $c_T$ & $\gamma_0$ & $\theta_{D,0}$ & $q$\\
          & &  & $\rm [ kg \ m^{-3}]$ & [GPa] & & $\rm [GPa \ K^{-1}]$ & $\rm [K ^{-1}]$ & $ \rm [K^{-2}]$ & [K] & & [K] & \\
         \hline
         MgO & Per & MGD & 3584$^{d}$ & 157$^{d}$ & 4.4$^{d}$& ... & ... & ... & ... & 1.45$^{d}$ & 430$^{d}$ & 3$^{d}$\\
         $\rm Mg(OH)_2$ & Br & BM3 & 2323$^{c}$ & 43.4$^{c}$ & 5.4$^{c}$ & -0.015$^{b}$ & 7.3e-5$^{b}$ & 3.6e-8$^{b}$ & 0$^{b}$ & ... & ... & ...\\
         FeO & Ws & MGD & 5864$^{d}$ & 157$^{d}$ & 4.4$^{d}$ & ... & ... & ... & ... & 1.45$^{d}$ & 430$^{d}$ & 3$^{d}$\\
         $\rm (Mg, Fe) Si O_3$ & Pv & MGD & 4078$^{a}$ & 257$^{a}$ & 4.0$^{a}$ & ... &  ... & ... & ... & 1.54$^{a}$ & 950$^{a}$ & 1.5$^{a}$\\
         $\rm FeH_x$ & Fe & Bel & 8334$^{d}$ & 174$^{d}$ & 5.3$^{d}$ & ... & ... & ... & ... & 1.36$^{d}$ & 470$^{d}$ & 0.489$^{d}$\\
         $\rm (Mg,Fe)SiO_3$ & pPv & MGD & 4020$^{e}$ & 205.4$^{a}$ & 5.069$^{a}$ & ... & ... & ... & ... & 1.495$^{a}$ & 995$^{a}$ & 1.97$^{a}$ \\
         \hline
        \end{tabular}
    \begin{tablenotes}
    \small
    \textit{a} \cite{Sun2018} \\
    \textit{b} \cite{Xia1998} \\
    \textit{c} \cite{Hermann2016} \\
    \textit{d} \cite{Alibert2014} \\
    \textit{e} adjusted to match Pv-pPv density contrast from \cite{Dorfman2014} \\
    \end{tablenotes}        
    \end{center}
\end{table*}
\begin{table}
    \caption[]{\label{tab:coeffTable1} Fit coefficients for water saturation content in Olivine polymorphs.}
    \begin{center}
        \begin{tabular}{ ccccccc }
         \hline
         $i$ & $j$ & $k$ & &  $c_{ijk}$ & & unit\\
         \hline
          & & & $\alpha$ & $\beta$ & $\gamma$ &\\
        \hline
           0 & 0 & 0 & -30.59 & 1192 & -567.2 & $10^{-4}$\\
           0 & 0 & 1 & 35 & -1.75 & 23.03 &  $10^{-3}$\\
           0 & 1 & 0 & 12.89 & -40.53 & 79.37 & $\rm 10^{-4}GPa^{-1}$\\
           0 & 1 & 1 & 1.218 & 0.0 & 0.0 & $\rm 10^{-3}GPa^{-1}$\\
           1 & 0 & 0 & 19.53 & -539.3 & 274 & $\rm 10^{-7}K^{-1}$\\
           1 & 0 & 1 & -23.88 & 0.0 & 0.0 & $\rm 10^{-6}K^{-1}$\\
           1 & 1 & 0 & -7.539 & 18.92 & -40.7 & $\rm 10^{-7}(GPa \ K)^{-1}$\\
           1 & 1 & 1 & 0.0 & 0.0 & 0.0 & $\rm (GPa \ K)^{-1}$
         
          \\ 
         \hline
        \end{tabular}
    \end{center}
\end{table}
Here we describe in detail the equations of state used for the different materials in the core, mantle, and surface ocean. 
All results acquired in this study have been obtained using pre-generated EoS tables to optimize computational performance. We use piece-wise logarithmic spacing for both the temperature and the pressure axis, that is, we divided each decade into $10^n$ grid points where we fixed $n=1$ for all materials except for pure water where $n=2$ was used. During the structure integration the corresponding parameters (s.a. $\rho$, $K_T$ or $\alpha_{\rm th}$) at given $P$ and $T$ were extracted from the tables by bi-linear interpolation using the four adjacent grid points in the $P-T$-plane.
\subsection{Dry and hydrous $\rm (Mg,Fe)_2 Si O_4$}\label{sec:olivine}

\subsubsection{Phase transitions}\label{sec:phasetrans}
$\rm (Mg,Fe)_2 Si O_4$ exhibits three major polymorphic structures denoted by $\alpha$-Ol, $\beta$-Ol and $\gamma$-Ol and hereafter generally referred to as Olivine polymorphs. To model the phase transitions of these phases, we performed linear fits from a set of experimental data for the transition pressure as a function of temperature (see Table \ref{tab:phaseTrans}). The transition pressure is given as:
\begin{ceqn}
\begin{align}\label{eq:phaseTrans}
\begin{split}
P_{\rm trans}(T) = a_{\rm Trans} T + b_{\rm Trans}.
\end{split}
\end{align}
\end{ceqn}
The corresponding values for the fit coefficients $a$ and $b$ are given in Table \ref{tab:coeff4}. These transitions depend on the Fe content in Olivine polymorphs and can be shifted by up to $\approx$ 0.5-2 GPa for Fe contents up to 25 mol $\%$. The stability field of the $\gamma$-phase reaches up to at least $\approx$ 25 GPa and likely higher (\cite{Miyahara2011, Miyahara2016}) and depends on the temperature. For simplicity we omit the effect of the Fe content on the phase transitions and assume a temperature independent stability field for $\gamma$-Ol up to 30 GPa to ensure that our prescription yields a strict upper bound for the water content. In light of the other simplifying assumptions for the composition (Section \ref{sec:comp}) these approximations are justified. Furthermore, we point out that for super-Earths the mantles and hence the mantle reservoirs are dominated by the lower mantles. Therefore, the exact value for the dissociation pressure of Olivine becomes irrelevant for larger planetary masses.
\begin{table}
    \caption[]{\label{tab:coeff4} Fit coefficients for phase transitions in Olivine polymorphs and Perovskite.}
    \begin{center}
        \begin{tabular}{ ccc }
         \hline
         $a_{\rm Trans}$ & $b_{\rm Trans}$ & transition \\
         $\rm [GPa/K]$ & $ \rm [GPa]$ & \\
         \hline
         1.999e-3 & 1.159e1 & $\alpha-\beta$ \\
         5.679e-3 & 1.089e1 & $\beta-\gamma$ \\
         8.93e-3 & 1.01e2 & $\rm Pv-pPv$ \\
          \\ 
         \hline
        \end{tabular}
        
    \end{center}
\end{table}
\begin{figure}[t]
\centering
\includegraphics[scale=0.375]{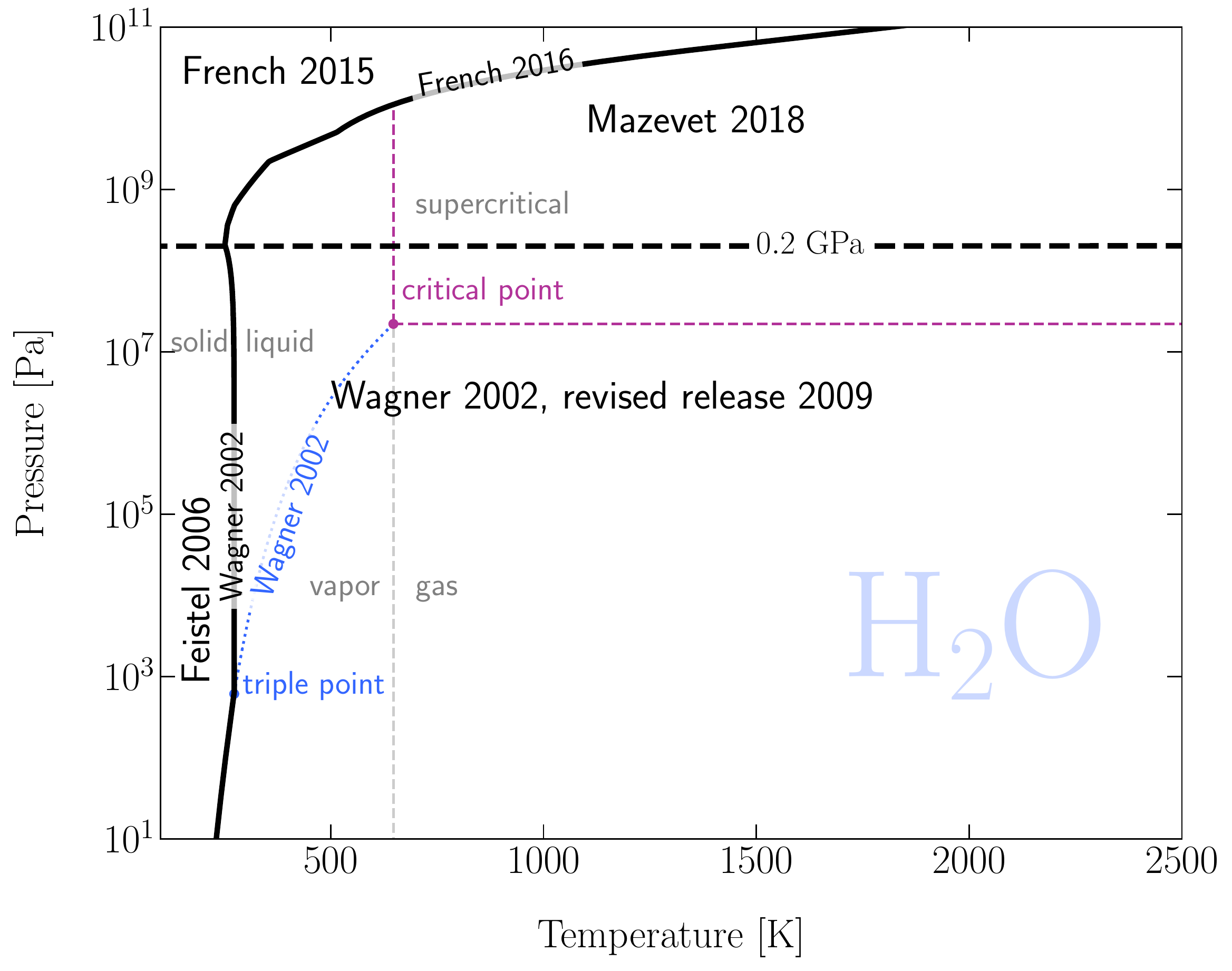}
\caption{Phase diagram of pure water showing the four distinct EoS regions used in this study (thick black curves). The black, solid curve is the solid-liquid phase boundary above the triple point and the sublimation curve below the triple point. The black, dashed horizontal line shows the pressure where we switch from the low pressure EoS to the high pressure one. The high pressure ices (\cite{French2015}) include ices VII and X. Other high pressure ice phases are not considered at this stage. The solid-liquid phase boundary is computed following \cite{Wagner2002} and \cite{French2016} for the low and high pressure regimes, respectively.}
\label{fig:water_phase_diagram}
\centering
\end{figure}
\begin{figure}[t]
\includegraphics[scale=0.84]{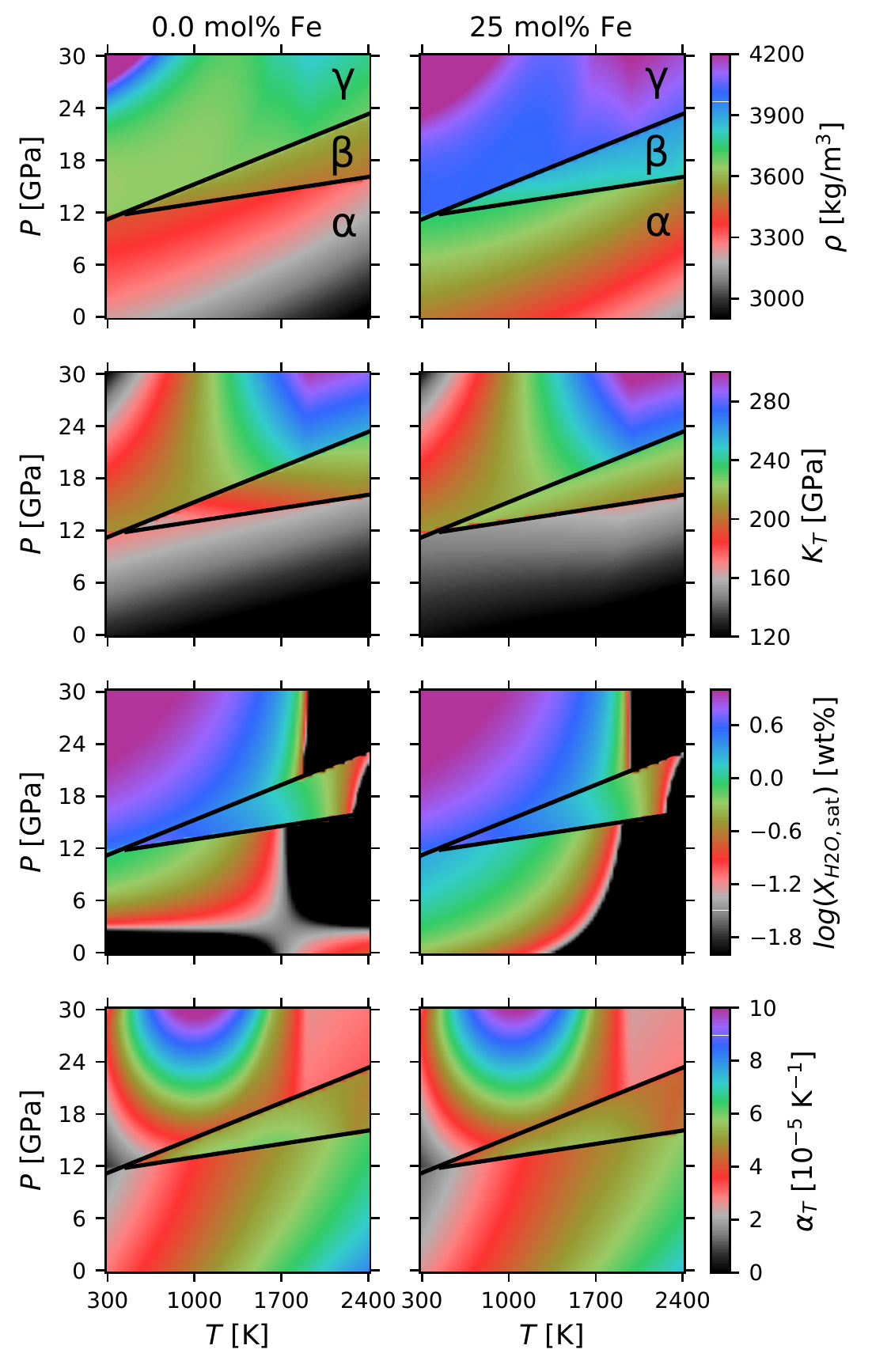}
\caption{Phase diagrams of $\rm (Mg, Fe)_2 Si O_4$. From top down: Density, bulk modulus, saturation water content, and thermal expansion coefficient as functions of temperature and pressure calculated from eq. \eqref{eq:bilinearmodel} \& \eqref{eq:BMEOS}. Shown are the Fe free case, that is $\xi_{\rm Fe}=0$ (left) and $\xi_{\rm Fe} = 0.25$ (right). The dependence of the phase transitions on the Fe and water content are neglected here. The pressure and temperature range in the plot roughly marks the validity range of the model fit. We note that we have assumed a stability field of the $\gamma$-phase of up to $\lesssim 30 \ \rm GPa$, which overestimates the dissociation pressure of Ringwoodite. We imposed this assumption to ensure that our model yields strict upper bounds on the water content.}
\label{fig:ParamMaps}
\centering
\end{figure}
\subsubsection{Thermoelastic properties}\label{sec:thermoelasticproperties}
In order to compute thermoelastic properties of water saturated Olivine polymorphs for different Fe contents $\xi_{\rm Fe}$ relevant for planetary interior modelling an adequate EoS is required. A large number of studies have investigated in depth the EoS parameters for the anhydrous case and some efforts have been made to prescribe hydration effects on these parameters (see references in Table \ref{tab:alpha}-\ref{tab:gamma}). The most commonly used analytical form of the EoS used in such studies is a third order Birch-Murnaghan equation. It relates the pressure $P$ to the temperature $T$ and the density $\rho$ according to:
\begin{ceqn}
\begin{align}\label{eq:BMEOS}
\begin{split}
P(\rho, T) = & \frac{3}{2} K_{T,0}(T) \left[ \left( \frac{\rho(T)}{\rho_{T,0}(T)} \right)^{7/3} -\left( \frac{\rho(T)}{\rho_{T,0}(T)} \right)^{5/3}\right] \\
& \times  \left( 1 - \frac{3}{4}(4-K_{T,0}'(T)) \left[ \left( \frac{\rho(T)}{\rho_{T,0}(T)}^{2/3}-1\right) \right]\right).
\end{split}
\end{align}
\end{ceqn}
$K_{T,0}$, $K_{T,0}'$, and $\rho_{T,0}$ are the reference values for the bulk modulus, pressure derivative of bulk modulus and density at zero pressure. The reference bulk modulus and the reference density are functions of temperature and given by:
\begin{ceqn}
\begin{align}
K_{T,0}(T) = K_0 + a_P(T-T_0)
\label{eq:bulk}
\end{align}
\end{ceqn}
\begin{ceqn}
\begin{equation}\label{eq:dens}
\rho_{T,0}(T) = \rho_0 \ \rm exp \mathit {\left( \int_{T0}^T \alpha_{T,0}(x) \rm d \mathit{x} \right)},
\end{equation}
\end{ceqn}
where $\rho_0$ and $K_0$ are the ambient density and the ambient bulk modulus (at $P_0$ = 0 bar and $T_0$ = 300 K). $\alpha_{T,0}$ is the isothermal expansion coefficient at zero pressure. It is expressed as:
\begin{ceqn}
\begin{align}\label{eq:alphaT}
\begin{split}
\alpha_{T,0}(T) = & \ a_T+\tfrac{\partial a_T}{\partial X_{\rm H_2O}} X_{\rm H_2O} \\ 
  +& \ (b_T+\tfrac{\partial b_T}{\partial X_{\rm H_2O}}X_{\rm H_2O}) T  \\
 -& \ (c_T+\tfrac{\partial c_T}{\partial X_{\rm H_2O}}X_{\rm H_2O}) T^{-2},
\end{split}
\end{align}
\end{ceqn}
where we have adopted linear correction terms to account for hydration. Here we neglect the effect of Fe content on $a_T, \ b_T$ and $c_T$. Furthermore we do not account for any effects of Fe and water content on $K_{T,0}'$. The reference parameters are computed following \cite{Mao2016} for the density and bulk modulus:
\begin{ceqn}
\begin{equation}\label{eq:rhoMao2016}
\rho_0(\xi_{\rm Fe}, X_{\rm H_2O}) = d_1 + d_2 \xi_{\rm Fe} + d_3 X_{\rm H_2O} + d_4 \rm{\xi_{\rm Fe}} \mathit{ X_{\rm H_2O}}
\end{equation}
\end{ceqn}
\begin{ceqn}
\begin{equation}\label{eq:KTMao2016}
K_0(\xi_{\rm Fe}, X_{\rm H_2O}) = k_1 + k_2 \xi_{\rm Fe} + k_3 X_{\rm H_2O} + k_4 \rm{\xi_{\rm Fe}} \mathit{ X_{\rm H_2O}}.
\end{equation}
\end{ceqn}
Here, $\rm \xi_{\rm Fe}$ and $X_{\rm H_2O}$ are to be inserted in mole and weight fraction, respectively. The parameter values for eq.\eqref{eq:bulk}, \eqref{eq:dens} and \eqref{eq:alphaT} used here are summarized in Table \ref{tab:coeffTable2} for the different phases along with the corresponding literature references. The coefficients for eq.~\eqref{eq:rhoMao2016} and \eqref{eq:KTMao2016} are listed in Table \ref{tab:coeffTable3}.
\\

All required parameters in eq.~\eqref{eq:BMEOS} can then be computed as functions of water and Fe content. The density, bulk modulus, thermal expansion coefficient, and water saturation content $X_{\rm H_2O, sat}$ are plotted for the relevant $P-T$ range in Fig.~\ref{fig:ParamMaps} for $\rm \xi_{\rm Fe} = 0$ (left panels) and for $\rm \xi_{\rm Fe} = 0.25$ (right panels). In Section \ref{sec:saturationmodel} we discuss how the saturation water content $X_{\rm H_2O, sat}(P,T)$ is computed.
\subsection{Hydration model for Olivine polymorphs}
The pressure in the compiled data set for water saturated Olivine polymorphs covers the entire stability range for each phase individually within the given temperature ranges. The latter is limited to the relevant regime for the Earth's upper mantle (up to $\gtrsim 2000 \rm \ K$). The colour maps in Fig.~\ref{fig:ParamMaps} provide an overview of the density, bulk modulus, water content and thermal expansion for the Fe free case and for $\xi_{\rm Fe}$ = 0.25. 
The overall effect of increasing temperature is a relatively strong decrease in the saturation water content $X_{\rm H_2O, sat}$ for all three phases. The water content reaches zero for $T < 2500 \rm \ K$ for all phases. This means that at higher temperatures the water content diminishes to zero and extrapolation to higher temperatures should therefore be unproblematic with regard to the hydration model. At low pressures $P \lesssim 5 \ \rm GPa$ and high temperatures $T \gtrsim 1500 \ \rm K$ the fit predicts an increase in water content with temperature. This behaviour might not be physical. We point out that the experimental data do not reach temperatures below $\approx 1000 \ \rm K$. This is because they have been performed in the context of the Earth's upper mantle for which temperatures between $\approx 1000-2000 \rm \ K$ are relevant. Therefore, to model low values of $\Delta T_{\rm TBL}$, it was inevitable to extrapolate to lower temperatures. It is, however, not clear how reliably the water content at such low temperatures is described by our model. 
Independent of the temperature, the model should be expected to yield unreliable results for Fe contents exceeding those covered by the data. Indeed, it can predict an increase in density upon hydration for $\rm \xi_{\rm Fe} \gtrsim 0.3$ in the $\alpha$-phase. At low temperatures ($T \lesssim 500 \ \rm K$), the model yields an increase in density upon hydration for the $\gamma-$phase above 25 GPa, which is clearly unrealistic. It is therefore strongly recommended to use the hydration model only for $0 \leq \xi_{\rm Fe} \lesssim 0.25$ in general and at elevated temperatures ($T > 500 \ \rm K$) if $P \gtrsim 25 \ \rm GPa$.

\subsection{Brucite}\label{sec:br}
For Brucite (Br) we use experimental fits for the 3rd order Birch-Murnaghan EoS (eq. \eqref{eq:BMEOS}-\eqref{eq:alphaT}). The correction terms $\partial{a_T}/\partial X_{\rm H_2O}$, $\partial{b_T}/\partial X_{\rm H_2O}$, and $\partial{c_T}/\partial X_{\rm H_2O}$ vanish in these cases. The EoS parameters used in this study along with the corresponding references are summarized in Table \ref{tab:coeffs_EoS}.

\subsection{Perovskite, Periclase, and Wüstite}\label{sec:pv_per_wüs}
For Perovskite (Pv), Periclase (Per), and Wüstite (Ws) we use experimental fits for the Mie-Gr{\"u}neisen-Debye EoS. In this case the pressure as a function of density and temperature is given as:%
\begin{ceqn}
\begin{align}\label{eq:MGDEOS}
P(\rho, T) =  P(\rho, T_0) + \Delta P(T).
\end{align}
\end{ceqn}
Here, $P(\rho, T_0)$ is the non-thermal pressure contribution. It is identical to the non-thermal contribution of the Belonoshko EoS:
\begin{ceqn}
\begin{align}\label{eq:MGDEOS_non_th}
\begin{split}
P(\rho, T) = & \frac{3}{2} K_{T,0}(T) \left[ \left( \frac{\rho(T)}{\rho_{T,0}(T)} \right)^{7/3} -\left( \frac{\rho(T)}{\rho_{T,0}(T)} \right)^{5/3}\right] \\
& \times  \left( 1 - \frac{3}{4}(4-K_{T,0}'(T)) \left[ \left( \frac{\rho(T)}{\rho_{T,0}(T)}^{2/3}-1\right) \right]\right).
\end{split}
\end{align}
\end{ceqn}
The $\Delta P$ denotes the thermal contribution to the pressure and is given as:
\begin{ceqn}
\begin{align}\label{eq:MGDEOS_th}
\Delta P = \gamma \rho(E(T) - E(T_0)),
\end{align}
\end{ceqn}
where:
\begin{ceqn}
\begin{align}\label{eq:MGDEOS_debye_integral}
E = \frac{9 n}{m} P \left( \frac{T}{\theta_D}\right)^3 \int_0^{\theta_D /T} \frac{x^3 e^x}{e^x-1} dx .
\end{align}
\end{ceqn}
The Gr{\"u}neisen parameter $\gamma(T,P)$ at elevated temperatures and pressures is computed via eq.~\ref{eq:gamma}. The Debye temperature is given as:
\begin{ceqn}
\begin{align}\label{eq:MGDEOS_theta_D}
\theta_D (T, P) = \theta_{D, 0} \left( \frac{\rho(T,P)}{\rho_0}\right)^\gamma.
\end{align}
\end{ceqn} 

The EoS parameters used in this study along with the corresponding references are summarized in Table \ref{tab:coeffs_EoS}. We model $\rm (Mg,Fe)O$ (Magnesiowüstite) in the lower mantle as a mixture of MgO and FeO using linear mixing (eq. \ref{eq:mixlaw}). For $\rm (Mg, Fe)SiO_3$ (Perovskite and post-Perovskite) we give the parameter values for the Fe free case in Table \ref{tab:coeffs_EoS}. The effect of the Fe content on the thermoelastic properties is modelled by adjusting $\rho_0$, $K_0$, $\theta_{D,0}$ and $\gamma_0$. For $\rho_0$ we taken the results for $3.8 \ \rm mol\%$ Fe and $6.1 \ \rm mol\%$ Fe from \cite{Sinmyo2014} for Pv. $K_0$, $\theta_{D,0}$, and $\gamma_0$ are adjusted using the values for 13 mol$\%$ Fe from \cite{Wolf2015} and the Fe free values from \cite{Sun2018}. The changes per $\rm mol \%$ Fe of these parameters are then $+31 \ \rm kg \ m^{-3}$, $-1.41 \rm \ GPa$, $+3.85 \ \rm K$, and $-0.0108$, respectively. For pPv, the effect of Fe has been investigated by \cite{Shieh2005}. However, comparing their results with the Fe free case from \cite{Sun2018} yields unrealistic results for Fe bearing pPv. For this reason we adopt the values from Pv above to model the Fe content in pPv. The transition pressure $P_{\rm trans}(T) = a_{\rm Trans}T + b_{\rm Trans}$ for $\rm Pv \rightarrow pPv$ has been taken from Fig. 3 of \cite{Townsend2016} (see also Table \ref{tab:coeff4}). \cite{Townsend2015} have found stable configurations of hydrous pPv containing at up to $2.31 \ \rm wt \%$ of water but have not investigated the saturation content as a function of pressure and temperature. Here we assume a constant maximum water content in pPv of $3 \ \rm wt \%$.
\subsection{Pure $\rm Fe$ and $\rm FeH_{x}$}\label{sec:iron}
For the iron in the cores of the modelled objects we proceed in analogy to \cite{Alibert2014} and use the EoS fit to molecular dynamics calculations from \cite{Belonoshko2010}. It has a similar form than the 3rd order Birch-Murnaghan EoS but instead of the temperature dependance of $K_0$, $K_0'$ and $\rho_0$ an additional term is introduced to account for the thermal pressure:
\begin{ceqn}
\begin{align}\label{eq:BelEOS}
\begin{split}
P(\rho, T) = & P_{th}(\rho, T)\\
& + \frac{3}{2} K_0 \left[ \left( \frac{\rho(T)}{\rho_0} \right)^{7/3} -\left( \frac{\rho(T)}{\rho_0} \right)^{5/3}\right] \\
& \times  \left( 1 - \frac{3}{4}(4-K_0') \left[ \left( \frac{\rho(T)}{\rho_0}^{2/3}-1\right) \right]\right),
\end{split}
\end{align}
\end{ceqn}
with
\begin{ceqn}
\begin{align}\label{eq:Pth}
P_{th}(\rho, T)= 3 R \gamma (T-T_0) m/\rho.
\end{align}
\end{ceqn}
 Here, $\gamma$ is the Grüneisen parameter, given by eq. \eqref{eq:gamma}, $R$ the gas constant and $m$ the molar mass of the modelled substance, here Fe. The EoS parameters are $T_0 = 300$ K, $K_0 = 174$ GPa, $K_0' = 5.3$, $\rho_0 = 8334 \ \rm kg / m^{3}$, $m=m_{Fe}=0.0558$ kg, $\gamma_0 = 2.434$ and $q = 0.489$. The effect of the hydrogen content in $\rm FeH_{\rm x}$ on the EoS parameters of pure Fe are taken from \cite{Thompson_2018}. For convenience we rewrite $\rm FeH_{\rm x} = \xi_{\rm Fe} Fe + (1-\xi_{\rm Fe} )H$ where $\xi_{\rm Fe} = 1-\xi_{\rm H}$ and $\xi_{\rm H} = \rm x/(1+x)$. We then express the change of density and bulk modulus as:
 \begin{ceqn}
\begin{align}\label{eq:drho_FeHx}
\rho_0(\xi_{\rm Fe}) = \rho_0(\xi_{\rm Fe}=1) + d (1-\xi_{\rm Fe})/\xi_{\rm Fe}
\end{align}
\end{ceqn}
\begin{ceqn}
\begin{align}\label{eq:KT0_FeHx}
K_0(\xi_{\rm Fe}) = K_0(\xi_{\rm Fe}=1) + k (1-\xi_{\rm Fe})/\xi_{\rm Fe}.
\end{align}
\end{ceqn}
To obtain the parameters $d$ and $k$ the change in density and bulk modulus between the pure Fe EoS and the values for $\rm FeH_{\rm x}$ from Table 2 in \cite{Thompson_2018} have been fitted as linear functions of the pressure and the hydrogen content. From this we extracted the values $d \approx -1620 \ \rm kg \ m^{-3}$ and $k \approx -40.2 \ \rm GPa$. With this simple model fit the densities in Table 2 of \cite{Thompson_2018} for low values of x, which are relevant for this study, can be reproduced with relative deviations of $\lesssim 1.2 \ \%$. The hydrogen content in the core itself, x, is obtained from chemical equilibrium between the core and the Mg-silicate mantle and is discussed in more detail in Section \ref{sec:core_hyd}.
\subsubsection{Pure water}
To model the most relevant phases of pure water in the surface oceans we employ different EoS for different $P-T$ regions. At pressures below $\approx 1$ GPa, the IAPWS formulation for liquid water (\cite{Wagner2002}) and water ice Ih (\cite{Feistel2006}) allow for as very reliable prescriptions of all relevant thermodynamic parameters. At higher pressures, where less or no experimental data are available, we switch to results obtained from molecular dynamics simulations. For the solid phase we use the thermodynamic potentials from \cite{French2015}. Their fit includes water ices VII and X, which are the most important high pressure water ice polymorphs. The transition from low pressure ice Ih to the high pressure ices occurs at $\approx 0.2 \rm \ GPa$ regardless of the temperature. Hence we switch from \cite{Wagner2002} to \cite{French2015} at 0.2 GPa. At high temperatures and pressures we switch to the results of \cite{Mazevet2018}. Their fit was performed such that it is consistent with the IAPWS at low pressures $P \lesssim 1 \rm \ GPa$. We switch from the IAPWS to Mazevet at 0.2 GPa. The transition between solid and liquid phases at all pressures are given by the water solidus. The solidus line is taken from \cite{Wagner2002} up to 1 GPa. For $P > 10 \rm \ GPa$ we take the results from \cite{French2016} and interpolate between 1-10 GPa. A summary of the phase diagram of pure water along with the EoS reference for each region used in this study is provided in Fig.~\ref{fig:water_phase_diagram}. For all phase regions the EoS is given in the form of a thermodynamic potential. This allows computing the adiabatic gradient directly from the EoS using:
\begin{ceqn}
\begin{align}\label{eq:adgrad_water}
\left( \frac{d T}{d P}\right)_{\rm ad} = \frac{P \alpha_{\rm th}}{c_{P} \rho},
\end{align}
\end{ceqn}
where $\rho$, $c_P$, $P$ and $\alpha_{\rm th}$ are the density, specific isobaric heat, pressure, and thermal expansion. The isobaric heat is computed from the entropy:
\begin{ceqn}
\begin{align}\label{eq:isobaric_heat}
c_P= T \left( \frac{\partial S}{\partial T}\right)_P.
\end{align}
\end{ceqn}
\section{Rough estimation of the ocean depth}\label{sec:estimate_ocean_depth}
\begin{figure}[t]
\includegraphics[scale=.6]{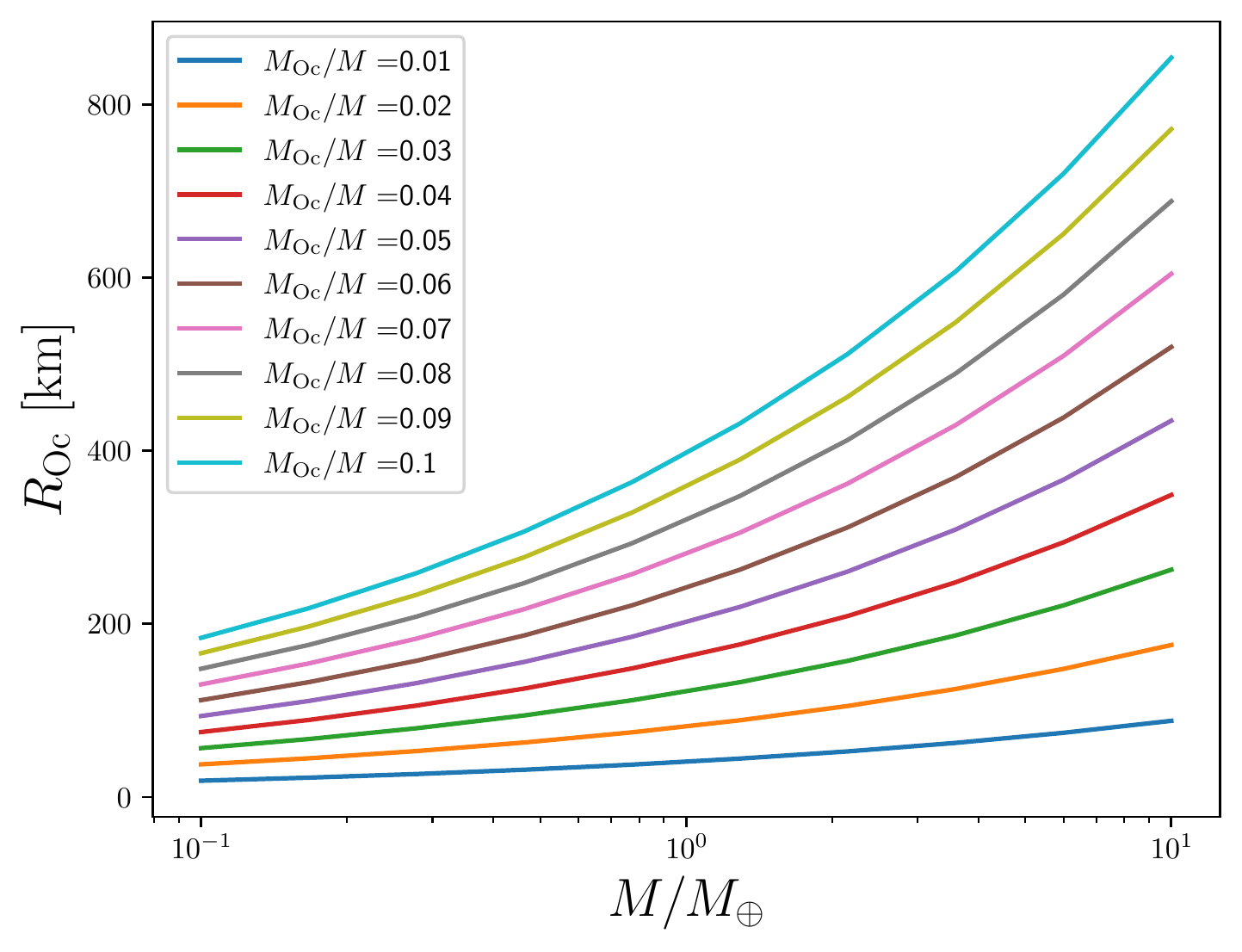}
\caption{Rough estimation of the ocean depth at given ocean mass fraction according to eq. \ref{eq:approx_ocean_depth}.}
\label{fig:estimate_R_Oc}
\centering
\end{figure}

Here we perform a very simple calculation to obtain a rough estimation on the behaviour of the ocean depth as a function of the total planetary mass for a fixed ocean mass fraction. Using the fact that for spherical objects we have $M \propto \left< \rho \right> R^3$, we express the ocean mass fraction $M_{\rm Oc}/M$ as:
\begin{ceqn}
\begin{align}\label{eq:ocean_mass_frac}
\frac{M_{\rm Oc}}{M} = \frac{[(R_0+R_{\rm Oc})^3-R_0^3]\left< \rho_{\rm Oc}\right>}{R_0 ^3 \left< \rho_0 \right>},
\end{align}
\end{ceqn}
where $R_0$ denotes the radius of the solid planet (iron core + silicate mantle) and $R_{\rm Oc}$ denotes the ocean depth as measured from $R_0$. Using the first order Taylor expansion $(1+q)^3 \approx 1 + 3q$ we can approximate $M_{\rm Oc}/M$ for $R_{\rm Oc}/R_0 << 1$ as:
\begin{ceqn}
\begin{align}\label{eq:approx_ocean_mass_frac}
\frac{M_{\rm Oc}}{M} \approx 3 \frac{R_{\rm Oc}}{R_0}\frac{\left< \rho_{\rm Oc} \right>}{\left< \rho_0 \right>},
\end{align}
\end{ceqn}
and hence:
\begin{ceqn}
\begin{align}\label{eq:approx_ocean_depth}
R_{\rm Oc}(M, M_{\rm Oc}) \approx \frac{1}{3} \left(\frac{3}{4 \pi} \left[ 1- \frac{M_{\rm Oc}}{M}\right]\frac{M}{\left< \rho_0 \right>}\right)^{1/3}  \frac{M_{\rm Oc}}{M}  \frac{\left< \rho_0\right>}{\left< \rho_{\rm Oc} \right>}.
\end{align}
\end{ceqn}
Eq. \ref{eq:approx_ocean_depth} is plotted for different ocean fractions $M_{\rm Oc}/M$ in Fig. \ref{fig:estimate_R_Oc} and assuming $\left< \rho_0 \right> = 5.5 \ \rm gcc$ and $\left< \rho_{\rm Oc} \right> = 1.0 \ \rm gcc$, which roughly corresponds to the values for an earth like planet. Of course, the densities would be strongly dependant on the bulk composition and the total mass of the planet. $\left< \rho_0 \right>$ and $\left< \rho_{\rm Oc} \right>$ could, in principle, be obtained from fits to our data. However, here we are merely interested in an order of magnitude estimation and the overall behaviour of the ocean depth as a function of the total mass. \ref{fig:estimate_R_Oc} shows a clear upwards trend of the ocean depth with total planetary mass at given ocean mass fraction. Hence, the fact that the same ocean mass fraction produces a deeper ocean for a higher total mass than for a lower total mass in Fig. \ref{fig:Delta_R2} can qualitatively be explained by eq. \ref{eq:approx_ocean_depth}.

\section{Sensitivity of $\rho$ to $\partial a_T/\partial X_{\rm H_2O}$ and $\partial b_T/\partial X_{\rm H_2O}$}\label{sec:sensitivity_aTbT}
To compute the isothermal expansion coefficient at zero pressure in eq. \eqref{eq:alphaT} we introduced linear correction terms to account for hydration effects. The corresponding coefficients $\partial a_T/\partial X_{\rm H_2O}$ and $\partial b_T/\partial X_{\rm H_2O}$ have been estimated by extrapolating existing data (\cite{Ye2009}). The uncertainty on these parameters are hence likely to be large and it is hence required to assess the sensitivity of our model to these parameters. We computed the effect on the density of saturated Olivine by varying $\partial a_T/\partial X_{\rm H_2O}$ and $\partial b_T/\partial X_{\rm H_2O}$ by $\pm 50 \%$. We find the density to be changed by up to $\approx4-6 \%$ for temperatures between $\approx 1000-1500  \ \rm K$ and $\approx 0-2 \%$ else. The enhanced sensitivity between $\approx 1000-1500\ \rm K$ comes from the fact that our hydration model is most sensitive to the water content in this regime. On one hand higher temperatures lead to lower water contents but at the same time stronger sensitivity to the water content according to eq. \eqref{eq:alphaT}. These two effects counteract each other and the resulting net effect is strongest between $\approx 1000-1500  \ \rm K$. This implies that more refined assessments of the hydration effects on Mg-silicates from experiments or first-principle simulations is crucial for obtaining reliable results from our model. However, for masses $M/M_\oplus \gtrsim 0.3$ the upper mantle contributes lesser to the overall structure the more massive the planets get and hence the sensitivity to $\partial a_T/\partial X_{\rm H_2O}$ and $\partial b_T/\partial X_{\rm H_2O}$ can be safely neglected for super-Earths. Furthermore, as for higher temperature profiles in the mantles, that is larger values of $\Delta T_{\rm TBL}$, the water content in the upper mantle diminishes the contribution of these parameters diminish.

\section{Data tables}

\begin{table}
    \caption[]{\label{tab:alpha} Saturation water content in $\rm \alpha-(Mg,Fe)_2 Si O_4$}
    \begin{center}
        \begin{tabular}{ ccccc }
         \hline
         $T$ & $P$ & $\xi_{\rm Fe}$ & $X_{\rm H_2O, sat}$ &
         Reference\\
          $[\rm{^\circ C}]$ & $[\rm{GPa}]$ &  & $[\rm{wt \%}]$ & \\
         \hline
          1100 & 2.5 & 0.0 & 0.0135 & \cite{Kohlstedt1996} \\ 
          1100 & 5.0 & 0.0 & 0.0496 & \cite{Kohlstedt1996} \\
          1100 & 6.5 & 0.0 & 0.026 & \cite{Kohlstedt1996} \\
          1100 & 8.0 & 0.0 & 0.0867 & \cite{Kohlstedt1996} \\
          1000 & 8.0 & 0.0 & 0.0886 & \cite{Kohlstedt1996} \\
          1100 & 8.0 & 0.0 & 0.0898 & \cite{Kohlstedt1996} \\
          1100 & 8.0 & 0.0 & 0.049 & \cite{Kohlstedt1996} \\
          1100 & 9.0 & 0.0 & 0.0984 & \cite{Kohlstedt1996} \\
          1100 & 10.0 & 0.0 & 0.107 & \cite{Kohlstedt1996} \\
          1100 & 12.0 & 0.0 & 0.151 & \cite{Kohlstedt1996} \\
          1100 & 13.0 & 0.0 & 0.109 & \cite{Kohlstedt1996} \\
          1100 & 13.7 & 0.0 & 0.76 & \cite{Chen2002} \\
          1000 & 6.0 & 0.0 & 0.0548 & \cite{Bali2008} \\
          1100 & 9.0 & 0.0 & 0.1223 & \cite{Bali2008} \\
          1100 & 2.5 & 0.0 & 0.0051 & \cite{Bali2008} \\
          1250 & 2.5 & 0.0 & 0.0051 & \cite{Bali2008} \\
          1325 & 2.5 & 0.0 & 0.0111 & \cite{Bali2008} \\
          1400 & 2.5 & 0.0 & 0.0173 & \cite{Bali2008} \\
          1100 & 6.0 & 0.0 & 0.1189 & \cite{Bali2008} \\
          1175 & 6.0 & 0.0 & 0.078 & \cite{Bali2008} \\
          1250 & 6.0 & 0.0 & 0.0607 & \cite{Bali2008} \\
          1325 & 6.0 & 0.0 & 0.0551 & \cite{Bali2008} \\
          1400 & 6.0 & 0.0 & 0.0489 & \cite{Bali2008} \\
          1175 & 9.0 & 0.0 & 0.1567 & \cite{Bali2008} \\
          1250 & 9.0 & 0.0 & 0.1984 & \cite{Bali2008} \\
          1325 & 9.0 & 0.0 & 0.1158 & \cite{Bali2008} \\
          1400 & 9.0 & 0.0 & 0.0386 & \cite{Bali2008} \\
          1400 & 5.0 & 0.043 & 0.0018 & \cite{Ferot2012} \\
          1400 & 13.5 & 0.062 & 0.168 & \cite{Ferot2012} \\
          1400 & 13.5 & 0.133 & 0.79 & \cite{Ferot2012} \\
          1175 & 2.5 & 0.074 & 0.0126 & \cite{Ferot2012} \\
          1250 & 2.5 & 0.074 & 0.0054 & \cite{Ferot2012} \\
          1325 & 2.5 & 0.068 & 0.0032 & \cite{Ferot2012} \\
          1175 & 2.5 & 0.083 & 0.0235 & \cite{Ferot2012} \\
          1250 & 2.5 & 0.081 & 0.0184 & \cite{Ferot2012} \\
          1325 & 2.5 & 0.112 & 0.0065 & \cite{Ferot2012} \\
          1400 & 2.5 & 0.061 & 0.00015 & \cite{Ferot2012} \\
          1175 & 5.0 & 0.071 & 0.0263 & \cite{Ferot2012} \\
          1250 & 5.0 & 0.078 & 0.0286 & \cite{Ferot2012} \\
          1250 & 5.0 & 0.08 & 0.0552 & \cite{Ferot2012} \\
          1325 & 5.0 & 0.09 & 0.0158 & \cite{Ferot2012} \\
          1400 & 5.0 & 0.053 & 0.0129 & \cite{Ferot2012} \\
          1175 & 7.5 & 0.073 & 0.2072 & \cite{Ferot2012} \\
          1250 & 7.5 & 0.077 & 0.1262 & \cite{Ferot2012} \\
          1325 & 7.5 & 0.066 & 0.0964 & \cite{Ferot2012} \\
          1400 & 7.5 & 0.073 & 0.083 & \cite{Ferot2012} \\
          1175 & 7.5 & 0.06 & 0.904 & \cite{Ferot2012} \\
          1250 & 7.5 & 0.087 & 0.0876 & \cite{Ferot2012} \\
          1325 & 7.5 & 0.097 & 0.0593 & \cite{Ferot2012} \\
          1400 & 7.5 & 0.128 & 0.0284 & \cite{Ferot2012} \\
          1175 & 9.0 & 0.094 & 0.142 & \cite{Ferot2012} \\
          1175 & 9.0 & 0.082 & 0.469 & \cite{Ferot2012} \\
          1250 & 9.0 & 0.083 & 0.2805 & \cite{Ferot2012} \\
          1400 & 9.0 & 0.068 & 0.0046 & \cite{Ferot2012} \\
          1000 & 1.5 & 0.0 & 0.00064 & \cite{Demouchy2003} \\
          1000 & 0.2 & 0.0 & 0.00028 & \cite{Demouchy2003} \\
          1060 & 0.2 & 0.0 & 0.00057 & \cite{Demouchy2003} \\
          1060 & 0.2 & 0.0 & 0.00055 & \cite{Demouchy2003} \\
          1100 & 0.2 & 0.0 & 0.00023 & \cite{Demouchy2003} \\
          \\ 
         \hline
        \end{tabular}
    \end{center}
\end{table}

\begin{table}
    Continuation of \textbf{Table \ref{tab:alpha}}
    \begin{center}
        \begin{tabular}{ ccccc }
         \hline
         $T$ & $P$ & $\xi_{\rm Fe}$ & $X_{\rm H_2O, sat}$ &
         Reference\\
          $[\rm{^\circ C}]$ & $[\rm{GPa}]$ &  & $[\rm{wt \%}]$ & \\
         \hline
          1110 & 0.2 & 0.0 & 0.00057 & \cite{Demouchy2003} \\
          1110 & 0.2 & 0.0 & 0.00036 & \cite{Demouchy2003} \\
          900 & 0.2 & 0.0 & 0.00013 & \cite{Demouchy2003} \\
          950 & 0.2 & 0.0 & 0.00042 & \cite{Demouchy2003} \\
          900 & 0.2 & 0.0 & 0.00038 & \cite{Demouchy2003} \\
          1000 & 0.3 & 0.091 & 0.0165 & \cite{Demouchy2010} \\
          1000 & 0.3 & 0.091 & 0.0053 & \cite{Demouchy2010}\\
          1100 & 0.3 & 0.091 & 0.0025 &\cite{Demouchy2010} \\
          1200 & 0.3 & 0.091 & 0.0071 & \cite{Demouchy2010}\\
          1600 & 6.3 & 0.0 & 0.0915 & \cite{Sokol2010}\\
          1600 & 6.3 & 0.0 & 0.095 & \cite{Sokol2010} \\
          1400 & 6.3 & 0.0 & 0.1 & \cite{Sokol2010}\\
          1200 & 6.3 & 0.0 & 0.104 & \cite{Sokol2010} \\
          1000 & 0.3 & 0.0 & 0.0112 & \cite{Zhao2004} \\
          1000 & 0.3 & 0.085 & 0.035 & \cite{Zhao2004} \\
          1000 & 0.3 & 0.12 & 0.0504 &\cite{Zhao2004} \\
          1000 & 0.3 & 0.149 & 0.0595 &\cite{Zhao2004} \\
          1000 & 0.3 & 0.153 & 0.0602 & \cite{Zhao2004} \\
          1000 & 0.3 & 0.168 & 0.0686 & \cite{Zhao2004} \\
          1000 & 0.3 & 0.169 & 0.0651 & \cite{Zhao2004} \\
          1050 & 0.3 & 0.0 & 0.0147 & \cite{Zhao2004} \\
          1050 & 0.3 & 0.085 & 0.0399 & \cite{Zhao2004} \\
          1050 & 0.3 & 0.12 & 0.0679 & \cite{Zhao2004} \\
          1050 & 0.3 & 0.149 & 0.0784 &\cite{Zhao2004} \\
          1050 & 0.3 & 0.153 & 0.0840 & \cite{Zhao2004} \\
          1050 & 0.3 & 0.168 & 0.0896 & \cite{Zhao2004} \\
          1050 & 0.3 & 0.169 & 0.0910 &\cite{Zhao2004} \\
          1100 & 0.3 & 0.0 & 0.0154 &\cite{Zhao2004} \\
          1100 & 0.3 & 0.085 & 0.0497 &\cite{Zhao2004}\\
          1150 & 0.3 & 0.12 & 0.0707 & \cite{Zhao2004} \\
          1100 & 0.3 & 0.149 & 0.0798 &\cite{Zhao2004} \\
          1100 & 0.3 & 0.153 & 0.0854 & \cite{Zhao2004} \\
          1100 & 0.3 & 0.168 & 0.1015 & \cite{Zhao2004} \\
          1150 & 0.3 & 0.0 & 0.0196 & \cite{Zhao2004} \\
          1150 & 0.3 & 0.085 & 0.0574 & \cite{Zhao2004} \\
          1150 & 0.3 & 0.12 & 0.0749 & \cite{Zhao2004} \\
          1150 & 0.3 & 0.149 & 0.105 & \cite{Zhao2004} \\
          1150 & 0.3 & 0.153 & 0.1036 &\cite{Zhao2004} \\
          1150 & 0.3 & 0.168 & 0.1253 &\cite{Zhao2004} \\
          1150 & 0.3 & 0.169 & 0.1204 & \cite{Zhao2004} \\
          1200 & 0.3 & 0.0 & 0.0203 &\cite{Zhao2004} \\
          1200 & 0.3 & 0.085 & 0.0686 & \cite{Zhao2004} \\
          1200 & 0.3 & 0.12 & 0.0924 & \cite{Zhao2004}\\
          1200 & 0.3 & 0.149 & 0.1078 & \cite{Zhao2004} \\
          1200 & 0.3 & 0.153 & 0.1141 & \cite{Zhao2004} \\
          1200 & 0.3 & 0.168 & 0.1379 & \cite{Zhao2004} \\
          1200 & 0.3 & 0.169 & 0.1428 & \cite{Zhao2004} \\
          1250 & 0.3 & 0.0 & 0.0245 &\cite{Zhao2004} \\
          1250 & 0.3 & 0.085 & 0.07 & \cite{Zhao2004} \\
          1250 & 0.3 & 0.12 & 0.1064 & \cite{Zhao2004} \\
          1250 & 0.3 & 0.149 & 0.1204 & \cite{Zhao2004} \\
          1250 & 0.3 & 0.153 & 0.1204 & \cite{Zhao2004} \\
          1250 & 0.3 & 0.168 & 0.1456 & \cite{Zhao2004} \\
          1250 & 0.3 & 0.169 & 0.1498 &\cite{Zhao2004} \\
          1300 & 0.3 & 0.0 & 0.0287 & \cite{Zhao2004}\\
          1300 & 0.3 & 0.085 & 0.0861 & \cite{Zhao2004} \\
          1300 & 0.3 & 0.12 & 0.1141 &\cite{Zhao2004} \\
          1300 & 0.3 & 0.149 & 0.1393 & \cite{Zhao2004} \\
          1300 & 0.3 & 0.153 & 0.1351 & \cite{Zhao2004}\\
          1300 & 0.3 & 0.168 & 0.1610 & \cite{Zhao2004} \\
          1300 & 0.3 & 0.169 & 0.1645 &\cite{Zhao2004} \\
          1100 & 12.95 & 0.03 & 0.29 & \cite{Chen2002} \\
          1100 & 13.0 & 0.041 & 0.37 & \cite{Chen2002} \\
          1100 & 13.05 & 0.037 & 0.71 & \cite{Chen2002}\\
          1100 & 13.1 & 0.032 & 0.64 & \cite{Chen2002} \\
          1100 & 13.1 & 0.029 & 0.55 & \cite{Chen2002} \\
          \end{tabular}
    \end{center}
\end{table}

\newpage
\begin{table}
    \caption[]{\label{tab:beta} Saturation water content in $\rm \beta-(Mg,Fe)_2 Si O_4$}
    \begin{center}
        \begin{tabular}{ ccccc }
         \hline
         $T$ & $P$ & $\xi_{\rm Fe}$ & $X_{\rm H_2O, sat}$ &
         Reference\\
          $[\rm{^\circ C}]$ & $[\rm{GPa}]$ &  & $[\rm{wt \%}]$ & \\
         \hline
          1100 & 12.5 & 0.118 & 2.03 &\cite{Litasov2008} \\ 
          1100 & 13.5 & 0.088 & 2.12 &\cite{Litasov2008} \\
          1200 & 12.5 & 0.118 & 1.65 & \cite{Litasov2008}\\
          1200 & 13.5 & 0.1 & 1.98 & \cite{Litasov2008} \\
          1400 & 14.5 & 0.133 & 1.6 & \cite{Litasov2008}\\
          1400 & 15.2 & 0.06 & 1.37 &\cite{Litasov2008} \\
          1400 & 17.0 & 0.059 & 0.63 & \cite{Litasov2008}\\
          1600 & 14.5 & 0.031 & 0.33 & \cite{Litasov2008}\\
          1600 & 15.2 & 0.04 & 0.31 & \cite{Litasov2008}\\
          1800 & 20.0 & 0.07 & 0.15 & \cite{Litasov2008} \\
          1800 & 16.0 & 0.058 & 0.18 & \cite{Litasov2008}\\
          1900 & 20.0 & 0.055 & 0.12 &\cite{Litasov2008}\\
          2000 & 16.5 & 0.051 & 0.07 & \cite{Litasov2008}\\
          1200 & 17.0 & 0.08 & 0.24 & \cite{Mrosko2015} \\
          1200 & 17.9 & 0.04 & 1.8 & \cite{Mrosko2015}\\ 
          1200 & 16.8 & 0.12 & 1.9 & \cite{Mrosko2015} \\
          1200 & 16.2 & 0.14 & 1.2 & \cite{Mrosko2015} \\
          1200 & 17.3 & 0.11 & 1.7 & \cite{Mrosko2015} \\
          1200 & 16.8 & 0.11 & 1.0 & \cite{Mrosko2015} \\
          1200 & 19.0 & 0.04 & 1.3 & \cite{Mrosko2015} \\
          1200 & 15.2 & 0.22 & 1.4 & \cite{Mrosko2015} \\
          1200 & 17.9 & 0.11 & 1.6 & \cite{Mrosko2015} \\
          1400 & 17.0 & 0.145 & 3.72 & \cite{Inoue2010}\\
          1400 & 16.5 & 0.139 & 2.28 & \cite{Inoue2010} \\
          1400 & 16.5 & 0.139 & 2.24 &\cite{Inoue2010} \\
          1400 & 16.5 & 0.14 & 1.88 & \cite{Inoue2010} \\
          1400 & 16.5 & 0.14 & 1.79 & \cite{Demouchy2005} \\
          1000 & 15.0 & 0.0 & 2.13 &\cite{Demouchy2005} \\
          1100 & 15.0 & 0.0 & 2.41 & \cite{Demouchy2005} \\
          1100 & 15.0 & 0.0 & 2.42 & \cite{Demouchy2005}\\
          1200 & 15.0 & 0.0 & 2.24 & \cite{Demouchy2005} \\
          1300 & 15.0 & 0.0 & 1.66 & \cite{Demouchy2005} \\
          1400 & 15.0 & 0.0 & 0.93 & \cite{Demouchy2005} \\
          1200 & 14.0 & 0.0 & 2.4 & \cite{Demouchy2005} \\
          1200 & 14.0 & 0.0 & 2.68 & \cite{Demouchy2005} \\
          1200 & 16.0 & 0.0 & 2.6 & \cite{Demouchy2005} \\
          1200 & 17.0 & 0.0 & 2.43 & \cite{Demouchy2005} \\
          1200 & 18.0 & 0.0 & 1.24 & \cite{Demouchy2005} \\
          1500 & 15.5 & 0.0 & 1.52 & \cite{Kawamoto1996} \\
          1360 & 15.5 & 0.0 & 3.13 & \cite{Kawamoto1996} \\
          1450 & 15.5 & 0.0 & 1.95 & \cite{Kawamoto1996}\\
          1425 & 13.5 & 0.0 & 2.89 & \cite{Kawamoto1996} \\
          1300 & 15.5 & 0.0 & 1.83 & \cite{Kawamoto1996} \\
          1400 & 16.5 & 0.0 & 2.61 & \cite{Kawamoto1996} \\
          1100 & 15.0 & 0.0 & 2.23 & \cite{Kohlstedt1996} \\
          1100 & 15.0 & 0.0 & 2.41 & \cite{Kohlstedt1996} \\
          1100 & 14.0 & 0.0 & 2.13 & \cite{Kohlstedt1996} \\
          1100 & 14.2 & 0.0 & 3.8 & \cite{Chen2002} \\
          1100 & 14.7 & 0.0 & 3.5 & \cite{Chen2002} \\
          1100 & 13.0 & 0.0 & 1.9 &\cite{Chen2002} \\
          1100 & 13.05 & 0.0 & 3.4 & \cite{Chen2002} \\
          1100 & 13.1 & 0.0 & 3.4 & \cite{Chen2002} \\
          1100 & 13.1 & 0.0 & 2.5 &\cite{Chen2002} \\
          1100 & 13.25 & 0.0 & 3.0 & \cite{Chen2002}\\
          1100 & 13.6 & 0.0 & 2.6 & \cite{Chen2002} \\
          1100 & 13.7 & 0.0 & 2.8 & \cite{Chen2002} \\
          1100 & 13.9 & 0.0 & 2.2 & \cite{Chen2002} \\
          1100 & 14.4 & 0.0 & 2.6 & \cite{Chen2002} \\
          \\ 
         \hline
        \end{tabular}
    \end{center}
\end{table}

\begin{table}
    Continuation of \textbf{Table \ref{tab:beta}}
    \begin{center}
        \begin{tabular}{ ccccc }
         \hline
         $T$ & $P$ & $\xi_{\rm Fe}$ & $X_{\rm H_2O, sat}$ &
         Reference\\
          $[\rm{^\circ C}]$ & $[\rm{GPa}]$ &  & $[\rm{wt \%}]$ & \\
         \hline
          2100 & 18.0 & 0.0 & 0.005 & \cite{Jacobsen2005} \\
          1400 & 17.0 & 0.0 & 0.015 & \cite{Jacobsen2005} \\
          1200 & 16.0 & 0.0 & 0.32 & \cite{Jacobsen2005} \\
          1300 & 16.0 & 0.0 & 0.6 & \cite{Jacobsen2005} \\
          1400 & 16.0 & 0.0 & 0.96 & \cite{Jacobsen2005} \\
          1400 & 16.0 & 0.0 & 1.06 & \cite{Jacobsen2005} \\
          1200 & 13.3 & 0.0 & 0.8 & \cite{Deon2010} \\
          1150 & 13.5 & 0.0 & 1.6 & \cite{Deon2010} \\
          1200 & 13.5 & 0.0 & 3.8 & \cite{Chen1998} \\
          1300 & 15.0 & 0.0 & 0.2212 & \cite{Casanova2000} \\
          1300 & 14.0 & 0.0 & 1.66 & \cite{Holl2008} \\
          1300 & 15.5 & 0.0 & 2.5 & \cite{Yusa1997} \\
          \end{tabular}
    \end{center}
\end{table}

\newpage
\begin{table}
    \caption[]{\label{tab:gamma} Saturation water content in $\rm \gamma-(Mg,Fe)_2 Si O_4$}
    \begin{center}
        \begin{tabular}{ ccccc }
         \hline
         $T$ & $P$ & $\xi_{\rm Fe}$ & $X_{\rm H_2O, sat}$ &
         Reference\\
          $[\rm{^\circ C}]$ & $[\rm{GPa}]$ &  & $[\rm{wt \%}]$ & \\
         \hline
          1400 & 17.0 & 0.224 & 1.69 & \cite{Inoue2010} \\
          1400 & 16.5 & 0.213 & 1.11 & \cite{Inoue2010} \\
          1400 & 16.5 & 0.213 & 1.25 & \cite{Inoue2010} \\
          1400 & 16.5 & 0.208 & 1.0 & \cite{Inoue2010} \\
          1400 & 16.5 & 0.208 & 1.1 & \cite{Inoue2010} \\
          1600 & 23.0 & 0.159 & 0.76 & \cite{Inoue2010} \\
          1600 & 32.1 & 0.135 & 0.71 & \cite{Inoue2010} \\
          1600 & 23.2 & 0.087 & 0.63 & \cite{Inoue2010} \\
          1300 & 19.0 & 0.0 & 2.2 & \cite{Inoue1998} \\
          1100 & 19.5 & 0.0 & 2.62 & \cite{Kohlstedt1996} \\
          1200 & 19.0 & 0.0 & 0.7817 & \cite{Casanova2000} \\
          1300 & 19.0 & 0.0 & 0.75 & \cite{Casanova2000} \\
          1250 & 20.0 & 0.0 & 1.77 & \cite{Thomas2015} \\
          1250 & 20.0 & 0.0 & 1.65 & \cite{Thomas2015} \\
          1250 & 20.0 & 0.0 & 2.5 & \cite{Thomas2015} \\
          1400 & 20.0 & 0.0 & 1.13 & \cite{Thomas2015} \\
          1400 & 20.0 & 0.0 & 0.85 & \cite{Thomas2015} \\
          1400 & 18.0 & 0.0 & 0.94 & \cite{Thomas2015} \\
          1400 & 18.0 & 0.0 & 1.11 & \cite{Thomas2015} \\
          1400 & 20.0 & 0.0 & 0.7892 & \cite{Smyth2003} \\
          1400 & 20.0 & 0.0 & 0.7892 & \cite{Smyth2003} \\
          1500 & 21.0 & 0.0 & 0.1988 & \cite{Smyth2003} \\
          1500 & 22.0 & 0.0 & 0.7358 & \cite{Smyth2003}\\
          1500 & 21.5 & 0.0 & 0.7354 & \cite{Smyth2003} \\
          1400 & 19.0 & 0.0 & 0.8557 & \cite{Smyth2003} \\
          1400 & 18.0 & 0.0 & 1.0661 & \cite{Smyth2003} \\
          1300 & 19.0 & 0.0 & 2.8 & \cite{Yusa2000} \\
          1300 & 19.0 & 0.0 & 2.2 & \cite{Chen1998} \\
          1200 & 17.0 & 0.15 & 0.14 & \cite{Mrosko2015} \\
          1200 & 17.9 & 0.13 & 0.7 & \cite{Mrosko2015} \\
          1200 & 16.8 & 0.18 & 1.2 & \cite{Mrosko2015}\\
          1200 & 17.3 & 0.18 & 0.7 & \cite{Mrosko2015} \\
          1200 & 16.8 & 0.17 & 0.5 & \cite{Mrosko2015} \\
          1200 & 15.2 & 0.31 & 0.7 & \cite{Mrosko2015} \\
          1200 & 17.9 & 0.17 & 0.6 & \cite{Mrosko2015} \\
          \\ 
         \hline
        \end{tabular}
    \end{center}
\end{table}
\begin{table}
    \caption[]{\label{tab:phaseTrans} Phase transition data for anhydrous $\rm Mg_2 Si O_4$}
    \begin{center}
        \begin{tabular}{ cccc }
         \hline
         $T \ [\rm{ ^\circ C}]$ & $P \ [\rm{GPa}]$ & transition & 
         Reference\\
         \hline
          1600 & 15.0  & $\alpha-\beta$ & \cite{Katsura1989}  \\
          1200 & 14.0 & $\alpha-\beta$  & \cite{Katsura1989} \\
          1100 & 14.88 & $\alpha-\beta$ & \cite{Suzuki2000} \\
          100 & 14.87 & $\alpha-\beta$ & \cite{Suzuki2000} \\
          900 & 14.95 & $\alpha-\beta$ & \cite{Suzuki2000} \\
          900 & 14.88 & $\alpha-\beta$ & \cite{Suzuki2000} \\
          875 & 15.01 & $\alpha-\beta$ & \cite{Suzuki2000} \\
          875 & 14.92 & $\alpha-\beta$ & \cite{Suzuki2000} \\
          985 & 15.88 & $\alpha-\beta$ & \cite{Suzuki2000} \\
          985 & 15.83 & $\alpha-\beta$ & \cite{Suzuki2000} \\
          750 & 15.32 & $\beta-\gamma$ & \cite{Suzuki2000} \\
          750 & 15.2 & $\beta-\gamma$ & \cite{Suzuki2000} \\
          1100 & 15.83 & $\beta-\gamma$ & \cite{Suzuki2000} \\
          1100 & 15.82 & $\beta-\gamma$ & \cite{Suzuki2000} \\
          800 & 16.3 & $\beta-\gamma$ & \cite{Suzuki2000} \\
          800 & 16.2 & $\beta-\gamma$ & \cite{Suzuki2000} \\
          1200 & 16.59 & $\beta-\gamma$ & \cite{Suzuki2000} \\
          1200 & 16.62 & $\beta-\gamma$ & \cite{Suzuki2000} \\
          650 & 14.98 & $\alpha-\gamma$ & \cite{Suzuki2000} \\
          650 & 14.83 & $\alpha-\gamma$ & \cite{Suzuki2000} \\
          650 & 15.32 & $\alpha-\gamma$ & \cite{Suzuki2000} \\
          650 & 15.16 & $\alpha-\gamma$ & \cite{Suzuki2000} \\
          800 & 16.3 & $\alpha-\gamma$ & \cite{Suzuki2000} \\
          800 & 16.18 & $\alpha-\gamma$ & \cite{Suzuki2000} \\
          700 & 16.37 & $\alpha-\gamma$ & \cite{Suzuki2000} \\
          700 & 16.2 & $\alpha-\gamma$ & \cite{Suzuki2000} \\
          900 & 17.26 & $\alpha-\gamma$ & \cite{Suzuki2000} \\
          900 & 17.18 & $\alpha-\gamma$ & \cite{Suzuki2000} \\
          1000 & 17.37 & $\alpha-\gamma$ & \cite{Suzuki2000} \\
          1000 & 17.32 & $\alpha-\gamma$ & \cite{Suzuki2000} \\
          930 & 18.91 & $\alpha-\gamma$ & \cite{Suzuki2000} \\
          930 & 18.85 & $\alpha-\gamma$ & \cite{Suzuki2000} \\
          700 & 19.5 & $\alpha-\gamma$ & \cite{Suzuki2000} \\
          700 & 19.34 & $\alpha-\gamma$ & \cite{Suzuki2000} \\
          1600 & 14.2 & $\alpha-\beta$ & \cite{Katsura2004} \\
          1900 & 15.4 & $\alpha-\beta$ & \cite{Katsura2004} \\
          1400 & 14.8  & $\alpha-\beta$ & \cite{Wu2012} \\
          1400 & 19.5  & $\beta-\gamma$ & \cite{Wu2012} \\ 
          1600 & 14.5  & $\beta-\gamma$ & \cite{Katsura1989} \\
          1600 & 19.2  & $\beta-\gamma$ & \cite{Katsura1989} \\
          1600 & 16.5 & $\beta-\gamma$ & \cite{Katsura1989} \\
          1600 & 13.2  & $\beta-\gamma$ & \cite{Katsura1989} \\
          1200 & 13.9  & $\beta-\gamma$ & \cite{Katsura1989} \\
          1200 & 12.6  & $\beta-\gamma$ & \cite{Katsura1989} \\
          1200 & 17.3  & $\beta-\gamma$ & \cite{Katsura1989} \\
          1200 & 18.6  & $\beta-\gamma$ & \cite{Katsura1989} \\
          1600 & 21.0  & $\beta-\gamma$ & \cite{Katsura1989} \\
          1200 & 19.0  & $\beta-\gamma$ & \cite{Katsura1989} \\
          1600 & 13.5  & $\beta-\gamma$ & \cite{Katsura1989} \\

          \\ 
         \hline
        \end{tabular}
    \end{center}
\end{table}
\newpage

\typeout{get arXiv to do 4 passes: Label(s) may have changed. Rerun}
\typeout{get arXiv to do 4 passes: Label(s) may have changed. Rerun}
\end{document}